\documentclass[11pt,journal,onecolumn]{IEEEtran}

\makeatletter
\renewcommand{\documentclass}[2][]{}
\makeatother

\def\ARXIVVERSION{1}
\documentclass[journal]{IEEEtran}

\usepackage{verbatim}
\usepackage{graphicx}
\usepackage[cmex10]{amsmath}
\usepackage{amssymb,amsfonts}
\usepackage{multirow}
\usepackage{array}
\usepackage{theorem}
\usepackage{cite}
\usepackage{makecell}
\usepackage{tabularray}
\usepackage[dvipsnames]{xcolor}
\usepackage{colortbl}
\UseTblrLibrary{booktabs} 

\newcommand\MYhyperrefoptions
{
	bookmarks=false, 
	bookmarksnumbered=true,
	pdfpagemode={UseOutlines},
	plainpages=false,
	pdfpagelabels=true,
	pdftitle={A Survey of Decentralized Physical Infrastructure Network, Research Directions, and Open Challenges},
	pdfsubject={Accepted manuscript for IEEE Communications Surveys and Tutorials; DOI: 10.1109/COMST.2026.3729997},
	pdfauthor={Ming Jiang, Erwu Liu, Xinyu Qu, Wei Ni, Zhongxiang Wei, and Ekram Hossain},
	pdfcreator={LaTeX},
}

\usepackage[\MYhyperrefoptions,breaklinks=true,colorlinks=true, allcolors=blue]{hyperref}
\usepackage{algorithm}
\usepackage{algorithmic}
\usepackage{caption}

\usepackage{bbm}
\usepackage{bm}

\newcommand{\greencheck}{\textcolor{green!70!black}{\scalebox{1.2}{$\checkmark$}}} 
\newcommand{\redcross}{\textcolor{red!70!black}{\scalebox{1.2}{$\times$}}} 
\newcommand{\bluecircle}{\textcolor{SkyBlue}{\scalebox{1.8}{$\circ$}}} 
\newcommand{\orangewave}{\textcolor{orange!85!black}{\scalebox{1.25}{$\bm{\sim}$}}} 
\newcommand{\redcircle}{\textcolor{red!70!black}{\scalebox{1.8}{$\circ$}}} 

\begin{document}

\title{
    A Survey of Decentralized Physical Infrastructure Network,
    Research Directions, and Open Challenges}

\author{
    Ming~Jiang,
    Erwu~Liu,
    Xinyu~Qu,
    Wei~Ni,
    Zhongxiang~Wei,
    and
    Ekram Hossain,

    \thanks{ 
    Corresponding author: Erwu Liu.}
    \thanks{M. Jiang is with the Shanghai Research Institute for Intelligent Autonomous Systems, Tongji University, Shanghai 201210, China (E-mail: jiangm\_leo@tongji.edu.cn).}
    \thanks{E. Liu and Z. Wei are with the College of Electronics and Information Engineering, Tongji University, Shanghai 201804, China (E-mail: erwu.liu@ieee.org, z\_wei@tongji.edu.cn).}
    \thanks{X. Qu is with the College of Computer Science and Technology, Tongji University, Shanghai 201804, China (E-mail: xinyuqu@tongji.edu.cn).}
    \thanks{W. Ni is with the School of Engineering, Edith Cowan University, Perth, WA 6027
    Australia (E-mail: wei.ni@ieee.org).}
    \thanks{E. Hossain is with the University of Manitoba, Winnipeg, MB R3T 2N2, Canada (E-mail: ekram.hossain@umanitoba.ca).}

}

\maketitle

\ifdefined\ARXIVVERSION
\begin{center}
\begin{minipage}{0.96\linewidth}
\footnotesize

Author-prepared preprint based on a manuscript accepted by \textit{IEEE Communications Surveys \& Tutorials}, 
doi: \href{https://doi.org/10.1109/COMST.2026.3729997}{10.1109/COMST.2026.3729997}.

\medskip
\end{minipage}
\end{center}
\fi

\begin{abstract}
    The Decentralized Physical Infrastructure Network (DePIN) represents a transformative paradigm that redefines the construction, operation, and governance of Information and Communication Technology (ICT) infrastructure in the Web 3.0 era.
    DePIN integrates physical resources, such as networking equipment, storage, and computing power, with decentralized digital governance, forming a self-incentivized ecosystem that is collaboratively built, shared, and governed by the community.
    It provides a foundational framework for future communication networks, facilitating decentralized edge intelligence, efficient resource sharing, and trustworthy coordination among heterogeneous devices.
    Focusing on the feasibility of this emerging paradigm,
    this paper examines the technology landscape in the pre-DePIN era and gaps between existing methodologies and the forthcoming decentralized infrastructure for Web 3.0.
    It provides a systematic and comprehensive survey of the background, core characteristics, technical architecture, and applications of DePIN across various vertical domains.
    The paper analyzes the DePIN technology stack from six layers: physical infrastructure, blockchain, interaction, trust, incentive, and application, 
    with special attention to their cross-layer feedback loops, implementation readiness, and deployment limitations. 
    To further bridge conceptual analysis and practical deployment, we propose a DePIN feasibility assessment framework covering technical, governance, and economic dimensions. Moreover, we highlight promising research directions, providing insights and guidance for further exploration and deployment of DePIN.

\end{abstract}

\begin{IEEEkeywords}
    DePIN, ICT Infrastructure, Decentralization, Feasibility 
\end{IEEEkeywords}

\section{Introduction}\label{sect:intro}
\subsection{Background and Definition}
\IEEEPARstart{N}{owadays}, key infrastructures, including telecommunications, energy, storage, and computing, are predominantly controlled by enterprises, often subject to oligopolistic dominance.
These dominant players may exploit user data through opaque practices to maximize profits, a problem particularly pronounced in emerging markets \cite{iotex_whitepaper_TS}.
While Software-Defined Networking (SDN) and Open Radio Access Network (O-RAN) have made some attempts to drive architectural openness and flexibility, the Information and Communication Technology (ICT) sector remains highly centralized and lacks dynamism.
This suggests that technological innovations focusing merely on virtualized and open architectures may not be sufficient to catalyze the paradigm shift from centralized control to decentralized, community-driven infrastructure.
Broad participation, properly incentivized through mindset, behaviour, and (financial) incentives, could be the missing part of the shift towards a more open, fair, flat network paradigm.
A new framework is needed to rethink the management of physical infrastructure and public utilities from a system perspective, integrating decentralized governance, incentive mechanisms, and cross-domain resource orchestration to achieve greater efficiency, transparency, and resilience.

\begin{figure}[!t]
    \centering
    \includegraphics[scale=0.3]{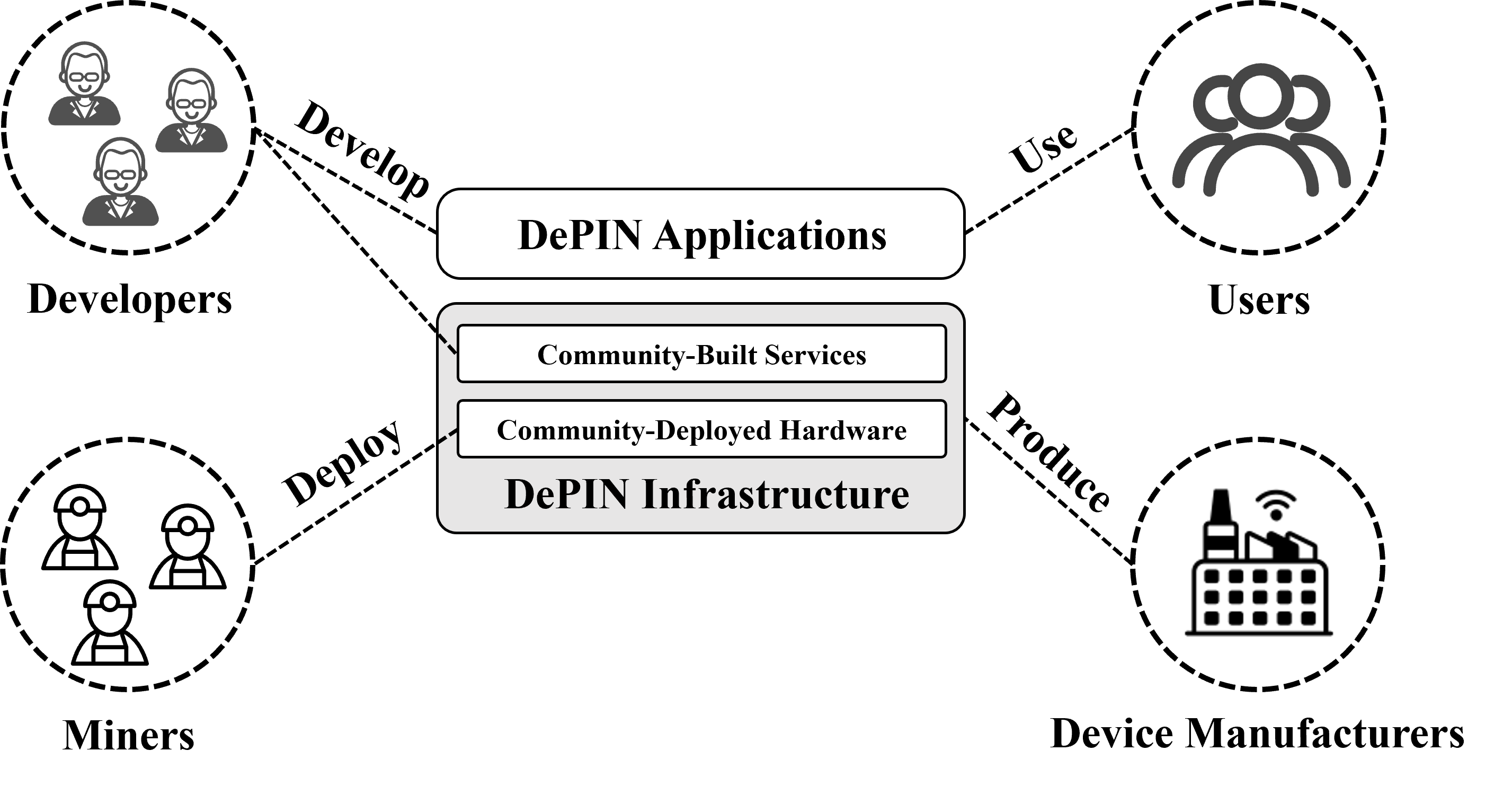}
    \caption{An example of the DePIN ecosystem \cite{iotex_whitepaper_TR}. DePIN stakeholders can be categorized into Device Manufacturers, Developers, Miners, and Users, contributing to production, deployment, application development, and service consumption.}
    \label{fig:example}
\end{figure}

\textit{Decentralized Physical Infrastructure Network (DePIN) represents a revolutionary framework capable of addressing these challenges and transforming the current landscape.
    Largely influenced by the success of Bitcoin and the digital currency,
    DePIN, a term coined by Messari in 2022 \cite{Messari_DePIN}, represents a novel network architecture that leverages decentralized technologies like blockchain as its underlying foundation to construct and operate physical infrastructure.}
The deployment, operation, and management of physical devices are collectively undertaken by distributed participants \cite{choi_DePIN_ICBC}. Through economic incentives, DePIN encourages participation and collaboration among stakeholders, ultimately forming a self-sustaining infrastructure network.

As illustrated in Fig. \ref{fig:example}, in a basic DePIN ecosystem, four key stakeholders are required \cite{iotex_whitepaper_TR}: device manufacturers, miners, developers, and users.
Device manufacturers are responsible for producing various DePIN devices, such as hotspots for network coverage, e.g., Helium hotspots~\cite{Helium}, decentralized storage nodes, e.g., Filecoin miners~\cite{Giacomelli_Filecoin_CrypTorino}, edge computing modules, or environmental sensing units.
Developers, driven by the demands of the decentralized market, establish the foundational infrastructure, develop core components and services, and create corresponding DePIN applications.
They issue virtual currencies with purchasing power in the market. Miners are incentivized to deploy their DePIN devices, while expanding the scope of DePIN applications, and earn rewards based on their contributions.
Users utilize the DePIN applications provided by the DePIN devices and pay tokens according to their usage.
By leveraging DePIN, the traditionally centralized ICT infrastructure can be decentralized in a technically and economically feasible manner. This approach transforms ICT systems from operator-driven to community-driven, enabling distributed ownership, transparent operation, and resource-efficient collaboration. Consequently, DePIN offers a viable path toward democratizing ICT resources and promoting an open, resilient, and user-centric digital ecosystem \cite{chen_digitaleconomy_IJCS}.

\begin{table}[!t]
    \centering
    \captionsetup{
    font={scriptsize}}
    \caption{Acronyms and definitions}
    \fontsize{6pt}{7pt}\selectfont
    \begin{tabular}{ l | l }
        \hline
        \makecell[l]{\textbf{Acronym}} & \makecell[l]{\textbf{Full name / Definition}} \\
        \hline
        AI                             & Artificial Intelligence                       \\
        AI-RAN                         & Artificial Intelligence-Radio Access Network  \\
        BBU                            & Baseband Unit                                 \\
        BCADD                          & Blockchain Address                            \\
        B-RAN                          & Blockchain-Enabled Radio Access Network       \\
        CDN                            & Content Delivery Network                      \\
        CKC                            & Chain-Key Cryptography                        \\
        CI/CD                          & Continuous Integration/Continuous Delivery    \\
        COTS                           & Commercial Off-The-Shelf                      \\
        DaaS                           & DePIN-as-a-Service                            \\
        DAG                            & Directed Acyclic Graph                        \\
        DAO                            & Decentralized Autonomous Organization         \\
        dApps                          & Decentralized Applications                    \\
        DePIN                          & Decentralized Physical Infrastructure Network \\
        D2D                            & Device-to-Device                             \\
        DeWi                           & Decentralized Wireless                        \\
        DFL                            & Decentralized Federated Learning              \\
        DID                            & Decentralized Identity                        \\
        DNS                            & Domain Name System                            \\
        DDoS                           & Distributed Denial of Service                 \\
        FL                             & Federated Learning                            \\
        GNN                            & Graph Neural Network                          \\
        ICT                            & Information and Communication Technology      \\
        IoT                            & Internet of Things                            \\
        LEO                            & Low Earth Orbit                               \\
        LLM                            & Large Language Model                          \\
        LoRA                           & Low-Rank Adaptation                            \\
        MCF                            & Multi-Core Fibers                             \\
        ML                             & Machine Learning                             \\
        MTBF                           & Mean Time Between Failures                   \\
        MTTR                           & Mean Time to Repair                         \\
        NFs                            & Network Functions                             \\
        NLP                            & Natural Language Processing                   \\
        OLT                            & Optical Line Terminal                         \\
        ONU                            & Optical Network Unit                          \\
        O-RAN                          & Open Radio Access Network                     \\
        OPEX                           & Operating Expense                            \\ 
        OTN                            & Optical Transport Network                    \\                      
        P2P                            & Peer-to-Peer                                  \\
        PoB                            & Proof-of-Bandwidth                            \\
        PoBT                           & Proof-of-Block-and-Trade                      \\
        PoC                            & Proof-of-Coverage                             \\
        PoFL                           & Proof-of-Federated Learning                   \\
        PoI                            & Proof-of-Intelligence                         \\
        PoRep                          & Proof-of-Replication                          \\
        PoR                            & Proof-of-Routes                               \\
        PoSt                           & Proof-of-Spacetime                            \\
        PoW                            & Proof-of-Work                                 \\
        PQC                            & Post-Quantum Cryptography                     \\
        P-RAN                          & Proximity Radio Access Network                \\
        PUF                            & Physical Unclonable Function                  \\
        QoS                            & Quality of Service                            \\
        RAG                            & Retrieval-Augmented Generation                \\
        RAN                            & Radio Access Network                          \\
        RL                             & Reinforcement Learning                        \\
        RU                             & Radio Unit                                    \\
        RWA                            & Real World Asset                              \\
        SDM                            & Space-Division Multiplexing                   \\
        SDN                            & Software Defined Networking                   \\
        SLA                            & Service Level Agreement                       \\
        TEE                            & Trusted Execution Environment                 \\
        TPM                            & Trusted Platform Module                       \\
        UAV                            & Uncrewed Aerial Vehicle                       \\
        VCs                            & Verifiable Credentials                        \\
        vRAN                           & Virtual Radio Access Network                  \\
        W3C                            & World Wide Web Consortium                     \\
        ZK                             & Zero-Knowledge                                \\
        ZKP                           & Zero-Knowledge Proof                         \\
        \hline
    \end{tabular}
    \label{tab:acronyms}
\end{table}

\textit{The key question is whether such a decentralized and highly automated/participatory network architecture is technically feasible and can meet the requirements for high reliability, high availability, security, and low latency obtained by close control and coordination.}

Operationally, DePIN realizes this community-driven model by allowing participants to contribute, consume, and maintain physical resources, while service interactions are recorded and settled through transparent decentralized contracts \cite{Lin_RW_1}. Tokenized rewards further link resource contribution, infrastructure maintenance, and service expansion, forming a feedback process that supports broader network coverage and service diversity \cite{Ling_BRAN_IEEEWC}. With the integration of Large Language Models (LLMs) and agentic Artificial Intelligence (AI), DePIN may further evolve toward adaptive and context-aware infrastructure coordination \cite{Verma_WLAN_COMST, Wu_LLM_TMC}.

\subsection{Development of DePIN}

The development of DePIN can be traced back to the late 1990s, marked by the rise of Peer-to-Peer (P2P) networks \cite{Wellman_P2P_IJURR}. Technologies, e.g., Napster and BitTorrent, demonstrated the potential of P2P communication, laying the foundation for subsequent decentralized networks \cite{liu_p2p_TIA}. The advent of Bitcoin in 2008 introduced blockchain technology, achieving value transfer and data verification without the need for centralized trust intermediaries \cite{zarrin_blockchain_Springer}. Its incentive mechanism and consensus algorithm inspire DePIN's economic framework \cite{sriman_pow_Springer}. In 2015, Ethereum introduced smart contract technology to enable decentralized systems to automatically execute predefined rules, expanding the application scenarios of decentralized systems \cite{hewa_smartcontracts_Access}. Projects, such as Helium~\cite{Helium}, Braintrust~\cite{braintrust}, and IoTeX \cite{iotex}, which redefined the relationships between enterprises, communities, and their resources, brought the initial concepts of DePIN into focus.

The existing centralized ICT infrastructure is insufficient to sustain the DePIN ecosystem.
China Mobile has collaborated with multiple sectors to establish the Hong Kong Web 3.0 Association. China Telecom is developing blockchain SIM cards tailored for DePIN \cite{niu_esim_Springer}. Spanish telecom company, Telefónica, has disclosed its investments and strategic positioning in the Web 3.0 domain, including DePIN.
Nokia explicitly identifies DePIN as an essential component of Web 3.0 in its technological vision \cite{NOKIA}. Ericsson, in collaboration with \textit{Uplink}, is delivering high-quality, decentralized network connectivity services worldwide \cite{Uplink}.
Similarly, the rise of the DePIN concept has also spurred numerous startups and projects, e.g., Helium \cite{Helium}, MetaBlox \cite{MetaBlox}, and Pollen \cite{Pollen}, which extend decentralization from the virtual blockchain layer to the physical infrastructure domain. By leveraging user-owned hardware resources, e.g., Wi-Fi hotspots and base stations, these projects integrate blockchain, distributed computing, and AI with communication hardware to construct community-driven, self-sustaining networks, thereby uncovering new opportunities for infrastructure growth.

\subsection{Related Surveys and Motivations of This Survey}

\begin{table*}[htbp]
    \centering
    \captionsetup{
    font={scriptsize}}
    \caption{Comparison of the definitions among different technological paradigms.
        The marker `\greencheck' represents full support of the feature; `\redcross' denotes a lack of the feature; and `\bluecircle' signifies that the feature depends on specific implementation contexts.}
    \label{tab:comparison}
    \fontsize{6pt}{7pt}\selectfont
    \begin{tabular}{lccccccc}
        \toprule
        Features    & Decentralization & Physical Infra. & Tokenomics  & Immersive   & Sovereignty & Sharing     & Real-world Utility \\ \hline \midrule
        DePIN       & \greencheck      & \greencheck     & \greencheck & \redcross   & \greencheck & \greencheck & \greencheck        \\
        Web 3.0     & \greencheck      & \redcross       & \greencheck & \redcross   & \greencheck & \bluecircle & \bluecircle        \\
        Metaverse   & \bluecircle      & \redcross       & \bluecircle & \greencheck & \greencheck & \redcross   & \redcross          \\      
        \bottomrule
    \end{tabular}
\end{table*}

As summarized in Table \ref{tab:comparison}, we first clarify the conceptual boundaries among three interconnected terminologies: DePIN, Web 3.0 and Metaverse.

$\bullet$ \textit{DePIN}: A paradigm for building and operating physical infrastructure networks through decentralized coordination. Its essence lies in the decentralization of physical resources coupled with digital incentive mechanisms (Tokenomics).

$\bullet$ \textit{Web 3.0}: A next-generation internet paradigm designed to mitigate platform centralization and data silos inherent in Web 2.0. By leveraging blockchain as the underlying layer, it emphasizes user-centric data sovereignty and asset ownership, facilitating permissionless value exchange through decentralized protocols.

$\bullet$ \textit{Metaverse}: An immersive digital environment powered by AR, VR, and blockchain technologies. It provides a persistent and interoperable ecosystem where users engage in social interaction, economic activities, and creative entertainment through digital identities.

\begin{table*}[t]
\centering
\captionsetup{
    labelfont={color=black},
    textfont={color=black},
    font={scriptsize}
}
\caption{Comparison with existing surveys on DePIN or related areas. The marker `\greencheck' denotes substantial coverage, `\orangewave' denotes partial or peripheral coverage, and blank entries indicate that the corresponding aspect is not included.}
\label{tab:related_works}
\fontsize{6pt}{7pt}\selectfont
\begin{tblr}{
    width = \linewidth,
    colspec = {X[0.5]X[1.5]X[1.5]X[1.5]X[1.5]X[1.5]X[1.5]X[2.5]},
    cells = {c,m,fg=black},
    row{1} = {font=\bfseries},
    row{2} = {font=\bfseries},
    column{8} = {l},
    cell{1}{1} = {r=2}{},
    cell{1}{2} = {c=6}{},
    cell{1}{8} = {r=2}{},
    hlines={black},
}
\textbf{Ref.} 
& \textbf{Covered Aspects (with DePIN or related areas)} 
&  &  &  &  & 
& \textbf{Survey Focus} \\

& \textbf{Architecture} 
& \textbf{Physical Infra.} 
& \textbf{Tokenomics} 
& \textbf{Tech. Stack} 
& \textbf{Deployment Feasibility} 
& \textbf{Ecosystem Landscape} 
& \\ \hline

\cite{Lin_RW_1}    & \greencheck & \orangewave        &              & \orangewave        &              & \greencheck & DePIN overview \\
\cite{Ballandies_RW_2}    & \orangewave       &               & \orangewave       & \orangewave        &              &              & Taxonomy of DePIN \\
\cite{Andrew_RW_3}    &              &               &              & \orangewave        & \orangewave       & \orangewave       & Classification method \\
\cite{Fan_RW_4}    & \orangewave       & \greencheck  & \orangewave       & \greencheck  &              &              & Modular DePIN infra. \\
\cite{sarkar_RW_5}    &              &               &              & \orangewave        & \orangewave       & \orangewave       & Generalised protocol \\
\cite{liu_RW_6}    &              &               & \greencheck &              &              & \greencheck & EconAgentic markets \\
\cite{zhu_RW_7}    &              &               & \orangewave       & \greencheck  & \orangewave       &              & Technologies for Web 3.0 \\
\cite{shen_RW_8}    & \greencheck & \orangewave        & \orangewave       & \greencheck  &              &              & AI for Web 3.0 \\
\cite{Hossain_RW_9}    &              &               &              & \orangewave        &              & \orangewave       & Decentralized applications \\
\cite{Wang_RW_10}   & \orangewave       &               &              & \greencheck  &              &              & Security in Metaverse \\
\cite{Zhang_RW_11}   &              & \orangewave        &              & \greencheck  &              & \orangewave       & Industrial Metaverse \\
Ours & \greencheck & \greencheck  & \greencheck & \greencheck  & \greencheck & \greencheck & Feasibility-oriented \\
\end{tblr}
\end{table*}

Recent surveys have contributed to the conceptual understanding of DePIN ~\cite{Lin_RW_1, Ballandies_RW_2, Andrew_RW_3}, 
as summarized in Table \ref{tab:related_works}.
Existing surveys have partially addressed the question
\textit{whether DePIN can be engineered as a technically feasible and operationally sustainable paradigm for future decentralized ICT infrastructure}.
For instance, Lin \textit{et al.}~\cite{Lin_RW_1} provided a broad overview of DePIN, including its architecture, design principles, applications, and market status.
Ballandies \textit{et al.}~\cite{Ballandies_RW_2} developed a systematic taxonomy covering the hierarchical structure of DePIN and the interrelationships among its components.
Andrew \textit{et al.}~\cite{Andrew_RW_3} proposed a decision-tree-based classification method to distinguish DePIN from other adjacent systems.
The main emphasis of these studies remains on overview, taxonomy, and classification.

Several surveys have examined design directions within DePIN.
Specifically,
Fan \textit{et al.}~\cite{Fan_RW_4} proposed a modular DePIN infrastructure, describing an emerging community-owned infrastructure comprising reusable modules.
Sarkar \textit{et al.}~\cite{sarkar_RW_5} introduced a generalized protocol for DePIN applications. It emphasizes device onboarding, witness-based validation, anomaly detection, and reward mechanisms.
Liu \textit{et al.}~\cite{liu_RW_6} presented EconAgentic, an LLM framework for modeling DePIN market evolution, stakeholder interactions, and indicators such as efficiency and stability.
These works each address a specific perspective rather than providing a systematic survey of DePIN as a decentralized ICT infrastructure. 

Related discussions can be found in the literature on Web 3.0 and Metaverse.
Surveys, e.g.,~\cite{zhu_RW_7, shen_RW_8, Hossain_RW_9, Wang_RW_10}, analyzed the technological foundations, challenges, and opportunities of Web 3.0, emphasizing decentralization, data sovereignty, and AI-driven applications. They highlighted blockchain as a core enabler, augmented by AI and edge computing to enhance scalability, security, and user experience.
Zhang \textit{et al.} \cite{Zhang_RW_11} reviewed decentralization in the industrial Metaverse, focusing on enabling technologies, e.g., blockchain, privacy-preserving computing, and digital twins, and open challenges in industrial applications.
These studies lack a comprehensive investigation of physical infrastructure, fundamentally distinguishing DePIN from purely digital blockchain-based networks.

This survey differs from the prior works in motivation and scope.
It focuses on the feasibility of DePIN deployment and investigates how DePIN may evolve from an emerging concept into practical decentralized ICT infrastructure.
It reviews a broader set of issues and solutions closely tied to this objective, including physical infrastructure transformation, tokenomics, systematic technology stack, and ecosystem landscape.

\subsection{Contributions and Organization}
This paper bridges the gap between current ICT paradigms and the emerging vision of DePIN. 
We provide a systematic analysis of the technological landscape, investigating how existing technologies can be leveraged to realize the decentralized infrastructure envisioned in Web 3.0.
Specifically, we dissect the DePIN technology stack across six critical dimensions \textemdash physical infrastructure, blockchain, interaction, trust, incentive, and application layers. For each layer, we evaluate its technical feasibility and implementation readiness.
We propose a structured DePIN feasibility assessment framework to examine whether DePIN systems can be practically deployed and sustainably operated in the real world.
We also identify open challenges and research directions, providing insights and a roadmap for the evolution of DePIN.

The key contributions of this paper include:
\begin{itemize}
    \item \textit{Gap Analysis and Evolutionary Roadmap:} This paper reviews the evolution from traditional ICT infrastructure to DePIN. We analyze the technological gaps between existing methodologies and the decentralized infrastructure envisioned for Web 3.0. We also identify the necessary improvements in current ICT infrastructures to meet the requirements of DePIN.
    \item \textit{Investigation of Core Attributes:} We provide a structured investigation into the development background and define six key characteristics of DePIN, ranging from the decentralized paradigm to ubiquitous intelligence and economic incentives. This establishes a tangible framework for understanding how DePIN differentiates from, yet integrates with, existing physical infrastructures.
    \item \textit{Six-Layer Technology Stack:} We analyze the DePIN technology stack across six layers: Physical Infrastructure, Blockchain, Interaction, Trust, Incentive, and Application, evaluating the technical feasibility of each layer by exploring how current technologies can be adapted to support decentralized coordination and examining the cross-layer feedback loops.
    \item \textit{DePIN Feasibility Assessment Framework:} We propose a structured feasibility assessment framework for DePIN deployment. 
    The framework evaluates DePIN systems from technical, regulatory and governance, and economic dimensions, and provides a practical reference for judging whether DePIN projects are technically deployable, governable, and economically sustainable.
    \item \textit{Identification of Challenges and Future Directions:} Key challenges are identified, including secure hardware, cross-chain and cross-domain collaboration, post-quantum security, token value management, and real-world deployment feasibility.
    We also outline promising research directions, including security-by-design hardware, AI-native networks, and DePIN-as-a-Service platforms, to guide future research endeavours.
\end{itemize}

\begin{figure}[!h]
    \centering
    \includegraphics[scale=0.5]{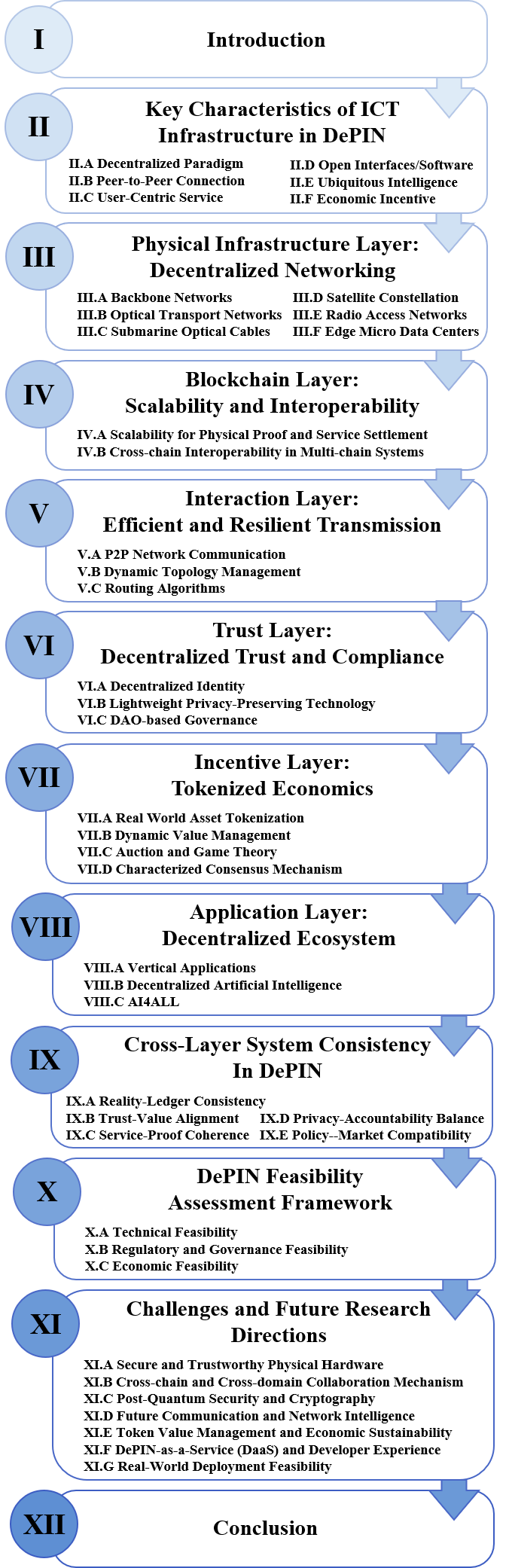}
    \caption{The structure of this survey, which covers conceptual foundations, system-level architecture, feasibility assessment framework, challenges, and future directions.}
    \label{fig:structure}
\end{figure}

As depicted in Fig. \ref{fig:structure}, Section \ref{sect:characteristic} discusses the six key characteristics of ICT infrastructure in DePIN.
Section \ref{sect:infrastructure} examines the physical infrastructure layer and its transformation under decentralized paradigms.
Section \ref{sect:blockchain} focuses on the blockchain layer, analyzing its scalability and cross-chain interoperability.
Section~\ref{sect:interaction} explores the interaction layer, addressing P2P communication, dynamic topology management, and intelligent routing.
Sections \ref{sect:trust} and \ref{sect:incentive} investigate the trust and incentive layers, where security, privacy, and tokenomics are discussed.
Section~\ref{sect:application} reviews the application layer, presenting representative DePIN use cases.
Section~\ref{sect:crosslayer} examines cross-layer system consistency and feedback loops.
Section~\ref{sect:feasibility} presents a feasibility assessment framework for DePIN deployment.
Section~\ref{sect:future} points to open challenges and research directions,
followed by conclusions in
Section~\ref{sect:conclusion}.
The key acronyms are summarized in Table~\ref{tab:acronyms}.

\section{Characteristics of ICT  in DePIN} \label{sect:characteristic}

This section delves into six key attributes of DePIN, i.e., \textit{Decentralized Paradigm}, \textit{Peer-to-Peer Communication}, \textit{User-centric Service}, \textit{Open Interfaces/Software}, \textit{Ubiquitous Intelligence}, and \textit{Economic Incentive},
along with its enabled layers, key enablers, DePIN approaches, and challenges;
see Table~\ref{tab:character}.

\begin{table*}[!ht]
\captionsetup{
    labelfont={color=black},
    textfont={color=black},
    font={scriptsize}
}
    \caption{Key characteristics of DePIN, elaborated on in Sec.~\ref{sect:characteristic} and realized through the technology stack analyzed in Secs.~\ref{sect:infrastructure}--\ref{sect:application}.}
    \label{tab:character}
    \arrayrulecolor{black}
    \fontsize{6pt}{7pt}\selectfont
    \begin{tblr}{
    width = \linewidth,
    colspec = {X[1]X[0.3]X[1.4]X[1.7]X[2.5]X[2.2]},
    cells = {c,m, fg=black},
    column{1} = {font=\itshape},
    column{2} = {c,m},
    column{5} = {j,m},
    row{1} = {c, font=\bfseries},
    cell{2}{2}={l},
    cell{2}{3}={l},
    cell{2}{4,6}={l},
    cell{3}{2}={l},
    cell{3}{3}={l},
    cell{3}{4,6}={l},
    cell{4}{2}={l},
    cell{4}{3}={l},
    cell{4}{4,6}={l},
    cell{5}{2}={l},
    cell{5}{3}={l},
    cell{5}{4,6}={l},
    cell{6}{2}={l},
    cell{6}{3}={l},
    cell{6}{4,6}={l},
    cell{7}{2}={l},
    cell{7}{3}={l},
    cell{7}{4,6}={l},
    hlines={black},
    vline{2-6} = {black},
    }
    Characteristic & Refs. & Enabled Layers  & Enabler Technologies & DePIN Approach  & Challenges  \\   
    \hline

    Decentralized Paradigm & \cite{gordon_decentralized_JITCAR, zhang_security_ACM, zainal_decentralized_CEE}
    &
    Physical Infra. Layer \newline Blockchain Layer \newline Trust Layer  &
    Blockchain \newline Smart contracts \newline Consensus Mechanisms \newline  Decentralized Identity &
    
    \bluecircle \texttt{} Decentralize physical infrastructure ownership and governance \newline
    \bluecircle \texttt{} Enable community-built and operated networks
    &
    Verifiable resource mapping, \newline Scalability trilemma, \newline Security and reliability  \\

    Peer-to-Peer Connection & \cite{tushar_p2pscale_elsevier, Xu_BRAN_arXiv, Sutton_PHY_COMST, Yang_P2P_NET, Zhu_privacy_TKDE, soto_p2p_elsevier}
    &
    Physical Infra. Layer \newline Blockchain Layer \newline Interaction Layer 
    & P2P protocol stacks \newline distributed routing & 
    \bluecircle \texttt{} Enable direct resource exchange among participants \newline
    \bluecircle \texttt{} Support communication and transactions without intermediaries
      & 
    Heterogeneous devices, protocols, and domains  \\

    User-Centric Service  & \cite{chen_Customized_INET, lacava_agent_TMC}
    &
    Interaction Layer \newline Trust Layer \newline Application Layer  & 
    Decentralized Identity \newline Edge intelligence \newline  Service orchestration \newline Decentralized applications & 
    \bluecircle \texttt{} Position users as resource providers and co-creators\newline
    \bluecircle \texttt{} Organize services around user-owned assets and local demand
        & 
    Complex dynamic identities, \newline Personalization with privacy protection \\

    Open Interfaces/Software & \cite{Polese_ORAN_COMST, abdalla_NGORAN_INET, Yue_dApps_COMST, wu_dapp_arXiv, dao_dApps_Springer}
    &
    Physical Infra. Layer \newline Interaction Layer \newline Application Layer & 
    Open APIs \newline Modular architecture \newline Open-source platforms  & 
    \bluecircle \texttt{} Decouple hardware and software via open interfaces \newline
    \bluecircle \texttt{} Enable interoperable services and applications
      & 
    Lack of unified standards, \newline Interoperability across vendors, \newline Code-governance vulnerabilities \\

    Ubiquitous Intelligence   & \cite{zaidi_AI4Network_Access, Hu_DML_COMST, Makhdoom_security_COMST, Khan_Light_IoTJ, Pan_AI_Network, tauseef_chatbots_IGI, Yang_AI_Network, Luo_semantic_WC}
    &
    Interaction Layer \newline  Application Layer & 
    AI/ML, LLM \newline Agentic AI \newline Edge intelligence & 
    \bluecircle \texttt{} Embed AI across devices, networks, and clouds
      & 
    Trustworthy \& privacy-preserving, \newline Model robustness \& explainability, \newline AI hallucinations       \\

    Economic Incentive  &  \cite{Lin_RW_1, Freni_Token_BRA, han_incentive_ACM, Ling_BRAN_IEEEWC}
    &
    Incentive Layer \newline Trust Layer  & 
    Tokenomics \newline Reputation system \newline DAO governance \newline Network effects & 
    \bluecircle \texttt{} Reward physical resource contributions with tokens\newline
    \bluecircle \texttt{} Drive self-reinforcing growth through network effects
    & 
    Token value management, \newline Resilience to manipulation \& attacks                  \\
\end{tblr}
\end{table*}

\vspace{-5mm}
\subsection{Decentralized Paradigm}

The ICT infrastructure in DePIN embodies a decentralized paradigm by distributing resources, data, and control across a network of independent nodes or participants
\cite{gordon_decentralized_JITCAR}.
The core technology is blockchain, which leverages distributed ledgers and consensus mechanisms to ensure data transparency, integrity, and immutability. By decentralizing trust, blockchain removes the need for centralized intermediaries, hence addressing security challenges inherent in traditional systems \cite{gordon_decentralized_JITCAR}. This decentralized approach is transformative in sectors demanding high trustworthiness and reliability, such as financial services, supply chain management, and data storage \cite{zhang_security_ACM}.
This paradigm empowers users with greater autonomy over their personal assets, aligning with the principles of privacy-by-design and self-sovereign identity \cite{zainal_decentralized_CEE}. 
By dispersing authority across the network, the ICT infrastructure in DePIN mitigates single points of failure and fosters innovation through open participation and censorship-resistant systems.
Meanwhile, realizing this paradigm requires verifiable mapping between on-chain logic and off-chain physical resources, while balancing scalability, security, and reliability.

\subsection{Peer-to-Peer Connection}

P2P networking is an enabler of scalable and efficient services in decentralized ICT infrastructure \cite{tushar_p2pscale_elsevier}.
P2P architectures eliminate single points of failure, reduce latency, and mitigate risks associated with third-party interference. This paradigm shift manifests in three core operational dimensions: P2P communication, P2P data sharing, and P2P transactions, each offering advantages over centralized models.

\subsubsection{P2P Communication}
\begin{itemize}
    \item Enable low-latency, direct connectivity between terminals within the same Radio Unit (RU) control range, while decentralizing critical core network functions, e.g., authentication and routing, to the edge \cite{Xu_BRAN_arXiv}.
    \item Use Cases: Instant communication for enterprise Internet of Things (IoT) systems and emergency networks \cite{Sutton_PHY_COMST}, e.g., disaster recovery scenarios where infrastructure is compromised. Dynamic mesh networks for ad hoc device collaboration without centralized coordination.
\end{itemize}

\subsubsection{P2P Data Sharing}
\begin{itemize}
    \item Facilitate secure, user-controlled data exchange between data centers or edge nodes, with permissions governed by smart contracts or decentralized identity protocols.
    \item Key Benefit: Eliminates reliance on central servers for data routing, enhancing privacy and resilience against censorship or tampering \cite{Yang_P2P_NET, Zhu_privacy_TKDE}.
\end{itemize}

\subsubsection{P2P Transactions}
\begin{itemize}
    \item Empower decentralized marketplaces in the sharing economy, e.g., energy, bandwidth, and compute resources, by allowing participants to autonomously set terms (price, quantity, counterparties) without central oversight \cite{soto_p2p_elsevier}.
    \item Example: In P2P energy trading, prosumers can dynamically negotiate transactions in real-time, optimizing local grid efficiency and renewable energy utilization \cite{wu_p2penergy_elsevier}.
\end{itemize}

Despite these benefits, scalable P2P coordination in DePIN must still cope with heterogeneous devices, protocol stacks, and cross-domain interoperability.

\subsection{User-Centric Service}

The user-centric paradigm is a fundamental shift in how ICT services are designed and delivered within DePIN, where individuals gain sovereign control over productive assets like data and energy through P2P mechanisms.
Unlike traditional operator-dominated models that deliver standardized services, DePIN enables users to act not only as service consumers but also as resource providers and service co-creators, actively participating in network deployment, e.g., establishing small base stations while collaborating with operators to create adaptive connectivity~\cite{chen_Customized_INET}.
The architecture achieves user-centricity through intelligent edge agents that continuously monitor and interpret individual requirements~\cite{lacava_agent_TMC}. 
It transforms network access from static service packages into responsive, need-based solutions, for enterprise applications requiring on-demand bandwidth allocation or emergencies demanding prioritized communications.
As a result, service provisioning is increasingly organized around user-owned assets and localized demand rather than one-size-fits-all operator policies.

\subsection{Open Interfaces and Software}

DePIN establishes a new paradigm of openness in the ICT infrastructure through standardized interfaces and open-source software.
At the infrastructure level, open standards like O-RAN decouple network elements, which allows vendors to develop compatible hardware components while empowering operators to avoid vendor lock-in through modular procurement strategies~\cite{Polese_ORAN_COMST}. This reduces operational costs and enables users to participate in network construction by customizing small base stations to meet customized requirements \cite{abdalla_NGORAN_INET}.

The principle of openness extends to the application layer through decentralized applications (dApps) that operate on blockchain platforms.
Here, dApps leverage P2P networks to provide serverless functionality while ensuring data integrity through cryptographic storage on distributed ledgers \cite{Yue_dApps_COMST}. These applications embody the ethos of open-source development, granting users autonomy to modify, package, and assert ownership without institutional constraints \cite{wu_dapp_arXiv}. This permissionless innovation environment is fertile \cite{dao_dApps_Springer}, with new DePIN services emerging from community-driven development, rather than corporate initiatives.
However, this openness also exposes DePIN to persistent challenges in standard unification, cross-vendor interoperability, and secure code governance across diverse open-source components.

\subsection{Ubiquitous Intelligence }

While decentralization is the central paradigm, sustainable growth does not just depend on hardware deployment, but also on effective data utilization. AI plays a key role in this regard.

On the network architecture side (AI4Network \cite{zaidi_AI4Network_Access}), intelligent ICT infrastructure can autonomously optimize the management and operation of networks by predicting traffic patterns and load conditions.
This enables dynamic resource allocation, e.g., storage, bandwidth \cite{Hu_DML_COMST}, and real-time threat detection \cite{Makhdoom_security_COMST}.
AI techniques, e.g., Federated Learning (FL), often combined with differential privacy, enable decentralized model training without exposing raw user data \cite{Khan_Light_IoTJ}.

On the network services side (Network4AI \cite{Pan_AI_Network}), AI significantly improves user experience.
For instance, Natural Language Processing (NLP), LLM, and chatbots empower decentralized applications with intuitive interfaces, making services accessible and user-friendly \cite{tauseef_chatbots_IGI}.
By analyzing historical user data, services can deliver personalized recommendations \cite{Yang_AI_Network}.
Furthermore, the integration of blockchain and AI, particularly LLM-powered semantic reasoning, supports the construction of decentralized knowledge bases that enhance large-scale intelligent information retrieval based on user intent \cite{Luo_semantic_WC}.

\begin{table*}[t]
    \centering
    \captionsetup{
    labelfont={color=black},
    textfont={color=black},
    font={scriptsize}
    }
    \caption{Summary of research on the physical layer of ICT infrastructure within DePIN.
    The marker `\greencheck' and `\redcross' indicate main DePIN adaptations and limitations, respectively.}
    \label{tab:Physical}
    \fontsize{6pt}{7pt}\selectfont
    \begin{tblr}{
        width = \linewidth,
        colspec = {X[1.5,c,m] X[0.9,c,m] X[4.4,j,m] X[4.6,j,m] X[0.9,c,m]},
        cells = {fg=black},
        row{1} = {c, font=\bfseries},
        hlines = {black},
        vline{2-5} = {black},
    }
        Type 
        & Refs. 
        & Main Adaptations in DePIN 
        & Main Limitations in DePIN 
        & Feasibility \\ \hline

        Backbone Networks
        & \cite{Hossain_Backbone_Network, Chen_Backbone_ICCCS, Wang_Backbone_IoTJ}
        & \greencheck \texttt{} Edge-distributed NF deployment for scalable service registration.
        & \redcross \texttt{} Decentralized routing convergence is difficult. \newline
          \redcross \texttt{} Unverifiable cross-domain SLA enforcement.
        & Medium \\

        Optical Transport Network
        & \cite{Lian_OTN_IoTJ, Helmy_OTN_TGCN, Sun_OTN_ECOC, Andriolli_OTN_OSN}
        & \greencheck \texttt{} Programmable OTN allows elastic allocation. \newline
          \greencheck \texttt{} Hybrid control combines decentralized switch.
        & \redcross \texttt{} Immature and fragmented hardware ecosystem.
        & Medium to high \\

        Submarine Optical Cables
        & \cite{Syauqi_SOC_Energy, Kaschel_SOC_CHILECON, mohsan_SOC_EarthScience}
        & \greencheck \texttt{} SDM and MCF improve transoceanic capacity.
        & \redcross \texttt{} High cost restricts permissionless participation. \newline
          \redcross \texttt{} Landing regulation limits decentralization.
        & Low to medium \\

        Satellite Constellation
        & \cite{Roth_SC_ASMS, torky_SC_Aerospace, Li_SC_JSTARS, torky_SC_LISS, Wang_SC_TAES, Mwase_SC_WCNC, Zhao_SC_TMC, Zhai_SC_TMC}
        & \greencheck \texttt{} Crowdsourced ground stations extend community operation access.
        & \redcross \texttt{} Spectrum, orbit, collision risks limit openness. \newline
          \redcross \texttt{} Visibility gaps hinder routing \& proofs.
        & Medium \\

        Radio Access Network
        & \cite{Bi_RAN_Network, Bi_P-RAN_Magazine, AIRAN_whitepaper, Khan_AIRAN_COMSOC, Wang_BRAN_DCN, Xu_BRAN_arXiv, Ling_BRAN_WirelessBlockchain, Cao_BRAN_TWC, Le_BRAN_IoTJ, Ling_BRAN_IEEEWC}
        & \greencheck \texttt{} COTS platforms reduce deployment barriers. \newline
          \greencheck \texttt{} Openness, AI control, and blockchain trust.
        & \redcross \texttt{} Coverage fraud and reward farming. \newline
          \redcross \texttt{} Strict latency, mobility, spectrum compliance.
        & Medium \\

        Edge Micro Data Centers
        & \cite{Bilal_MDC_CN, Cao_MDC_TII, Bruschi_MDC_NFVSDN, Lin_HASFL_TMC, Chakraborty_MDC_TSC, Scano_MDC_Access}
        & \greencheck \texttt{} Edge sites as community resource providers. \newline
          \greencheck \texttt{} Green SLA rewards energy-aware operation.
        & \redcross \texttt{} Fragmented scheduling \& accounting. \newline
          \redcross \texttt{} SLA fulfillment remains difficult.
        & Medium to high \\
    \end{tblr}

    \vspace{1mm}
    \parbox{\linewidth}{
    \color{black}
    \fontsize{6pt}{7pt}\selectfont
    \raggedright
    \textit{Note:} Feasibility is assessed from a DePIN deployment perspective. 
    `Low' indicates severe capital, licensing, or verifiability constraints; 
    `Medium' indicates partial deployability with remaining compliance barriers; 
    and `High' indicates relatively mature and open conditions for deployment.
    }
    
\end{table*}

\subsection{Economic Incentive}

DePIN's economic incentive model is anticipated to create a self-sustaining ecosystem where aligned economic interests drive network growth and security \cite{Lin_RW_1}. By implementing token-based rewards, the system ensures that honest participation becomes the most profitable strategy for nodes, while malicious behaviors incur prohibitively high costs \cite{Freni_Token_BRA}. Participants are incentivized not only through token rewards but also through governance rights in community decision-making~\cite{han_incentive_ACM}, fostering a collaborative environment that accelerates technological innovation and application adoption.
The model's effectiveness is exacerbated by powerful network effects \cite{Ling_BRAN_IEEEWC}, where each new participant increases the network's overall value. This creates a virtuous cycle: incentives drive user participation in infrastructure development, which improves network services, attracting more users and further strengthening the ecosystem.

Nevertheless, sustaining this incentive layer in DePIN relies on effective token value management and resilience against manipulation, Sybil attacks, and other strategic behaviors.

\section{Physical Infrastructure Layer: Decentralized Networking}\label{sect:infrastructure}

This section examines the physical infrastructure transformations required under DePIN.
We analyze how standard ICT components, ranging from backbone networks to edge data centers, are re-architected to support trustless collaboration and token-incentivized deployment, as summarized in Table~\ref{tab:Physical}.

\subsection{Backbone Networks}
Backbone networks comprise core routers, high-speed optical fiber links, regional nodes, and network switching equipment \cite{Dharmaweera_Backbone_COMST}.
Under DePIN, backbone networks face new challenges, including the need for rapid access by decentralized nodes, the flow of multi-source heterogeneous data, highly available topological reconfiguration, and dynamically autonomous network management \cite{Chen_Backbone_ICCCS}.
Backbone networks suitable for DePIN must feature open architectural designs, flexible resource orchestration capabilities, and access and service mechanisms that accommodate distributed participants.

In DePIN, most current studies suggest that network functions (NFs) be deployed in a distributed manner and extended to the edge network to support the dynamic scaling of NF instances \cite{Hossain_Backbone_Network, Chen_Backbone_ICCCS}. This evolution helps alleviate the load on the core network, reduces pressure on backbone infrastructures,
and overcomes the limitations of centralized NF registration and discovery mechanisms.

Blockchain technology is widely cited as a key enabler for autonomous collaboration among different organizations~\cite{Chen_Backbone_ICCCS, Wang_Backbone_IoTJ}. By recording service information on the blockchain and utilizing smart contracts to achieve consensus and data consistency across all participating nodes, it ensures data immutability and effectively addresses the trust issues among heterogeneous network entities, thereby reducing dependence on a single authoritative body.
Notably, Gao \textit{et al.} \cite{Gao_Backbone_ICCC} proposed an innovative concept in which base stations and core NFs are directly deployed on uncrewed aerial vehicle (UAV) platforms, creating a self-contained mobile network system.

\subsection{Optical Transport Networks}
Optical transport networks (OTN) serve as the core communication infrastructure supporting high-speed transmission of large data. Within DePIN, these networks can evolve toward more elastic and programmable resource-sharing models, enabling multi-stakeholder collaboration and decentralized interconnectivity.
Helmy \textit{et al.} \cite{Helmy_OTN_TGCN} pioneered this transformation by decentralizing the passive optical network.
Unlike traditional architectures where the optical line terminal (OLT) dictates all scheduling, their proposed framework empowers optical network unit (ONU) to manage uplink access autonomously. This effectively removes the central authority from the physical layer, significantly reducing the latency overhead caused by request-grant cycles in centralized controllers.

Out-of-band signaling is repurposed to establish direct, verified communication among ONUs, allowing P2P data exchange without traversing a central OLT, which is a prerequisite for decentralized local traffic offloading.
To overcome bandwidth limitations, Lian \textit{et al.} \cite{Lian_OTN_IoTJ} proposed a fine-grained dynamic bandwidth adjustment method for optical transport networks.
Leveraging a parameter-server architecture, the network can dynamically increase bandwidth during parameter download and upload phases, reducing transmission time.
While full decentralization poses challenges in resource optimization \cite{Andriolli_OTN_OSN}, Andriolli \textit{et al.} advocated that a hybrid architecture where the control plane is governed by the decentralized paradigm while the data plane retains centralized high performance switching represents the most viable path for DePIN-ready optical backbones.

\subsection{Submarine Optical Cables}
Submarine cables represent the most capital-intensive component of the internet, traditionally monopolized by telecom giants. DePIN introduces a paradigm of community-financed and strictly governed global connectivity.
Submarine optical cables require several enhancements to meet the demands of decentralized networking \cite{Syauqi_SOC_Energy}.
First, the adoption of advanced technologies, such as space-division multiplexing (SDM) and multi-core fibers (MCF), is essential to increase transmission capacity and energy efficiency, accommodating the growing data volume resulting from multi-stakeholder collaboration.
Second, the development of distributed intelligent operation and maintenance and autonomous management is crucial \cite{Kaschel_SOC_CHILECON}. 
Integrating environmental sensing capabilities, such as distributed fiber-optic sensing, can enhance the multifunctional infrastructure value of submarine cable systems \cite{mohsan_SOC_EarthScience}.

Current submarine cables face bottlenecks in transoceanic P2P high-speed transmissions, where traditional centralized control mechanisms hinder dynamic path reconfiguration and flexible direct interconnection among parties \cite{Ye_SOC_INFOCOM}.
One potential solution is to construct large-scale underwater optical exchange centers to interconnect submarine cables from different vendors. This physical interoperability allows DePIN to route traffic based on real-time performance and token costs, alleviating congestion.
Combining advanced encryption standard with blockchain technology \cite{Bruhwiler_SOC_Blockchain} transforms data packets into verifiable assets. Each packet carries a digital signature that accompanies the transmission to enable authenticity verification at the receiver. This ``Proof of Transit'' is critical for DePIN, as it provides the cryptographic evidence needed to trigger smart contract payments to infrastructure owners, ensuring a fair and tamper-proof revenue stream.

\subsection{Satellite Constellation}
Although traditional satellite networks have been controlled by centralized and monopolistic entities, DePIN can lower entry barriers by turning satellite infrastructure into a permissionless, community-driven system. 
Through token economics, globally distributed ground stations, orbital assets, and secure consensus nodes can be coordinated within one infrastructure network, where ground station deployers act as DePIN miners, satellite data and bandwidth become tokenized service assets, and orbital nodes serve as space-based trust anchors.

\subsubsection{Crowdsourced Ground Station}
Crowdsourced ground stations provide a practical DePIN pathway for satellite networks by enabling distributed users and small organizations to contribute reception, telemetry, relay, and edge processing capabilities.
Existing studies reveal the feasibility of such a shift.
Garcia-Cabeza \textit{et al.}~\cite{Garcia_Starlink_COMM} conducted the first large-scale empirical analysis of Starlink's Direct Satellite-to-Device network using millions of crowdsourced measurements collected from Android user devices across the U.S., demonstrating a strong correlation between constellation deployment progress and the volume of community-generated signal observations.
Similarly, the starlinkstatus.space platform~\cite{Bülo_Starlink_ICC} aggregated nearly 1.7 million performance measurements from 309 voluntary contributors across 29 countries, demonstrating that a globally distributed, self-organized community can monitor satellite network behavior with no centralized coordination.

\begin{figure}[!t]
    \centering
    \includegraphics[scale=0.3]{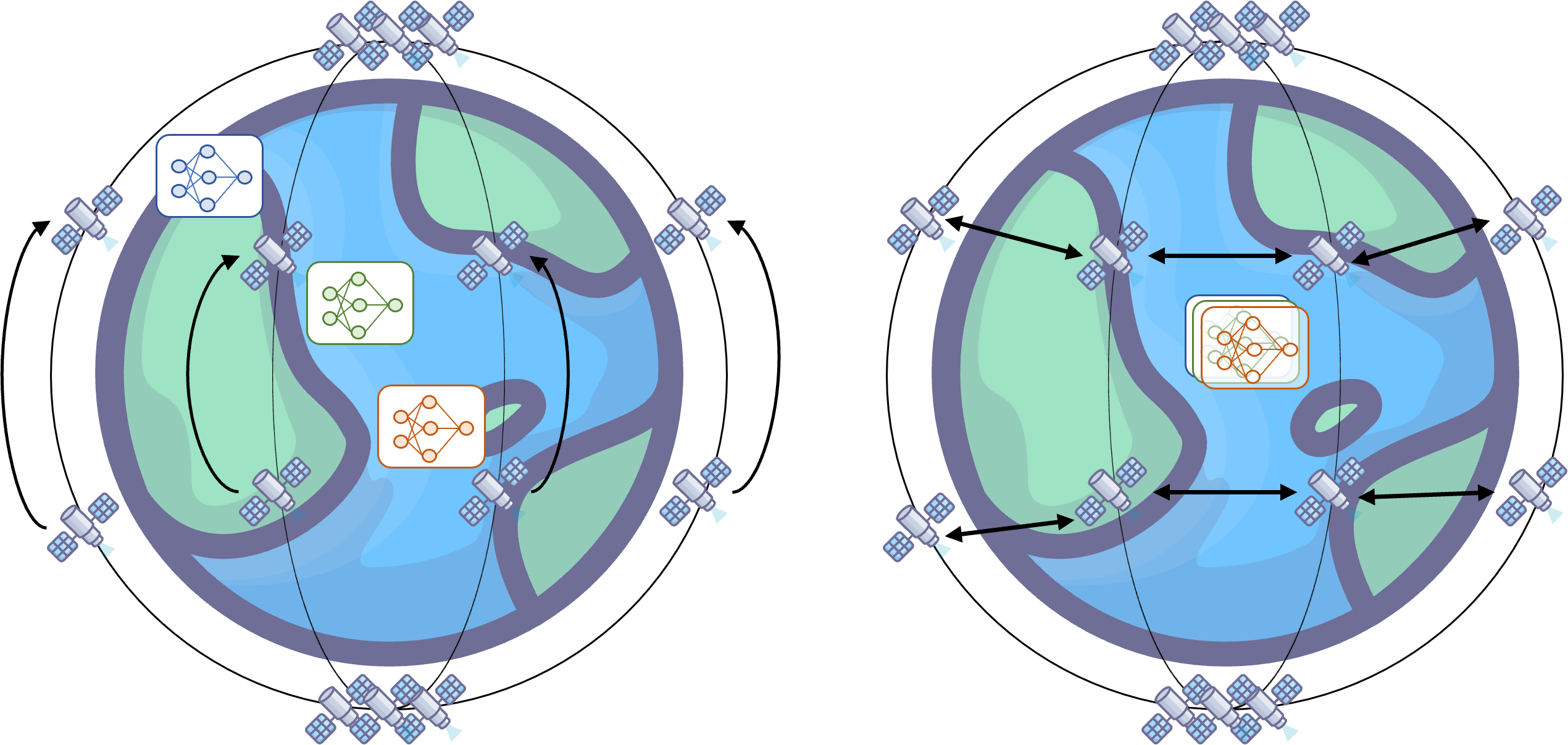}
    \caption{A DFL architecture in satellite networks, where black arrows indicate aggregation directions: the left side illustrates intra-orbit aggregation; the right shows inter-orbit aggregation.}
    \label{fig:Satellite}
\end{figure}

\subsubsection{Decentralized Architecture and Learning}
Satellite systems can align with DePIN through autonomous in-orbit coordination and distributed intelligence.
To manage highly dynamic LEO constellations, Roth \textit{et al.}~\cite{Roth_SC_ASMS} proposed a hybrid SDN-P2P architecture that clusters satellites and assigns onboard SDN controllers for local coordination, which supports DePIN by distributing control-plane decisions among space nodes.
Intra-cluster routing uses proactive load balancing, whereas inter-cluster routing relies on geographic heuristics to the destination region~\cite{torky_SC_Aerospace}, providing a basis for decentralized service routing under dynamic satellite topology.
Li \textit{et al.}~\cite{Li_SC_JSTARS} transformed LEO satellite constellations into a distributed cloud computing architecture, where onboard computing satellites and inter-satellite links can serve as DePIN resource contributors.
In this architecture, a geostationary satellite coordinates LEO worker nodes, with predictable orbital topology pre-stored to reduce routing latency.

As DePIN satellites generate massive amounts of heterogeneous data, transmitting all raw data to the Earth is constrained by bandwidth and latency.
Decentralized Federated Learning (DFL) is essential to address this bottleneck and realize true on-orbit distributed intelligence, as illustrated in Fig.~\ref{fig:Satellite}.
Mwase \textit{et al.}~\cite{Mwase_SC_WCNC} proposed a fully onboard DFL framework that selects satellites according to contextual relevance and maintains consensus under dynamic orbital topology, enabling satellites to act as decentralized learning contributors without centralized raw-data collection.
To address the resolution heterogeneity of satellite imagery, Zhao \textit{et al.}~\cite{Zhao_SC_TMC} first applied a super-resolution model locally on each satellite before aggregation, improving the consistency of heterogeneous satellite contributions before collaborative training.
To improve communication efficiency, Zhai \textit{et al.}~\cite{Zhai_SC_TMC} introduced a multi-hop offloading scheme leveraging the Ring Allreduce algorithm within orbital planes, aligning aggregation with constellation geometry and reducing communication burden for distributed DePIN learning.
For more related works, please refer to \cite{Zhai_DML_TMC, Duan_Metaverse_TSC, Kawamoto_Traffic_IoTJ} and the surveys \cite{Oh_SC_ICAIIC, Fontanesi_SC_COMST, Matthiesen_SC_Network}, which provide complementary insights into resource control, satellite FL, and AI-enabled satellite communication.

\subsubsection{Blockchain Empowerment}
For satellite networks in DePIN, blockchain provides secure authentication and, more critically, enables asset tokenization and programmable service coordination.
Torky \textit{et al.} \cite{torky_SC_LISS} treated satellites as blockchain nodes, where satellite bandwidth and sensor data are digitized into space digital tokens and verified through consensus mechanisms.
Wang \textit{et al.} \cite{Wang_SC_TAES} integrated blockchain into LEO satellite constellations for IoT authentication. The method employs a consortium blockchain to record device registration and ensure certificate transparency, extends Merkle Patricia Trees to build a decentralized certificate repository, and applies certificateless encryption to generate key pairs for smart devices.
The study \cite{torky_SC_LISS}  proposed a novel blockchain consensus, ``Proof of Space Transaction," which ensures secure and tamper-proof inter-satellite communications using cryptographic evidence and verifier-free validation.

\subsubsection{Satellite-Based Trust Anchors}
A potential role of satellite systems in DePIN is to provide trusted reference points that are difficult to manipulate under censorship or jurisdictional pressure. 
Since decentralized P2P systems can improve resilience when coordination and governance are no longer concentrated in a few entities~\cite{Daniel_IPFS_COMST}, LEO satellites may extend this logic to the physical layer by hosting validator functions, threshold signing services, or multisignature control points that are geographically separated from terrestrial infrastructure~\cite{Paneri_P2P_SNPD}. 
This points to a long-term DePIN trust architecture that combines logical decentralization with physical and geopolitical separation.

\subsection{Radio Access Networks}
\begin{figure}[!t]
    \centering
    \includegraphics[scale=0.3]{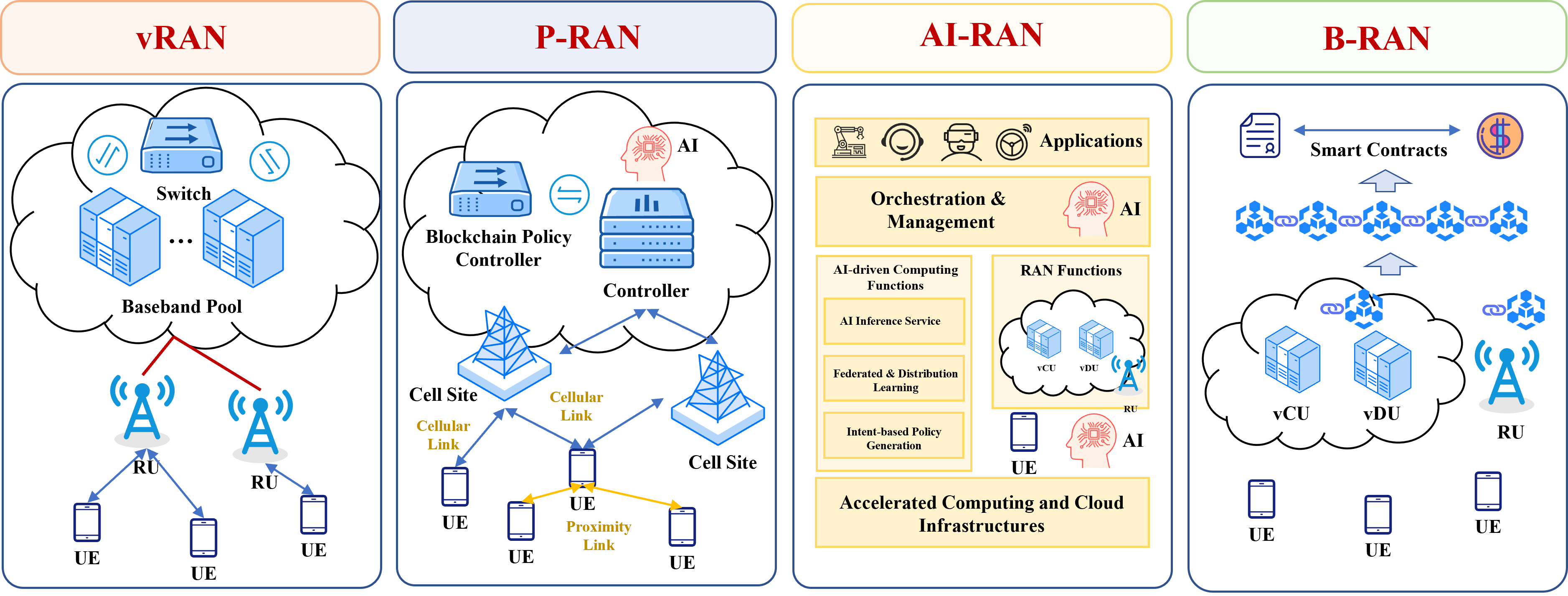}
    \caption{The schematic architectures of four types of RAN: vRAN, P-RAN, AI-RAN, and B-RAN.}
    \label{fig:RAN}
\end{figure}

Radio Access Network (RAN) is a vital component of mobile communication systems. In DePIN, it is unique, as it is deployed by prosumers yet must provide carrier-grade service. This necessitates a shift from proprietary hardware to open, verifiable, and incentive-compatible architectures.
Fig. \ref{fig:RAN} illustrates the schematic architecture of four types of RANs that may bridge this gap: Virtual RAN (vRAN), Proximity RAN (P-RAN), AI RAN (AI-RAN), and Blockchain-Enabled RAN (B-RAN). Table \ref{tab:RAN} compares these architectures.

\subsubsection{vRAN}
In vRAN, a Baseband Unit (BBU) based on software virtualization technology can operate on Commercial Off-The-Shelf (COTS) servers. This enables the separation of hardware and software components and the use of standardized hardware. Network flexibility and reliability are enhanced while hardware costs are significantly reduced \cite{Foukas_vRAN_SIGCOMM}. This approach aligns with a user-centric design philosophy, which is essential for access networks oriented towards DePIN.

\subsubsection{P-RAN}
P-RAN was first proposed in \cite{Bi_RAN_Network} to extend the coverage range and enhance the system capacity of wireless networks by integrating traditional cellular networks with proximal networks, such as Device-to-Device (D2D) technology in smartphones. The core concept involves leveraging smartphones as relay nodes to expand the coverage area of cellular base stations, following a software-defined network architecture. Driven by the token economy, more nodes are encouraged to join to expand the network \cite{Bi_P-RAN_Magazine}.
P-RAN faces challenges, such as a lack of user privacy protection, relay device certification, and the need for dynamic management of relay discovery, link quality, interference, and handovers~\cite{Xia_Localization_TWC}.

\subsubsection{AI-RAN}
The core vision of AI-RAN is to establish a more adaptive, intelligent, efficient, and multifunctional RAN, thereby enhancing network performance, enabling new business opportunities, and advancing the future of network infrastructure.
This vision has been articulated in a white paper \cite{AIRAN_whitepaper} released by the AI-RAN Alliance, a consortium of industry leaders, research institutes, and academia.
This aligns with the fundamental concept of DePIN.
The architecture and operational mechanisms of AI-RAN remain an open issue.
Within the wave of DePIN, AI-RAN has the potential to amplify collaboration among vendors, operators, and cloud service providers, thereby enabling more efficient utilization of network and spectrum resources through diverse approaches \cite{Khan_AIRAN_COMSOC, Batool_AMADRL_IoTJ}.
By developing a blockchain-based DePIN protocol, multiple stakeholders can contribute to and share network resources without relying on trusted third parties, while smart contracts automate resource allocation and payments, ensuring fair compensation for contributors.

\begin{table*}
    \centering
    \captionsetup{
    font={scriptsize}}
    \caption{Comparison of different RANs within DePIN} \label{tab:RAN}
    \fontsize{6pt}{7pt}\selectfont
    \begin{tblr}{
        width = \linewidth,
        colspec = {X[1]X[3]X[2]X[1.3]X[1.2]X[2.4]X[1]},
        row{1} = {c},
        cell{1}{1} = {r=2}{c},
        cell{1}{2} = {r=2}{c},
        cell{1}{3} = {r=2}{c},
        cell{1}{4} = {r=2}{c},
        cell{1}{5} = {r=2}{c},
        cell{1}{6} = {r=2}{c},
        cell{1}{7} = {r=2}{c},
        cell{3}{1} = {r=2}{c},
        cell{3}{2} = {r=2}{j},
        cell{3}{3} = {r=2}{l},
        cell{3}{4} = {r=2}{c},
        cell{3}{5} = {r=2}{c},
        cell{3}{6} = {r=2}{l},
        cell{3}{7} = {r=2}{c},
        cell{5}{1} = {r=2}{c},
        cell{5}{2} = {r=2}{j},
        cell{5}{3} = {r=2}{l},
        cell{5}{4} = {r=2}{c},
        cell{5}{5} = {r=2}{c},
        cell{5}{6} = {r=2}{l},
        cell{5}{7} = {r=2}{l},
        cell{7}{1} = {r=2}{c},
        cell{7}{2} = {r=2}{j},
        cell{7}{3} = {r=2}{l},
        cell{7}{4} = {r=2}{c},
        cell{7}{5} = {r=2}{c},
        cell{7}{6} = {r=2}{l},
        cell{7}{7} = {r=2}{c},
        cell{9}{1} = {r=2}{c},
        cell{9}{2} = {r=2}{j},
        cell{9}{3} = {r=2}{l},
        cell{9}{4} = {r=2}{c},
        cell{9}{5} = {r=2}{c},
        cell{9}{6} = {r=2}{l},
        cell{9}{7} = {r=2}{c},
        hlines,
        vline{2-7} = {},
        }
        \textbf{Type} & \textbf{Core Concept}                             & \textbf{User Role}                       & \textbf{Hardware Dependency} & \textbf{Intelligence Level} & \textbf{DePIN Reflection}                              & \textbf{Refs.}                                                                                                  \\ \\ \hline

        vRAN          & Virtualized BBU with hardware-software decoupling & Passive consumer                         & Low                          & Medium                      & \bluecircle \texttt{} Hardware decoupling \newline \bluecircle \texttt{} virtualization      & \cite{Foukas_vRAN_SIGCOMM}                                                                                      \\ \\
        P-RAN         & Coverage extension via proximal relay nodes       & Active relay; Contributor                & Medium                       & Low to medium               & \bluecircle \texttt{} User autonomy \newline \bluecircle \texttt{} Economic incentives       & \cite{Bi_RAN_Network, Bi_P-RAN_Magazine}                                                                        \\ \\
        AI-RAN        & AI-driven architecture                            & Active contributor                       & Medium                       & High                        & \bluecircle \texttt{} Shared infrastructure \newline \bluecircle \texttt{} Intelligent RAN   & \cite{AIRAN_whitepaper, Khan_AIRAN_COMSOC, Batool_AMADRL_IoTJ}                                                                      \\ \\
        B-RAN         & Blockchain-based decentralized management         & Governance; \newline Service co-provider & Medium                       & High                        & \bluecircle \texttt{} Decentralized trust \newline \bluecircle \texttt{} Autonomous services & \cite{Wang_BRAN_DCN, Xu_BRAN_arXiv, Ling_BRAN_WirelessBlockchain, Cao_BRAN_TWC, Le_BRAN_IoTJ, Ling_BRAN_IEEEWC} \\ \\
    \end{tblr}
\end{table*}

\subsubsection{B-RAN}
Blockchain technology is increasingly integrated into RAN architectures to promote the open, green, smart, and flexible development of RAN within DePIN.
For instance, Wang \textit{et al.} \cite{Wang_BRAN_DCN} proposed a blockchain-enabled fog RAN architecture that integrates blockchain technology across different network layers to address the centralization, security, and efficiency challenges of conventional RAN. In this design, a consensus mechanism is employed for physical-layer authentication; smart contracts are leveraged to implement authorization-free access control; and a blockchain-based distributed platform is established for dynamic multi-resource management.
BeMutual was introduced in \cite{Xu_BRAN_arXiv} to support P2P communication between users under the same RU by blockchain-enabled mutual authentication. The system assigns each user and network element a globally unique blockchain address (BCADD), derived as a hash of the user's public key. BCADD serves as an anonymous identity and is bound to the physical address, enabling decentralized mutual authentication at the RAN level, where users can locally complete identity verification and generate secure P2P communication keys.
Moreover, Ling \textit{et al.} \cite{Ling_BRAN_IEEEWC} introduced a blockchain-enabled architecture that establishes a large-scale, trustworthy, and efficient RAN among mutually untrusted entities. They employed a hash-based access mechanism, where users gain access by solving hash puzzles, and hash time-locked contracts ensure fair transactions between service providers and clients. The dual-layer architecture of the mainchain and sidechains balances scalability with security.

\subsection{Edge Micro Data Centers}
Edge micro data centers are essentially small, distributed data centers deployed at the network edge, with the core advantage of delivering low-latency and high-efficiency services \cite{Bilal_MDC_CN}.
Edge micro data centers naturally align with the physical layer foundation of DePIN, functioning as ``nodes'' or ``physical resource contributors'' within the network. Under the DePIN model, these distributed micro data centers are no longer owned and operated by a single centralized entity, e.g., large cloud service providers, but are jointly built and managed by decentralized community members \cite{Cao_MDC_TII}.
However, current economic and technological conditions are more suitable for a hierarchical and moderately decentralized system, rather than a fully decentralized deployment \cite{Bruschi_MDC_NFVSDN}.
From the perspective of renewable energy integration, Chakraborty \textit{et al.} \cite{Chakraborty_MDC_TSC} emphasized deploying local renewable energy in edge micro data centers to reduce dependence on centralized power grids. They improved energy efficiency through overload reservation, virtual machine consolidation, and migration. A green service level agreement (SLA) was further introduced, allowing customers to tolerate limited overload when renewable energy is insufficient in exchange for incentives such as discounts or carbon tax reductions.
On the operational level, Scano \textit{et al.} \cite{Scano_MDC_Access} suggested using Kubernetes orchestration and P4-based network telemetry to achieve coordinated optimization of computing and networking resources.

\subsection{Summary and Lessons Learned}

The transformation of the physical infrastructure layer under DePIN signifies a fundamental departure from the traditional centralized ICT architectures. Although extensive literature \cite{Espinel_SDN_COMST, Wang_MEC_WFIoT, Zhang_Metaverse_COMST} has envisioned the future of infrastructure under decentralized paradigms, a hybrid model \cite{Sinaeepourfard_Hybrid_SusTech} remains the more practical deployment approach, given current technological, economic, and regulatory constraints.
The hybrid model provides a pragmatic phased path for DePIN deployment by combining centralized efficiency in the initial build-out with decentralized incentives for long-term expansion. Centralized investment can first establish the core backbone; stable cash flow can later support progressive decentralization.

Existing challenges include: 1) immature mechanisms for linking token incentives with verifiable physical resource contributions; 2) expanded attack surfaces caused by open participation; and 3) insufficient coordination frameworks for multi-party infrastructure operation. Future breakthroughs may rely on: 1) consensus algorithms tailored to physical resource measurement and verification; 2) standardized trusted execution environment (TEE) and lightweight identity systems for edge devices; and 3) efficient edge intelligence networks for coordinating computing, storage, and transmission resources.

\section{Blockchain Layer: Scalability and Interoperability}\label{sect:blockchain}
The blockchain layer functions as the technical foundation for decentralized ICT infrastructure within DePIN. While enabling trustworthy collaboration, blockchain faces critical bottlenecks. Foremost among these are scalability
and cross-chain interoperability.
This section examines these bottlenecks to assess how existing technologies can be repurposed for decentralized infrastructure implementation.

\subsection{ Scalability for Physical Proof and Service Settlement}
In DePIN, the core bottleneck lies in managing physical proofs. Since proofs of physical services depend on heterogeneous environments, the evidence may be delayed, incomplete, or disputed.  
Scalability requires mechanisms that can record proof commitments, preserve auditable evidence, support delayed verification, and settle rewards without blocking service execution~\cite{Assen_performance_ICBC}.
Table \ref{tab:Scalability} summarizes representative Layer-1 and Layer-2 scalability techniques for DePIN systems.

\begin{table*}[t]
    \centering
    \captionsetup{
    labelfont={color=black},
    textfont={color=black},
    font={scriptsize}
    }
    \caption{Summary of Scalability Technologies on the Blockchain Layer of ICT Infrastructure within DePIN.}
    \label{tab:Scalability}
    \fontsize{6pt}{7pt}\selectfont
    \begin{tblr}{
        width = \linewidth,
        colspec = {
            X[1.25,c]
            X[0.8,c]
            X[0.8,c]
            X[2.8,l]
            X[1.7,c]
            X[1.8,c]
            X[1.7,c]
        },
        cells = {m, fg=black},
        row{1} = {c, font=\bfseries},
        hlines = {black},
        vline{2-7} = {black},
    }
        Technology 
        & Type 
        & Refs. 
        & Proof Handling Mode in DePIN 
        & On-chain Evidence Footprint 
        & Finality / Dispute Delay 
        & Settlement Suitability \\ \hline

        Sharding 
        & Layer-1 
        & \cite{Mao_Sharding_SRDS, Chen_sharding_IoTJ, Hong_Sharding_INFOCOM, Yang_sharding_DLT}
        & Regional or service-domain proof streams
        & Medium
        & Medium
        & High \\

        DAG 
        & Layer-1 
        & \cite{Wang_DAG_ACM, Qu_DAG_TMC, Li_DAG_IoTJ, Yang_DAG_TVT}
        & Asynchronous sensing and witness events
        & Medium
        & Low
        & Medium \\

        State Channels 
        & Layer-2 
        & \cite{Mashru_SC_ICC}
        & Repeated small value service sessions
        & Low
        & Low in normal case; \newline High under dispute
        & Medium \\

        Sidechains 
        & Layer-2 
        & \cite{Yin_Sidechain_TDSC, Fang_Sidechain_IoTJ, Deng_Sidechain_IoTJ}
        & Application-specific infrastructure services
        & Low on main chain; \newline High on sidechain
        & Medium
        & High \\

        Rollups 
        & Layer-2 
        & \cite{Thibault_rollups_Access, Ye_rollups_SP, Rammouz_Helium_WiMob, Khoa_OR_Access, Mounica_ZK_ICAMLDA, Bai_ZK_IoTJ}
        & Batch aggregation of service records
        & Medium
        & Medium to high
        & High \\
    \end{tblr}
\end{table*}

\subsubsection{Sharding}
Sharding is a technique that partitions blockchain state and transaction processing into multiple parallel shards, reducing the need for every node to validate every transaction~\cite{Mao_Sharding_SRDS}.
Unlike typical blockchain sharding, which mainly partitions digital transactions and on-chain states, DePIN sharding should also organize physical proof streams according to geographic regions or service domains.
This provides a way to distribute physical proof workloads, but the key constraint is that physical proofs are not immediately verifiable when submitted.
If shard block production waits for full physical proof verification, block intervals may expand from seconds to minutes or hours, undermining the throughput.

A more practical DePIN sharding design is to decouple block production from reward finality: shards can maintain short block intervals for recording proof commitments or service receipts, while rewards are finalized only after evidence is available, validated, and passed the dispute window.
Under a conservative sequential verification-dispute model, if $T_p$ denotes the proof generation and validation latency and $T_w$ denotes the challenge window, then economic finality should satisfy $T_\text{final} \ge T_p + T_w$, where $T_\text{final}$ refers to irreversible reward settlement rather than block inclusion~\cite{Hong_Sharding_INFOCOM, Yang_sharding_DLT}.
This implies that physical proof latency should delay reward finality rather than shard block production.

Hong \textit{et al.}~\cite{Hong_Sharding_INFOCOM} proposed a hierarchical sharding model tailored to address node heterogeneity, a common situation in DePIN. This model leverages two types of shards: i-shards handle intra-shard transactions, and b-shards validate cross-shard transactions. This design reduces redundant sub-transactions, supports DePIN node heterogeneity, and improves security through Byzantine fault tolerance.
Yang \textit{et al.}~\cite{Yang_sharding_DLT} proposed Co-Sharding for large-scale IoT applications, where sub-chain ledgers associated with geographic regions are organized as shards and coordinated for cross-regional communication.
This benefits DePIN deployment by reducing storage requirements for lightweight nodes and improving query throughput across regions.
Nguyen \textit{et al.} \cite{Nguyen_Sharding_TMC} introduced MetaShard, a dynamic sharding mechanism combined with a Proof-of-Engagement consensus. They formulated the shard management as an NP-complete problem decomposed into subproblems via binary search, and solved using the Lagrangian multiplier method.

\subsubsection{Directed Acyclic Graph (DAG)}
DAGs have been investigated as an alternative to linear blockchain structures, enabling parallel transaction processing and asynchronous confirmation.
Such properties make DAGs suitable for DePIN, which involves frequent microtransactions and data transfers from edge devices \cite{Wang_DAG_ACM}. By allowing lightweight devices to engage in real-time communication and micropayments without block confirmation, DAGs provide an efficient and scalable foundation for DePIN \cite{Qu_DAG_TMC}.

Existing DAG-based studies provide insights for DePIN Layer-1 design.
For instance,
Li \textit{et al.} \cite{Li_DAG_IoTJ} proposed an efficient DAG-based blockchain architecture that employs a weight-chain rule for global ordering and a tree-structured gossip protocol with FL-based dynamic networking, enabling parallel consensus while maintaining security against view-splitting attacks.
Yang \textit{et al.} \cite{Yang_DAG_TVT} introduced a lightweight DAG-based blockchain with social relationship driven data pruning and hierarchical historical data trimming. 
Nodes store only data from their subscribed ``topic groups'', while older historical replicas are selectively pruned to reduce storage overhead without sacrificing data integrity and security.

From a DePIN perspective, DAG is useful for handling infrastructure events generated asynchronously before final settlement. 
By organizing sensing reports, witness responses, and service receipts under a partial order rather than forcing immediate global sequencing, DAG ledgers can tolerate delayed or intermittent submissions from heterogeneous devices.

\subsubsection{State Channels}
State channels offload repeated interactions from the main chain and submit only the final state for settlement.
This is relevant for DePIN, which often involves frequent, small-value, and repeated interactions.
Prior work~\cite{Mashru_SC_ICC} examined state-channel protocols for blockchain-based IoT networks, where off-chain computation and on-chain arbitration are combined to improve throughput and reduce transaction cost.

In DePIN state channels, off-chain updates should bind payment states to evidence of service delivery. A payment update may be tied to a signed service record, according to the specific infrastructure domain.
The final signed state records how balances should be settled, but it may not explain whether the corresponding service was delivered as claimed.
DePIN state-channel protocol must specify what evidence is retained off-chain, how long it remains auditable, and how it can be used during arbitration.
Without such rules, state channels may reduce on-chain cost while weakening service accountability.

\subsubsection{Sidechains}
Sidechains are independent blockchains that run in parallel with a main chain and maintain their own consensus and validation mechanisms.
This architecture offers distinct structural advantages in DePIN by enabling application-specific customization and isolation. DePIN services, which are typically diverse, require specialized operational logic and token models. The sidechain architecture allows DePIN projects to deploy dedicated sidechains to accommodate specific requirements for consensus algorithms, transaction fees, or data processing. 
This flexibility allows DePIN applications to optimize operations according to their business logic~\cite{Yin_Sidechain_TDSC}.

Recent studies provide useful evidence for this direction.
To manage massive IoT access and local model training, two crucial DePIN functions requiring high throughput and low latency, Fang \textit{et al.} \cite{Fang_Sidechain_IoTJ} utilized sidechains to offload transaction verification and model aggregation from the main chain. Sidechains are deployed for device management and model aggregation, with a leader node overseeing interaction to ensure secure cross-chain exchange and streamlined device authentication.
Deng \textit{et al.}~\cite{Deng_Sidechain_IoTJ} introduced a secure sidechain design that offloads verification from the main chain, avoids main-chain forking, and supports edge IoT devices. By decoupling consensus, it enables asset/data interoperability and supports scalable management of heterogeneous DePIN nodes.

\subsubsection{Rollups}
Rollups perform off-chain computation and compression of large transaction volumes, submitting only a cryptographic summary, e.g., zero-knowledge (ZK) proof (ZKP), to the main chain. 
In DePIN, rollups can translate asynchronous physical service records into verifiable settlement states. 
The distributed and time-varying nature of physical contributions requires rollup batches to preserve auditable proof commitments, so that verification and reward finalization can occur after service delivery.
This creates different design trade-offs: Optimistic Rollups are better suited to contestable evidence, whereas ZK Rollups are more effective when validity can be proven efficiently before settlement~\cite{Thibault_rollups_Access}.

Optimistic Rollups provide a natural fit for DePIN services that generate frequent records but cannot verify every physical contribution before state updates~\cite{Ye_rollups_SP}.
For DePIN, the challenge window governs the balance between reward liquidity and settlement safety.  
Finality should balance fraud prevention and reward timeliness to avoid premature fraudulent settlement and delayed honest incentives.
Optimistic Rollups for DePIN need a staged settlement process in which submitted proofs receive provisional recognition, while final reward release depends on the auditability of the corresponding physical evidence~\cite{Rammouz_Helium_WiMob}. 
The challenge window should be determined by the evidence latency and verification cost of the underlying service, not by a generic rollup setting. 
Fraud proofs are effective only when the challenger’s expected compensation is sufficient to cover the cost of reconstructing and submitting the required physical evidence~\cite{Khoa_OR_Access}.

\begin{table}[t]
\centering
    \captionsetup{
    labelfont={color=black},
    textfont={color=black},
    font={scriptsize}
    }
\caption{Comparison between digital state transitions and physical attestations in ZK circuits.}
\label{tab:ZKcircuits}
\scriptsize
    \begin{tblr}{
        width   = \columnwidth,
        colspec = {
            X[0.5,l]
            X[1.5,l]
            X[1.5,l]
        },
        cells   = {valign=m},
        row{1}  = {font=\bfseries},
        hlines = {black},
        vline{2-3} = {black},
    }

    Aspect &
    Digital state transition &
    Attestation in DePIN \\ \hline

    Input &
    Discrete and cryptographic &
    Continuous and noisy \\
    State relation &
    Exact and reproducible &
    Uncertain \\
    Circuit encoding &
    Exact arithmetic constraints &
    Quantization with tolerance bounds \\
    Limitation &
    Circuit/prover cost &
    Witness reliability and physical truthfulness 

    \end{tblr}
\end{table}

ZK Rollups provide DePIN with a privacy-preserving pathway to scalability by proving rule compliance over committed service records without exposing raw data. 
This is attractive for sensitive information, e.g., location traces, device identities, and telemetry records, where direct disclosure may compromise user privacy or commercial confidentiality.

The main difficulty encountered is that DePIN evidence is not clean digital state because of noisy environments.
For example, radio frequency propagation measurements vary with fading and interference, while GPS coordinates are affected by position errors, drift, or spoofing.
These continuous and environmental measurements should be quantized through range constraints, tolerance bounds, or multi-source consistency checks~\cite{Fioravanti_ZKP_OJCS} before they can be encoded in R1CS- or Plonk-compatible circuits.
Such transformations increase witness construction complexity and arithmetic constraints.
Moreover,
a valid ZKP proves that committed inputs satisfy a predefined relation; it does not establish the truthfulness of the underlying physical event~\cite{Wan_ZKP_TDSC}.
Table~\ref{tab:ZKcircuits} compares deterministic digital state transitions with noisy physical attestations in ZK circuits.

ZK Rollups in DePIN separate physical evidence attestation from proof generation.
Devices, witnesses, TEEs, or oracles first collect, sign, and commit physical measurements.
The committed evidence is transformed into circuit compatible representations, after which off-chain provers generate validity proofs for deterministic predicates~\cite{Bai_ZK_IoTJ}.
Existing studies also provide some complementary design approaches.
Particularly,
Tao \textit{et al.} \cite{Tao_ZKP_Access} used independently generated proofs and deterministic cross-verification to corroborate the same physical observation, reducing reliance on a single noisy report.
Ernstberger \textit{et al.} \cite{Ernstberger_ZKP_SP} proposed a floating point SNARK construction that is compliant with IEEE 754 standard, enabling accurate encoding of continuous location measurements.

\subsection{Cross-chain Interoperability in Multi-chain Systems}
The current blockchain ecosystem exhibits a fragmented landscape, where different blockchain networks (e.g., Ethereum, Filecoin, and Solana) operate independently, resulting in isolated ``data islands” and ``value islands” \cite{Duan_Cross_JAS}. For DePIN, this ``chain island” effect poses obstacles. Cross-chain interoperability is to break down these barriers and offer secure and trusted exchange of information and value,
mobilizing physical infrastructure assets  across heterogeneous networks.

\subsubsection{Cross-chain Data Transfer}
Cross-chain data transfer constitutes the primary challenge that must be addressed within the cross-chain technology framework. Its core lies in how to transfer state data or ledger records from the source blockchain to the target blockchain in a trustworthy and secure manner, thereby enabling interoperability and resource sharing across heterogeneous chains \cite{Zheng_Cross_CSP}.

The mainstream approach is to employ relay chains for centralized storage of block headers, thereby standardizing inter-chain state verification.
Wang \textit{et al.} \cite{Wang_Cross_ICISCAE} proposed a relay chain-based cross-chain identity scheme that uses verifiable credentials instead of complex simplified payment verification proofs. 
During cross-chain operations, user data is transferred to a multi-signature address managed by the relay chain for legitimacy verification and automated custody.
Meanwhile, the Cosmos project exemplifies the use of a relay-chain paradigm in cross-chain data transfer \cite{Cosmos_Cross}. Its core architecture, i.e., the Hub-and-Zone model, designates the Hub as the central hub for interchain communication, through which data can be indirectly transmitted across Zones once channel connections are established.

During the transfer process, a trusted third party needs to verify the state of the source chain on behalf of the receiver, referred to as a ``validator". Source-chain nodes are only permitted to establish such validators with authorization from the project team.
For example, Augusto \textit{et al.} \cite{Augusto_Cross_ICBC} employed gateways as validators, requiring each to be pre-authorized and certified while dynamically electing a coordinator gateway as the central control node to ensure the robustness of cross-chain communication.
Projects, such as \textit{Wormhole} \cite{Wormhole}, \textit{Nomad} \cite{Nomad}, and \textit{deBridge} \cite{deBridge}, adopt similar approaches to ensure the security and trustworthiness of cross-chain data transfer.

\begin{figure}[!t]
    \centering
    \includegraphics[scale=0.3]{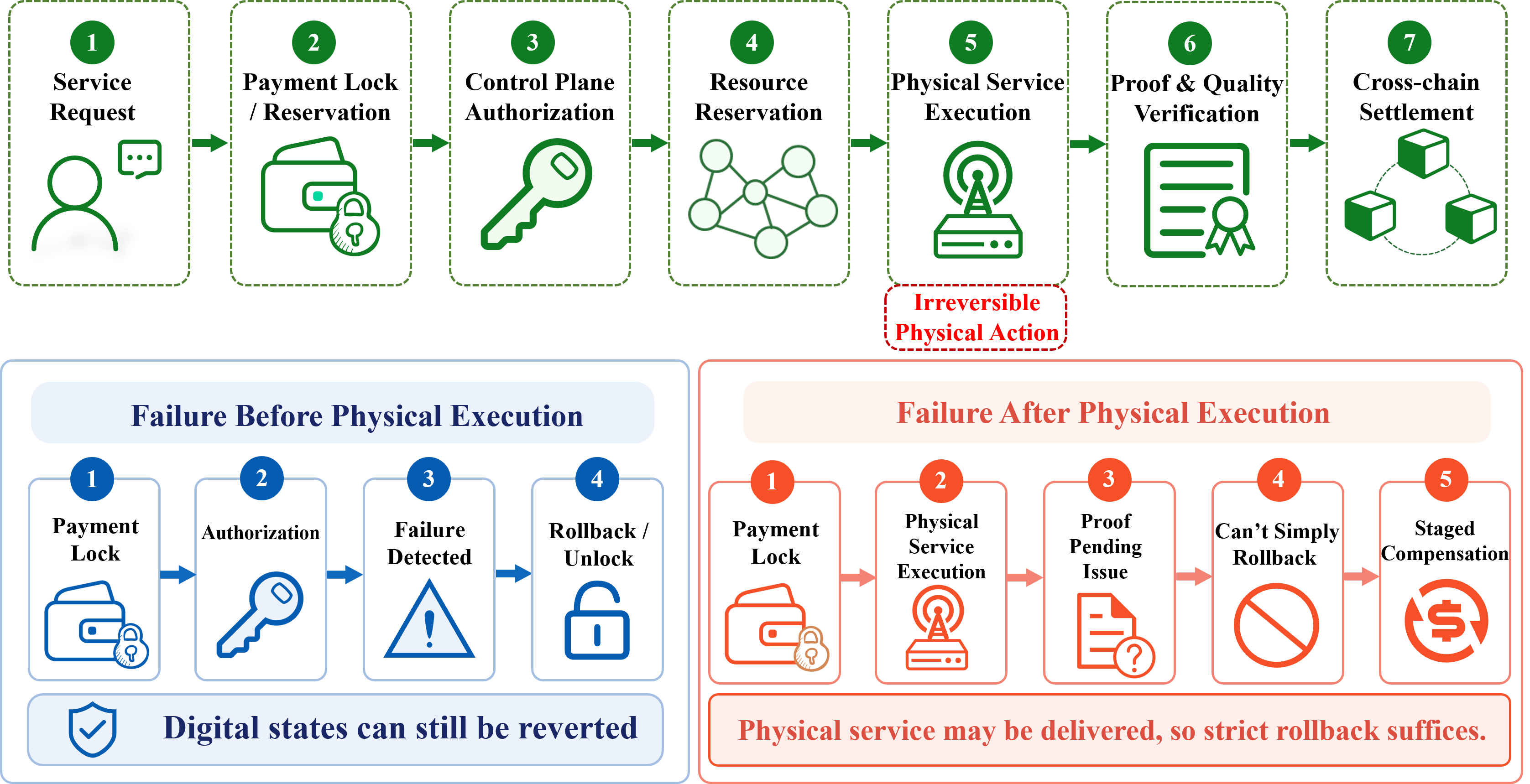}
    \caption{Atomicity in DePIN cross-chain services: digital rollback before physical execution and compensation after irreversible actions~\cite{Zhu_atomic_TrustCom, Shi_atomic_INFOCOM}.}
    \label{fig:atomicity}
\end{figure}

        \subsubsection{Atomicity of Cross-chain Operations}
        Cross-chain atomicity ensures that operations spanning multiple heterogeneous blockchains either succeed entirely or fail completely~\cite{Han_atomicity_DLT}.
        In DePIN, the atomic unit extends to a service-based process that may jointly involve payment settlement, state transition, proof generation, control plane authorization, and physical service execution.
        DePIN interactions are tightly coupled with off-chain devices and persistent infrastructure services, as illustrated in Fig. \ref{fig:atomicity}.

        \textit{A key challenge of atomicity in DePIN is the inherent irreversibility of physical actions.}
        Unlike digital assets that can be cryptographically locked and reverted upon failure,
        physical service execution, once initiated, cannot be undone.
        This asymmetry between the reversibility of on-chain states and the irreversibility of off-chain physical actions makes DePIN vulnerable to one-sided failures, e.g., successful payment without service delivery, or completed service execution without cross-chain settlement.
        Current studies, while addressing atomicity in the context of cross-chain asset exchange, provide insights that can be extended to DePIN.
        Zhu \textit{et al.}~\cite{Zhu_atomic_TrustCom} proposed a fault-tolerant atomic swap scheme that supports multiple concurrent swap paths, enabling automatic failover to alternative paths upon single-path failures and improving the robustness of cross-chain transactions.
        Although this design remains asset-centric, its key insight \textemdash decoupling digital path confirmation from actual execution commitment \textemdash translates meaningfully to DePIN. Confirming a viable cross-chain settlement path before triggering physical services preserves a fallback window before irreversible physical actions occur.
        Meanwhile, ARC~\cite{Zhang_atomic_EdgeCom} and RAC-Chain~\cite{Xie_atomic_TMCCA} show that fault tolerant cross-chain coordination in asynchronous environments can be strengthened through relay chains, asynchronous consensus, global state-machine replication, and explicit finality protocols.
        A relevant work is represented by BrokerAS~\cite{Shi_atomic_INFOCOM}, which
        relaxes strict all-or-nothing semantics through asset partitioning, staged execution, refund procedures, and collateral-based accountability. This indicates that when physical rollback is infeasible, DePIN atomicity should incorporate staged settlement and compensation mechanisms rather than relying solely on binary transaction success or failure.

        \textit{Another key challenge lies in the nature of multi-chain, multi-stage service orchestration.}
        A complete DePIN service pipeline is inherently sequential and cross-domain.
        This requires a form of compositional atomicity, where dependent service, verification, and settlement actions must remain consistent across stages. 
        Lu \textit{et al.}~\cite{lu_atomicity_arxiv} proposed a protocol for atomic execution of general transactions whose operations may span several chains, which is aligned with the needs of DePIN service workflows.
        Robison \textit{et al.}~\cite{Robinson_atomicity_BRA} showed that atomic cross-chain execution can be organized as a composable function call tree, where inter-contract and inter-blockchain function calls are executed synchronously and atomically under coordination contracts and threshold-signature verification.
        This model is insightful for DePIN scenarios where a service process may involve chained steps across blockchains.
        Cai \textit{et al.}~\cite{Cai_atomicity_SRDS} moved beyond weaker notions of financial atomicity by targeting complete atomicity through a layered execution framework, where state transitions across chains are cached, synchronized, and serially committed.

\subsection{Summary and Lessons Learned}

This section has examined scalability and cross-chain interoperability as two fundamental challenges for the blockchain layer of DePIN. Both directly affect whether DePIN can support large-scale, multi-party, and multi-service deployment. 
From the scalability perspective, Layer-1 techniques, such as sharding and DAG, improve parallel transaction and data processing, whereas Layer-2 approaches, including state channels, sidechains, and rollups, offload or aggregate workloads to reduce on-chain congestion and settlement costs. From the interoperability perspective, cross-chain data transfer and atomicity mechanisms aim to mitigate the “chain island” problem and enable secure, verifiable coordination across heterogeneous blockchains.

Several open challenges remain. 
1) sharding and DAG-based designs must balance performance gains with security guarantees, especially under adversarial conditions and delayed physical-proof verification; 
2) Layer-2 solutions introduce trade-offs among latency, system complexity, data availability, and trust assumptions. Hybrid designs, such as DAG-based processing within shards or reputation-based sidechain consensus, may provide promising directions for DePIN-oriented scalability;
3) cross-chain interoperability still lacks unified standards, as heterogeneous consensus mechanisms, data structures, and finality models complicate reliable cross-chain communication and settlement.

\begin{table*}[t]
    \centering
    \captionsetup{
    labelfont={color=black},
    textfont={color=black},
    font={scriptsize}
    }
    \caption{Comparison of interaction layer mechanisms in DePIN, where the marker `\bluecircle' indicates representative mechanisms and `\redcross' denotes main limitations in DePIN.}
    \label{tab:Interaction}
    \fontsize{6pt}{7pt}\selectfont
    \begin{tblr}{
        width = \linewidth,
        colspec = {X[1]X[1.5]X[0.8]X[3.7]X[3.9]X[4.6]},
        cells = {m, fg=black},
        column{1-3} = {c},
        column{4,6} = {l},
        column{5} = {j},
        row{1} = {c, font=\bfseries},
        row{1} = {c, font=\bfseries},
        cell{2}{1} = {r=2}{c},
        cell{4}{1} = {r=2}{c},
        cell{6}{1} = {r=3}{c},
        hlines = {black},
        vline{2-6} = {black},
    }
        Type 
        & Aspects 
        & Refs. 
        & Representative Mechanisms 
        & Main Role in DePIN 
        & Main Limitations in DePIN \\ \hline

        P2P Network
        & Decentralized Network Protocols
        & \cite{Zhou_BEDNS_WFIoT, Srinivasa_IoT_IMC, Lygerou_Honeypot_IJIS}
        & \bluecircle \texttt{} Blockchain-enabled DNS \newline
          \bluecircle \texttt{} Decentralized IoT honeypot \newline
          \bluecircle \texttt{} P2P protocol stack
        & Map service identifiers to endpoints and support secure service discovery
        & \redcross \texttt{} Heterogeneous protocol stacks \newline
          \redcross \texttt{} Scalable identity resource mapping \\

        & Streaming Media Transmission
        & \cite{Hao_stream_TNSM, Farahani_stream_GLOBECOM, Ding_stream_MM}
        & \bluecircle \texttt{} Trust-enhanced P2P broadcast \newline
          \bluecircle \texttt{} Hybrid P2P-CDN
        & Support continuous service delivery and traceable contribution records
        & \redcross \texttt{} Real-time QoS is vulnerable to churn \\

        Topology
        & Topology Structures
        & \cite{Xu_PoR_Network, Gouveia_topology_ICNP}
        & \bluecircle \texttt{} SnowedNet \newline
          \bluecircle \texttt{} Kollaps
        & Provide scalable organization for large-scale decentralized node networks
        & \redcross \texttt{} Logical topology may not reflect physical infrastructure states \\

        & Topology Optimization
        & \cite{Li_topology_TCOM, Ali_topology_Access, Palmieri_topology_ICDCS}
        & \bluecircle \texttt{} DRL-based topology search \newline
          \bluecircle \texttt{} Graph-based link refinement
        & Adapt network structure
        & \redcross \texttt{} Rely on accurate observations \newline
          \redcross \texttt{} Generalization is limited \\

        Routing
        & Decentralized Path Discovery
        & \cite{Seyyedabbasi_route_Micro, Mostafizi_route_ITS, Li_route_ICRA}
        & \bluecircle \texttt{} ACO-based multi-agent routing \newline
          \bluecircle \texttt{} Priority-aware path coordination \newline
          \bluecircle \texttt{} Decentralized cooperative routing
        & Enable path construction without global control
        & \redcross \texttt{} Path quality fluctuates under churn \\

        & Intent-aware Intelligent Routing
        & \cite{Anh_Quang_route_ICC, Ma_route_CN}
        & \bluecircle \texttt{} Intent-based routing optimization \newline
          \bluecircle \texttt{} Predictive adaptive routing
        & Translate service intents into routing policies
        & \redcross \texttt{} Policy enforcement is complex \newline
          \redcross \texttt{} Coordination between controllers and edge nodes \\

        & Robust Routing
        & \cite{Maheswari_robust_ICICES, Murali_robust_TNSM, Lu_route_Automatica, Guo_route_IoTJ}
        & \bluecircle \texttt{} Blockchain-based registry \newline
          \bluecircle \texttt{} DRL-based secure routing \newline
          \bluecircle \texttt{} Attack-resilient routing
        & Improve route authenticity, attack resistance, and resilience
        & \redcross \texttt{} Attack coverage is limited \newline
          \redcross \texttt{} Auditable routing verification at scale \\
    \end{tblr}
\end{table*}

\section{Interaction Layer: Efficient and Resilient Transmission}\label{sect:interaction}
The interaction layer functions as the vital nexus bridging off-chain physical infrastructure with the on-chain digital world in DePIN.
Unlike centralized cloud architectures, where service delivery is supported by managed data centers and stable backbone networks, DePIN operates on heterogeneous community-operated devices whose availability, performance, and trustworthiness vary over time~\cite{Lin_RW_1}.
The interaction layer must jointly support three coupled flows: service data flows, control and coordination flows, and evidence flows.
This makes standard networking protocols insufficient for DePIN.
The following discussion investigates P2P communication, topology management, and routing from this perspective.

\subsection{P2P Network Communication}
In DePIN, core mechanisms, e.g., transmission, topology, and scheduling, rely on P2P protocols to coordinate distributed physical resources and support service-related information exchange \cite{pacitti_dep2p_Springer}. 
Streaming media transmission provides a representative scenario for evaluating the real-time reliability and scalability of such protocols in DePIN.

\subsubsection{Decentralized Network Protocols}
Internet communication has relied on protocols, e.g., TCP/IP, UDP, and HTTP. 
These protocols provide the basic connectivity substrate for DePIN.
Their original design does not fully address the operational requirements of decentralized physical infrastructure.

Recent research highlights the shift toward decentralized protocols within the DePIN framework.
By integrating blockchain technology, such protocols can distribute critical coordination information across the network, mitigating single points of failure and strengthening resilience against cyberattacks \cite{Khan_6G_WCM}. 
Decentralized protocols also provide a mechanism for DePIN to connect communication endpoints with verifiable physical service claims.
A representative example is the blockchain-enabled Domain Name System (DNS) \cite{Zhou_BEDNS_WFIoT}.
By storing domain bindings on blockchains, it ensures immutability and prevents DNS hijacking, single points of failure, distributed denial of service (DDoS) attacks, and spoofing.
In DePIN, this mechanism can be a building block for mapping service identifiers to physical or logical resources, supporting nearby service discovery without centralized registries and verifying the legitimacy of service endpoints.

The explosive growth of IoT devices exacerbates the need for DePIN protocol security.
A study \cite{Srinivasa_IoT_IMC} revealed that, in 2021, the number of protocol-misconfigured devices reached 1.8 million.
Protocol misconfiguration, weak authentication, and exposed services can be entry points for attacks against the decentralized infrastructure. 
Honeypots support early warning in DePIN protocol security by attracting malicious traffic and revealing attack patterns before real devices are compromised.
They were adapted in \cite{Lygerou_Honeypot_IJIS} to transform mobile phones into virtual IoT devices operating over modern IoT protocols.
The system adopts a modular plug-and-play design, enables dynamic switching among virtual IoT honeypots on the same device, supports a mobile device as a honeypot agent, and decentralizes the logical layer.

\subsubsection{Streaming and Continuous Service Transmission}
P2P communication is important for streaming media, where real-time delivery, privacy, and scalability must be jointly supported. 
From the DePIN perspective, streaming media represents a broader class of continuous or session-based services, in which efficient transmission should be coupled with verifiable records of service continuity and quality.

The hybrid P2P-Content Delivery Network (CDN) architecture has proven effective for large-scale, high-quality streaming \cite{Farahani_stream_GLOBECOM}. 
It integrates P2P delivery, CDN support, edge computing, and online learning to adaptively select segment sources, thereby improving cost efficiency, scalability, and quality of service (QoS).
For DePIN, this architecture is relevant to decentralized streaming services, in which community-operated nodes can provide peer-side delivery capacity and CDN resources mitigate fluctuations in peer availability.
It offers insights for continuous service transmission in DePIN, where local peer collaboration, infrastructure fallback, and QoS-aware adaptation need to be jointly considered.

The migration of P2P streaming toward DePIN intensifies concerns about privacy and trust. 
This may distort contribution attribution, reputation evaluation, and reward settlement, since delivery records can be linked to economic incentives~\cite{He_P2P_TPDS}. 
To address this challenge, Ding \textit{et al.}~\cite{Ding_stream_MM} integrated blockchain into a hybrid P2P-CDN stream-on-demand architecture. 
Their system combines P2P-based parallel chunking with CDN redundancy to maintain high-resolution streaming; blockchain and ZKP provide privacy-preserving verification and traceable transaction records.

\subsection{Dynamic Topology Management}
Unlike traditional enterprise networks, the P2P networks in DePIN are decentralized and highly dynamic, where nodes can join or leave at any time \textemdash a phenomenon known as node churn. 
In DePIN, 
frequent topology changes may further compromise network connectivity and stability, potentially leading to routing failures or network collapse. 
Designing topology management that can maintain stability and resilience under high churn conditions is of critical importance.

\subsubsection{Topology Structures}
\begin{figure}[t]
    \centering
    \includegraphics[width=0.3\textwidth]{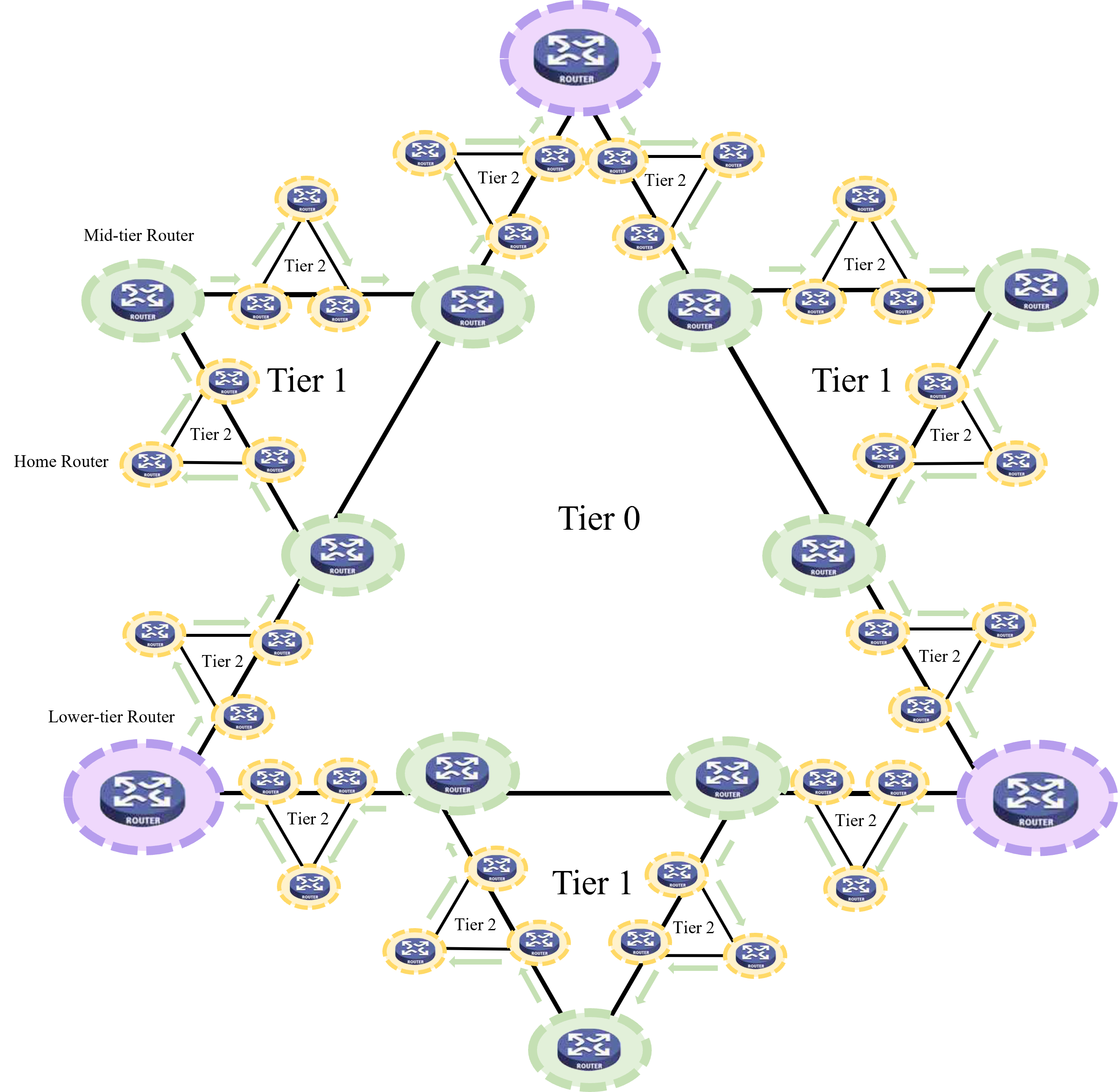}
    \caption{SnowedNet: A decentralized topology architecture based on hyperdimensional n-simplex fractal \cite{Xu_PoR_Network}}
    \label{fig:SnowedNet}
\end{figure}

To enhance network scalability, Xu \textit{et al.} \cite{Xu_PoR_Network} introduced SnowedNet, a decentralized topology architecture constructed upon a hyperdimensional $n$-simplex fractal. SnowedNet establishes a spatiotemporal coordinate framework tailored for Web 3.0, where blockchain technology is employed to realize discrete-time operation and consensus. As depicted in Fig. \ref{fig:SnowedNet}, the network is constructed based on a Koch fractal, where the numbers of nodes and links grow iteratively and, in principle, can expand without bound. Each node establishes a specific topological relation determined by its port degree; for instance, when the port degree is $D$, the node is associated with $D$ connection arms, each maintaining an identical hop distance within the topology.

To bridge the gap between theory and simulation, Gouveia \textit{et al.} \cite{Gouveia_topology_ICNP} proposed Kollaps, a decentralized network topology simulation framework. 
By adopting a decentralized architecture and a simplified simulation model, Kollaps addresses the challenges of evaluating large-scale distributed systems. 
It offers a flexible, scalable, accurate, and easy-to-use simulation tool, making it suitable for experimental scenarios that involve dynamic network conditions and large-scale containerized applications.

\subsubsection{AI/ML-based Topology Control}
Conventional P2P networks rely on periodic ``keep-alive” mechanisms to mitigate node churn, incurring significant maintenance overhead. Machine learning (ML) can be employed to analyze node behaviors and topological dynamics, thereby enabling adaptive adjustments to the network topology.
From the DePIN perspective, topology control should consider physical resource states, node reliability, energy availability, and the potential impact of topology changes on service verification and incentive allocation.
Li \textit{et al.} \cite{Li_topology_TCOM} proposed a graph search algorithm based on deep reinforcement learning (RL), where a graph neural network (GNN) is employed to efficiently approximate topology scores for performance evaluation. Leveraging RL strategies, the algorithm explores the topology space to identify optimal solutions.
Ali \textit{et al.} \cite{Ali_topology_Access} extended this idea to topology optimization by leveraging a graph attention network to identify critical links and employing two cooperative agents, one adding and the other removing links, to iteratively refine the topology.
In addition, Palmieri \textit{et al.} \cite{Palmieri_topology_ICDCS} investigated the impact of different network topologies on decentralized learning and revealed several key factors: the role of hub nodes in knowledge dissemination, the complex influence of network connectivity, and the inhibitory effect of community structures on information propagation.

\subsection{Routing Algorithms}
Traditional routing algorithms, such as the Dijkstra algorithm for shortest-path computation \cite{Makariye_route_ICIOT} or the Bellman-Ford algorithm which discovers cheaper routes through iterative edge relaxation \cite{Karthikeyan_route_ICSSAS}, typically rely on global network knowledge and centralized control.
In the highly decentralized, volatile environment of DePIN, acquiring global information is difficult and contradicts the decentralized ethos.
DePIN routing mechanisms must evolve from optimizing for ``shortest path'' to becoming ``QoS-verifiable and incentive sensitive,'' ensuring reliable service delivery in the absence of central authority.

\subsubsection{Decentralized Path Discovery}
Seyyedabbasi \textit{et al.} \cite{Seyyedabbasi_route_Micro} developed an improved ant colony optimization method for decentralized IoT routing. 
By incorporating residual energy, buffer size, and traffic load, this scheme adapts pheromone- and distance-based routing to resource constrained DePIN edge nodes. 
Its support for multiple source-destination pairs further suits concurrent data transmission in decentralized IoT systems.
Li \textit{et al.} \cite{Li_route_ICRA} proposed a decentralized path discovery method termed Prioritized Communication Learning. This approach leverages imitation learning to acquire implicit priorities and constructs dynamic communication topologies based on these priorities to guide decentralized agents in cooperative path planning.

Routing and path discovery in decentralized networks, such as those in DePIN ecosystems, are evolving toward more intelligent and adaptive paradigms. 
RL agents can be deployed at the network edge or within nodes, enabling them to observe real-time physical states.
Combined with GNN, these observations can be used to derive routing strategies that jointly optimize transmission efficiency, service continuity, and the generation of auditable records for later verification.

\begin{table*}[t]
    \centering
    \captionsetup{
    labelfont={color=black},
    textfont={color=black},
    font={scriptsize}
    }
    \caption{Representative threat models in DePIN. The marker `\redcircle' denotes main impacts, while `\bluecircle' denotes defense directions.}
    \label{tab:Threat}
    \fontsize{6pt}{7pt}\selectfont
    \begin{tblr}{
        width = \linewidth,
        colspec = {X[0.8]X[3]X[1.8]X[1.8]},
        cells = {m, fg=black},
        column{1} = {c, font=\itshape},
        column{2} = {j},
        column{3-4} = {l},
        row{1} = {c, font=\bfseries},
        hlines = {black},
        vline{2-4} = {black},
    }
    Threat Model 
        & Adversarial Capability 
        & Main Impact on DePIN 
        & Defense Direction \\ \hline

    Identity Inflation
        & Create many DIDs, clone credentials, or use one physical node to register multiple reward-eligible identities.
        & \redcircle \texttt{} Inflated rewards; \newline
        \redcircle \texttt{} Overestimated resource capacity; \newline
        \redcircle \texttt{} Unfair contribution weight.
        & \bluecircle \texttt{} One root identity per device; \newline
        \bluecircle \texttt{} PUF-based uniqueness; \newline
        \bluecircle \texttt{} TEE attestation; \newline
        \bluecircle \texttt{} On-chain revocation. \\

    Hardware Compromise
        & Compromise device, firmware, or manufacturer trust anchors to obtain fraudulent credentials.
        & \redcircle \texttt{} Fraudulent onboarding; \newline
        \redcircle \texttt{} Firmware compromise; \newline
        \redcircle \texttt{} Key leakage; \newline
        \redcircle \texttt{} Manufacturer risk.
        & \bluecircle \texttt{} Secure boot; \newline
        \bluecircle \texttt{} Credential issuer audit; \newline
        \bluecircle \texttt{} Revocation logs. \\

    Service Falsification
        & Spoof location, fabricate coverage, replay measurements, manipulate environmental signals, or submit forged resource evidence.
        & \redcircle \texttt{} Unreliable service maps; \newline
        \redcircle \texttt{} Unfair settlement.
        & \bluecircle \texttt{} Witness-based validation; \newline
        \bluecircle \texttt{} Multi-source evidence correlation. \\

    Network Disruption
        & Perform DDoS, eclipse attacks, routing manipulation, prefix hijacking, selective forwarding, or censorship of evidence submissions.
        & \redcircle \texttt{} Proof delay; \newline
        \redcircle \texttt{} Evidence censorship; \newline
        \redcircle \texttt{} Honest node reward loss.
        & \bluecircle \texttt{} Anti-eclipse design; \newline
        \bluecircle \texttt{} Redundant P2P paths; \newline
        \bluecircle \texttt{} Fallback submission channels. \\

    Verification Collusion
        & Bribe or control witnesses, oracle operators, measurement nodes, or off-chain verifiers that validate physical service claims.
        & \redcircle \texttt{} False-proof acceptance; \newline
        \redcircle \texttt{} Verifier bias.
        & \bluecircle \texttt{} Verifier staking; \newline
        \bluecircle \texttt{} Diverse verification constraints; \newline
        \bluecircle \texttt{} Randomized witness sampling. \\

    Mechanism Abuse
        & Manipulate reputation, exploit DAO voting, coordinate reward farming, or delay revocation and penalties.
        & \redcircle \texttt{} Revocation delay; \newline
        \redcircle \texttt{} Trust degradation; \newline
        \redcircle \texttt{} Reputation gaming; \newline
        \redcircle \texttt{} Long-term trust degradation.
        & \bluecircle \texttt{} Contribution aware reputation; \newline
        \bluecircle \texttt{} Voting safeguards; \newline
        \bluecircle \texttt{} Dispute transparency. \\
    \end{tblr}
\end{table*}

\subsubsection{Intent-Aware Intelligent Routing}
Current routing optimizes packet-level transmission, whereas DePIN requires intent-aware service orchestration across heterogeneous physical nodes. 
Future routing may shift toward intent declaration, where users specify service requirements, such as latency, privacy, cost, reliability, and geographic proximity, and networks autonomously select suitable paths \cite{Leivadeas_route_COMST}. 
Such path selection should further consider physical-resource accountability, service verifiability, and decentralized coordination.

Anh Quang \textit{et al.} \cite{Anh_Quang_route_ICC} proposed an intent-based semi-distributed routing optimization system for complex environments, which can be adapted to DePIN by balancing centralized oversight with edge autonomy. In this system, the central controller uses global intent and application QoS requirements to determine available overlay links and their priorities, while edge routers balance traffic over these links according to real-time network conditions. 
Furthermore, to address traffic burstiness, the intent-driven routing paradigm based on prediction and proactive deployment proposed by Ma \textit{et al.} \cite{Ma_route_CN} anticipates future traffic from user intents, performing resource planning in advance, thereby enhancing the stability and user experience of DePIN services.

\subsubsection{Robust Routing Against Attacks}
The open and dynamic nature of DePIN increases its exposure to attacks. 
Ensuring communication stability and security becomes dependent on the robustness of the underlying DePIN-specific routing mechanisms.
To counter DDoS attacks, Maheswari \textit{et al.} \cite{Maheswari_robust_ICICES} introduced a bidirectional verification handshake that leverages the unidirectional behavior of ``Hello Flood'' attackers to identify and filter malicious nodes. 
This mechanism is relevant, as it ensures that secure associations are formed only among trusted nodes, defending against adversaries who may deploy fraudulent physical nodes to compromise network integrity.
Lu \textit{et al.} \cite{Lu_route_Automatica} utilized the decentralization and immutability of blockchain to build a decentralized property ledger for IP prefixes. This ledger enables routers to authenticate protocol announcements, mitigating prefix hijacking while strengthening the security of DePIN core routing.
Additionally, Guo \textit{et al.} \cite{Guo_route_IoTJ} proposed an RL-based scheme applicable to DePIN for enhancing attack resilience, where the reward function accounts for malicious behaviours. 
This mechanism, demonstrated to be resilient against gray hole and DDoS attacks in simulations, presents a promising technical avenue for enhancing the security and utility of routing protocols within DePIN's ever-changing threat environments.

\subsection{Summary and Lessons Learned}
Table \ref{tab:Interaction} summarizes representative mechanisms in DePIN, including P2P network communication, dynamic topology management, and routing algorithms, together with their roles and main limitations.
In DePIN, the interaction layer is expected to evolve from basic node connectivity into an adaptive coordination layer for heterogeneous physical nodes. By combining P2P communication, dynamic topology management, intelligent routing, blockchain-enabled trust and incentives, and AI techniques such as RL and GNN~\cite{Li_graph_RBE}, this layer supports autonomous coordination, efficient resource utilization, and trustworthy multi-party collaboration without centralized control.
Several challenges remain: 1) efficient and secure integration of heterogeneous network protocol stacks; 2) coordinated management of physical-logical dual-layer topologies; 3) dynamic defense mechanisms to enhance robustness against topology disruptions; and 4) verifiable and auditable routing for service validation and settlement.

\section{Trust Layer: Decentralized Trust and Compliance}\label{sect:trust}
This section introduces technologies enabling trust
in the ICT infrastructure in DePIN, including decentralized identity (DID), privacy-preserving technology, and Decentralized Autonomous Organization (DAO)-based governance.

\textit{Threat model in DePIN.}
A representative DePIN threat model includes six categories, with the details presented in Table \ref{tab:Threat}.
These threat models reveal that DePIN security cannot be reduced to conventional blockchain security.
Even when ledger integrity and transaction finality are guaranteed, the system may still fail if device identities are fabricated or duplicated, hardware roots are compromised, or service evidence is falsified.
The trust layer should establish cross-layer trust mechanisms that connect hardware authenticity, service verifiability, data protection, and accountable governance into a unified framework.

\subsection{Decentralized Identity}
This threat model reframes DID in DePIN as a mechanism for hardware-rooted identity binding, rather than as a standalone tool for decentralized authentication.
Recent research on DID generally centers around the World Wide Web Consortium (W3C) and Verifiable Credentials (VCs) standards \cite{Avellaneda_DID_COMSM, Mazzocca_DID_COMST}.
In the context of DePIN, which involves a large number of physical devices and dynamic network environments, the central problem is how to bind a digital identifier to a unique, operational, and accountable physical device.
Otherwise, a single physical device may create multiple DIDs and claim multiple rewards, leading to a hardware-level Sybil attack. 

\begin{table*}[t]
    \centering
    \captionsetup{
    labelfont={color=black},
    textfont={color=black},
    font={scriptsize}
    }
    \caption{Cryptographic primitives for physical binding, where the marker `\greencheck' indicates benefits and `\redcross' indicates limitations.}
    \label{tab:bind}
    \fontsize{6pt}{7pt}\selectfont
    \begin{tblr}{
        width = \linewidth,
        colspec = {X[1.6]X[1.6]X[4.3]X[2.3]},
        cells = {m, fg=black},
        column{1} = {c, font=\itshape},
        column{2} = {l},
        column{3} = {l},
        column{4} = {l},
        row{1} = {c, font=\bfseries},
        hlines = {black},
        vline{2-4} = {black},
    }
    Mechanism 
        & Main Security Property 
        & Benefits and Limitations in DePIN
        & Main Cost Dimension: Level \\ \hline

    PUF
        & Hardware-level uniqueness
        & \greencheck \texttt{} Device-specific identity root; \newline
          \greencheck \texttt{} Resistance to modeling attacks; \newline
          \greencheck \texttt{} Distinguishability between devices under environment variations; \newline
          \redcross \texttt{} No direct evidence of location, firmware integrity, or service delivery.
        & Hardware integration/ enrollment: High \\

    TEE/TPM
        & Remote attestation and key protection
        & \greencheck \texttt{} Firmware/software measurement and DID public key binding; \newline
          \greencheck \texttt{} Protected execution and isolated key storage; \newline
          \redcross \texttt{} Relies on endorsement keys or attestation services from vendors.
        & Attestation/vendor trust: Medium to high \\

    Secure Elements/Hardware Security Modules
        & Private key protection and anti-extraction
        & \greencheck \texttt{} Support for persistent cryptographic identities; \newline
          \redcross \texttt{} Binds keys to a container, not to the whole device; \newline
          \redcross \texttt{} No guaranteed uniqueness or migration resistance.
        & Hardware provisioning: Medium \\

    ZKP/Anonymous Credentials/Cryptographic Accumulators
        & Privacy-preserving uniqueness proof
        & \greencheck \texttt{} Selectively disclosed hardware-rooted eligibility; \newline
          \greencheck \texttt{} Privacy-preserving separation between DIDs and reward accountability; \newline
          \redcross \texttt{} Inherited trust from enrollment, not independent physical validation.
        & Computation/verification: Medium to high \\

    \end{tblr}
\end{table*}

\subsubsection{Identity Binding with Physical Entities}
The effectiveness of DID depends on securely binding on-chain credentials to real-world entities, ensuring trustworthy and secure network operations. However, a DID by itself proves control over a cryptographic key, but it does not prove that the key corresponds to a unique physical device.
This distinction is critical for DePIN,
where an attacker may use a single physical device to generate multiple DIDs that appear as independent contributors, thereby claiming excessive rewards without providing proportional resources.

This hardware Sybil threat is serious since DePIN incentives are tied to scarce physical capabilities.
An identity binding mechanism must answer two questions:
whether the identity is controlled by a legitimate device, and
whether the device is unique rather than a duplicated identity instance.

Existing DID and VC mechanisms provide a starting point for the first question.
VCs are issued by trusted third parties, e.g., manufacturers, via distributed ledgers, containing cryptographic signatures and attribute claims~\cite{Gebresilassie_DID_WF-IoT}. 
During verification, VCs can be bound to a unique physical identifier, e.g., hardware fingerprint, linking real-world properties to digital identities and ensuring authenticity within decentralized networks \cite{Garzon_DID_6GNet}. 
Xiong \textit{et al.} \cite{Xiong_DID_IoTJ} proposed a dual-chain architecture: DID Chain + M Chain, where the DID Chain stores immutable identity operation records, and the M Chain maintains an index of the latest block heights, reducing large-scale query latency to 0.5 ms.

\subsubsection{Cryptographic Primitives for Physical Binding}
Several cryptographic primitives can address the second question by binding digital identities to hardware-rooted properties.
Since each primitive offers only partial protection, a practical DePIN identity system should evaluate its defense scope, security limitations, and main cost; see Table \ref{tab:bind}.

\begin{itemize}
    \item A Physical Unclonable Function (PUF) provides each device with a unique hardware identity by exploiting manufacturing variations, which generate challenge response mappings that are difficult to clone~\cite{AL_PUF_ACS}.
    In DePIN onboarding, a secret derived from a PUF can be bound to a reward eligible root DID, while the blockchain stores only commitments or verification results rather than raw PUF responses.
    This reduces the risk that one physical device creates multiple independent root identities.  
    However, PUFs cannot verify device location, service quality, or firmware integrity.
    PUFs are affected by instability, aging, enrollment overhead, and modeling attacks. 
    The major cost comes from hardware integration and enrollment, including secure enrollment, helper data management, calibration, and hardware standardization.
    
    \item TEE and Trusted Platform Module (TPM) provide remote attestation capabilities. 
    Compared with PUFs, remote attestation verifies the device firmware state and the protection of private keys in an isolated or hardware environment~\cite{Cirne_TEE_COMST, Ott_TEE_ACM}. 
    A DePIN device can generate an attestation quote binding its firmware state, public key, and DID document to a fresh nonce. This allows verifiers to confirm both the trusted hardware root and the approved software state. 
    Similar to PUFs, TEE/TPM attestation strengthens device-level trust but cannot independently prove physical coverage, location, or service delivery. 
    It also depends on vendor endorsement keys and remains exposed to rollback and side-channel risks.
    Its overhead mainly stems from attestation and vendor trust.
    The cost level is medium to high because the deployment requires quote verification, vendor-backed attestation, and device lifecycle management.

    \item Secure elements and hardware security modules can protect private keys and reduce the risk of key extraction or software-based cloning. 
    They are useful for infrastructures requiring persistent cryptographic identities. 
    However, they mainly bind an identity to a protected key container rather than to a complete physical DePIN device. 
    Without tamper-resistant integration and auditable enrollment, they cannot guarantee device uniqueness or prevent identity migration.
    In this case, the cost is mostly associated with hardware provisioning, which is typically medium for element deployment but can increase when module grade protection is required.

    \item ZKP, anonymous credentials, or cryptographic accumulators can help preserve privacy while enforcing uniqueness constraints. 
    For example, a device may prove that it owns a valid hardware-based credential without revealing its full serial number or manufacturer identity. 
    Together with a nullifier style uniqueness constraint, such ZK-based credentials can prevent the same hardware root from registering multiple identities within the same DePIN domain~\cite{Rosenberg_zkcreds_SP}. 
    This is useful in DePIN devices which may need pseudonymous or pairwise DIDs for privacy, while being limited to one reward eligible hardware root. 
    However, ZKPs do not create physical trust by themselves. 
    Their cost is dominated by computation and verification, including circuit design, proof generation, and verifier implementation.
\end{itemize}

\subsubsection{Dynamic Multi-Role Identities}
In DePIN, the roles of devices and users, such as miners, providers, consumers, and developers, may dynamically shift in response to changing environmental conditions. 
In such scenarios, DID systems must support real-time identity verification and adaptive role-based permission management based on device status, network conditions, and user demands. 
To address these challenges, Mazzocca \textit{et al.} \cite{Mazzocca_DID_COMST} proposed three types of decentralized identifiers (Anywise, Pairwise, and N-wise) to support identity interactions under varying trust models. They also introduced mitigation strategies, such as multi-factor authentication and key rotation, to defend against key threats in dynamic environments, including key leakage and credential misuse.
Saif \textit{et al.} \cite{Saif_DID_BCCA} adopted a hierarchical tree structure for role inheritance and permission delegation. Their hierarchical factory model, consisting of an access control manager, a super factory, and client factories, supports recursive sub-factory creation and enables fine-grained decentralized access control for complex organizations.
Ning \textit{et al.} \cite{Ning_DID_IoTJ} introduced a modular identity modeling and addressing framework that emphasizes low coupling and modular independence. By leveraging a microservices architecture, their design supports flexible identity module extension and employs semantic web-based modeling to define identities through attributes rather than fixed identifiers.

\subsubsection{Cross-Network Interoperability}
Cross-network DID supports identity authentication, semantic interoperability, and resource coordination across heterogeneous and autonomous DePIN networks. 
With sharded blockchain, identity management nodes can be organized into user, shard, and cross-domain management layers, allowing domain-specific DIDs to operate independently while establishing cross-network trust through the main chain and inter-shard communication. 
Simulations demonstrated a significant improvement in throughput and reduced latency (cross-domain verification $<$ 2 seconds) \cite{Liu_DID_IoTJ}.
Zecchini \textit{et al.} \cite{Zecchini_DID_DAPPS} proposed a cross-chain identity management framework to address the standardization challenges of identity and data interoperability across blockchains. At its core, the framework introduces StateRelay, a blockchain relay that stores only block headers with the latest user state updates and verifies data validity through a binary search-based interactive challenge protocol, reducing storage and transaction costs; cf. traditional relays.

\subsection{Lightweight Privacy-Preserving Technology}
DePIN involves a large number of edge devices with limited computational capabilities, requiring privacy-preserving technologies to be lightweight and efficient. Below, we list several lightweight algorithms along with representative studies.

\subsubsection{Cryptography-Based Lightweight Algorithms}
Cryptography-based lightweight algorithms aim to achieve a delicate balance between computational efficiency and cryptographic robustness. It is important in the context of DePIN.
Kumar \textit{et al.} \cite{Aneesh_Light_TCE} proposed a hybrid encryption framework that integrates three key technologies. The framework employs quantum key generation to establish secure initial keys and leverages the no-cloning theorem \cite{Epstein_nocloning_TIT} to ensure unconditional security during key distribution within the blockchain network. The use of chaotic mapping encryption and post-quantum cryptographic primitives enables millisecond-level latency, satisfying the requirements of real-time encryption.
Symmetric encryption algorithms such as SIMON/SPECK \cite{Ertaul_Light_CSCE}, asymmetric algorithms such as elliptic curve cryptography \cite{Lara-Nino_Light_ADHOC}, and various hybrid schemes and protocols are viable alternatives. For detailed implementations, please refer to surveys \cite{Latif_Light_GCWOT, Khan_Light_IoTJ}.

\begin{figure}[!t]
    \centering
    \includegraphics[scale=0.3]{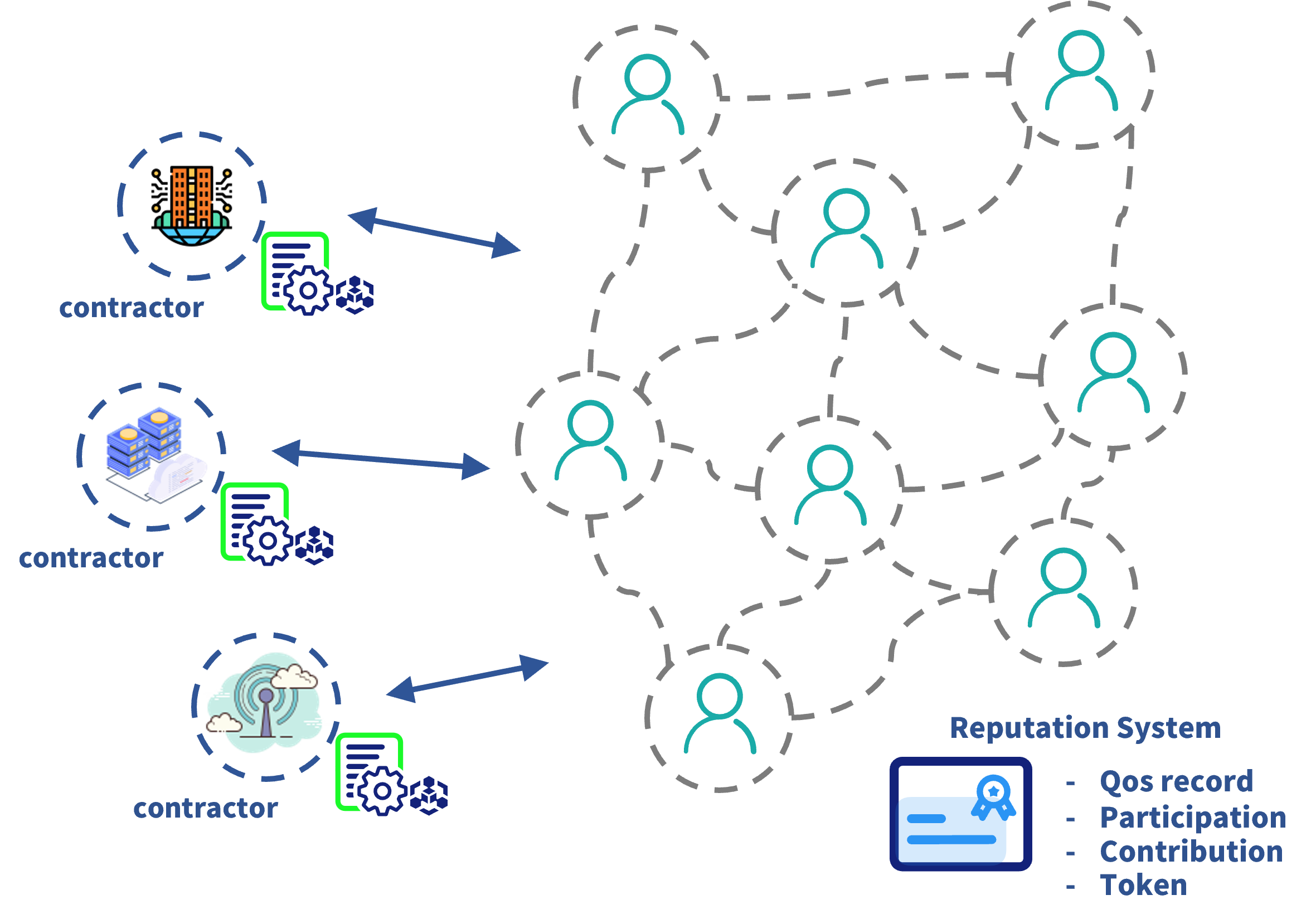}
    \caption{DAO regulation mechanism for decentralized ICT infrastructure.}
    \label{fig:DAO}
\end{figure}

\subsubsection{Lightweight Differential Privacy Federated Learning}
In heterogeneous edge node environments, localized modeling with differential privacy enhancement represents an important research direction.
To deal with that, Fan \textit{et al.} \cite{Fan_Light_IoTJ} combined blockchain and differential privacy through a bilinear pairing-based signature scheme with batch verification, reducing communication and computation overhead. A multidimensional subjective logic model was used for worker reputation evaluation, enabling high-quality node selection and filtering malicious or low-quality data.
To balance privacy noise and model performance, Song \textit{et al.} \cite{Song_Light_BigData} proposed a lightweight differential privacy FL approach based on random gradient perturbation. The core idea involves selectively injecting noise into the gradients of a subset of mini-batches during the local training phase on federated clients. Simulations show that this method achieves 5\% higher accuracy than benchmarks under privacy preservation, with its lightweight perturbation offering increasing time efficiency as local datasets grow.
The study \cite{Zhang_Light_TIFS} introduced a secure aggregation framework that leverages two non-colluding, honest-but-curious servers, enabling natural extensions to multi-server or blockchain-coordinated DePIN environments.
By integrating secret sharing and shuffling algorithms, the framework proposed in \cite{Chen_Light_TETC} achieves lightweight, decentralized privacy preservation for raw data, without relying on trusted third parties, while maintaining low overhead and strong resistance to collusion attacks.

\subsubsection{Lightweight Protocols Based on ZKP}
Within DePIN, the lightweight design of ZKPs is emerging as a critical research frontier for enabling privacy-preserving authentication and secure verification. Given the potentially pervasive deployment of DePIN across edge nodes, mobile devices, and resource-constrained environments, conventional ZKP protocols, characterized by high computational and communication overhead, can be infeasible for real-world implementation \cite{Sun_Light_Network, Khandait_Light_GCCIT}. 
Hence, lightweight ZKPs must be optimized for computational efficiency and minimized communication costs, enabling rapid proof generation and verification while reducing bandwidth and storage to accommodate the heterogeneous capabilities of DePIN nodes. Methods like LiteZKP \cite{Boo_Light_SJ}, Ligetron \cite{Wang_Light_SP}, and ZK VC \cite{Zhang_Light_arXiv} have been proposed to address this issue.

To support inter-system trust and interoperability across DePIN's diverse subsystems and sub-networks, lightweight ZKP schemes should exhibit modularity, cross-domain operability, and composability. This is critical for seamless integration with DID, smart contracts, and multi-chain ecosystems \cite{Tran_Light_ICISN}. Of particular importance are
protocols that avoid trusted setup or rely on universally trusted parameters, thereby enhancing verifiability and security robustness.
Nevertheless, true decentralization and long-term security require more than technical improvements. The next subsection examines how DAO-based governance can empower the community to build a transparent, verifiable, and secure decentralized ecosystem.

\subsection{DAO-based Governance}
DAO is a smart contract-based governance structure that uses programmatic rules and on-chain consensus to enable transparent auditing, autonomous decision-making, and dynamic incentives for physical resources and services \cite{Santana_DAO_TFSC}. DAO enhances system verifiability and institutional trust, as illustrated in Fig. \ref{fig:DAO}.
The DAO governance mechanism of decentralized ICT infrastructure must integrate the dual complexities of on-chain smart contracts and off-chain physical device operations \cite{Reijers_DAO_Topoi} to establish a dynamic, adaptive, and trustworthy governance framework. Central to ensuring consistency between on-chain governance and off-chain execution is a trust-bridging mechanism \cite{Cappiello_blockchain_Book}, which enables verifiable transmission of policy decisions formulated through on-chain proposal, negotiation, and voting to physical devices, thereby ensuring atomicity and auditability of execution.

From the architecture's point of view, a two-tier DAO structure facilitates functional specialization and organizational coordination. The upper-layer Resource Operation DAO governs network-level decisions, such as topology optimization, computational distribution, and access policies, through collective voting and policy-making \cite{Goldberg_DAO_JBR}. The lower-layer User Experience DAO, by contrast, centers on service consumers, allowing them to evaluate service latency, availability, and quality \cite{Li_DAO_TIV}, and influence operator incentives or penalties through governance feedback loops, thus achieving a ``use-evaluate-improve'' cycle.
Furthermore, a dynamic reputation system introduces temporally and behaviorally aware trust modeling \cite{Saito_DAO_Frontiers, Kaal_DAO_JRFM}. By aggregating historical service performance, governance participation, and user feedback, the reputation system dynamically adjusts voting power, resource prioritization, and participation eligibility. More than an incentive layer, this reputation system serves as a decentralized trust fabric, enhancing robustness and resistance to manipulation while supporting the long-term adaptability of DAO governance.

\subsection{Summary and Lessons Learned}
Compared with \cite{Wang_FL_TDSC, Bai_Trust_Blockchain}, this section emphasizes the specific requirements of trust technologies under the DePIN architecture, particularly their role in enabling secure, compliant, and feasible decentralized infrastructure deployment.

DePIN may involve a large number of decentralized physical infrastructure nodes and network devices, requiring robust authentication, traceability, and hardware-rooted identity binding to ensure device trustworthiness, authenticity, and compliance \cite{iotex_whitepaper_TR}. Unlike conventional DID systems that mainly prove control over cryptographic keys, DID in DePIN should further bind digital identifiers to unique and accountable physical devices, thereby mitigating device-level Sybil attacks and fraudulent resource claims.
Since roles may change dynamically, DePIN should support dynamic and cross-chain identities. Meanwhile, privacy-preserving mechanisms must remain lightweight for edge devices, and DAO-based governance should bridge on-chain decisions with off-chain operations while preserving user-centric accountability.

Future directions include: 1) compliance modeling for heterogeneous networks and devices, such as unified data privacy protocols for IoT devices and communication networks \cite{Xiong_trust_IoTJ}; 
2) interpretable regulatory mechanisms based on smart contracts and explainable AI to support transaction compliance verification and smart contract conformity analysis~\cite{sahu_trust_arXiv}.

\begin{figure}[!t]
    \centering
    \includegraphics[scale=0.3]{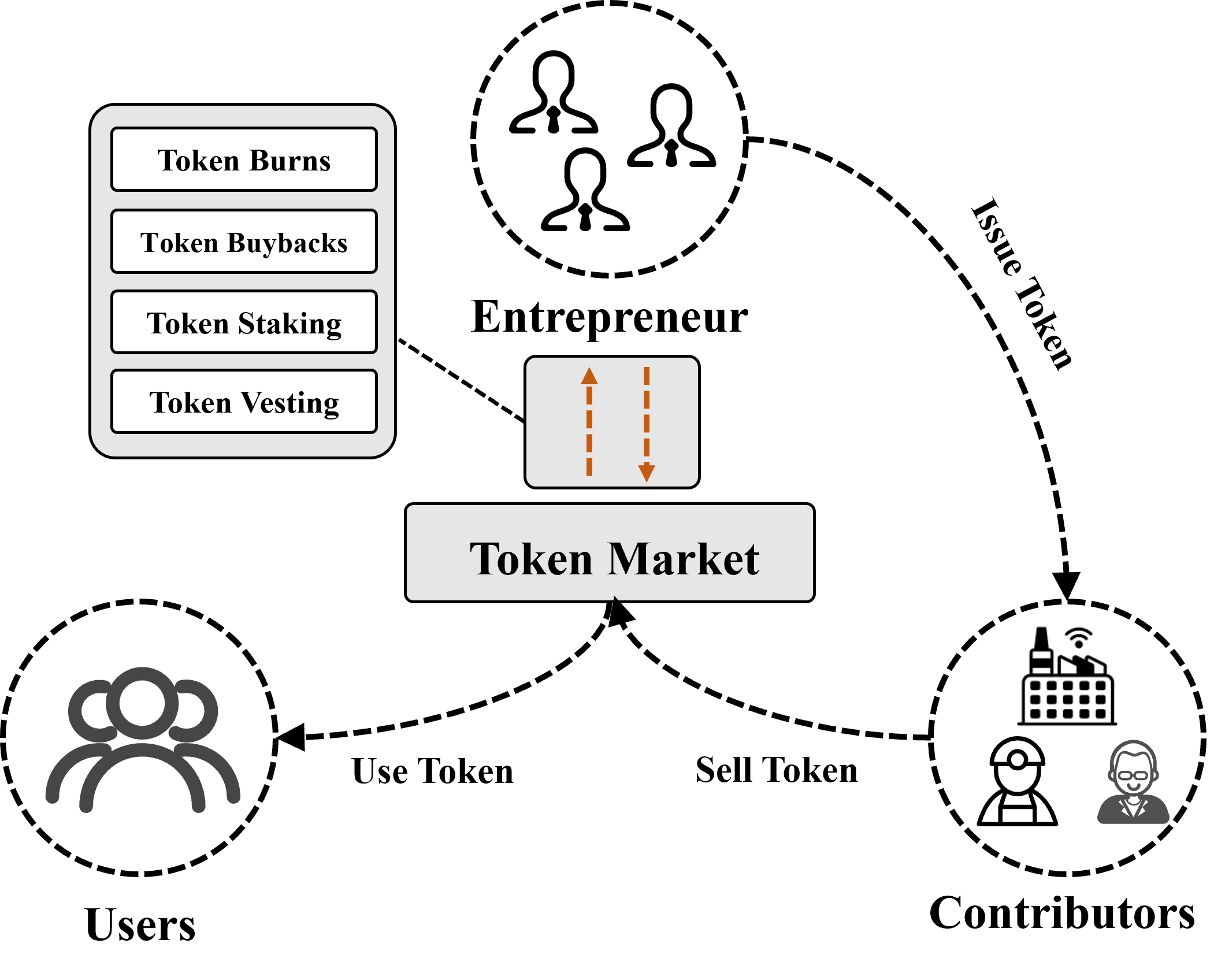}
    \caption{
    Token flow in a DePIN system. Black arrows show token circulation, while red arrows denote value management mechanisms, including burn, buyback, staking, and vesting.}
    \label{fig:Token}
\end{figure}

\section{Incentive Layer: Tokenized Economics}\label{sect:incentive}
The incentive layer serves as a pillar of DePIN. Through token economics, it coordinates participant behavior and drives the construction, maintenance, and utilization of decentralized infrastructure.
This section presents a framework for tokenized economic mechanisms in the ICT infrastructure under DePIN.

\subsection{Real World Asset Tokenization}

Real-world asset (RWA) tokenization bridges the gap between physical and digital economies, with token value directly anchored to underlying tangible or intangible assets, in contrast to purely digital crypto-native assets \cite{xia_RWA_arXiv}.
DePIN and RWA tokenization are not independent concepts but rather mutually reinforcing and co-evolving paradigms. DePIN leverages token incentives to aggregate physical resources, e.g., GPUs, storage, sensors, into decentralized networks, while RWA tokenization provides an efficient mechanism for capturing and utilizing the intrinsic value of these resources in the digital domain \cite{Gong_RWA_Book}.
This dual tokenization of both infrastructure and the services it provides enables DePIN networks to establish a robust ``production-consumption” economic loop, directly transforming physical productivity into on-chain assets that are programmable, tradable, and permissionlessly transferable on a global scale.

\subsubsection{Tokenization of Hardware Ownership (A Case Study of PinLink)}
PinLink is an innovative platform that integrates RWA tokenization with DePIN networks \cite{PinLink}.
The model establishes a dual-revenue mechanism to address the cost disadvantage DePIN projects face when competing with centralized incumbents.
Hardware owners can generate income in two ways: 1) by leasing computational resources to AI developers, similar to other DePIN projects; 2) by selling tokenized hardware ownership shares to passive investors. This business model integrates ownership and service value into a unified token economy. PinLink improves financial flexibility for hardware owners while lowering service costs for consumers, thereby enhancing network competitiveness.

\subsubsection{Tokenization of Data Streams (A Case Study of DIMO)}
DIMO is a decentralized mobility platform designed to return ownership of vehicle data to car owners \cite{DIMO}.
Users contribute data such as driving behavior, vehicle health, and location information to the network by connecting external hardware or built-in onboard systems. In return, they are rewarded with DIMO tokens.
The aggregated data is organized into a decentralized marketplace, accessible to third parties, e.g., insurance companies, automakers, and developers.
This approach transforms traditionally centralized ``data streams” into tradable and valuable RWAs,
a key step to data tokenization.

\subsection{Dynamic Value Management}

The dynamic value management mechanism under token economics is an economic framework that adjusts the token supply in real time, in response to network growth, productivity fluctuations, and user demand, as illustrated in Fig. \ref{fig:Token}. Its objective is to maintain system equilibrium and optimize overall performance \cite{Tan_Token_Economics}. At its core, this approach treats the token economy as a controllable dynamic system, aiming to address the limitations of traditional fixed-supply models in terms of incentive alignment and value management.

Four types of mechanisms are involved: token burns, token buybacks, token staking, and token vesting. Particularly, Lin \textit{et al.} \cite{Lin_Token_JFEC} proposed a dynamic platform economic model where tokens are issued to finance service capacity investments, with a supply-and-demand driven ``token burns''  mechanism ensuring token value sustainability. Their model highlights that financial constraints can cause underinvestment due to owner-user conflicts, which blockchain's transparency helps mitigate.
Akcin \textit{et al.} \cite{Akcin_Token_arXiv} introduced a control theory-based dynamic token value mechanism, advocating for token rewards to scale with system growth while balancing consumer demand and inflation, using a ``token buybacks" system where new tokens reward service providers and repurchased tokens from consumer payments are burned.
Li \textit{et al.} \cite{Gogol_Token_arXiv} conducted a comprehensive analysis of liquid staking and restaking frameworks, offering theoretical guidance for how ICT infrastructure within DePIN can leverage ``token staking'' mechanisms to enhance capital efficiency, strengthen network security, increase asset flexibility, and attract broader participation.
Li \textit{et al.} \cite{Li_Token_RFS} leveraged ``token vesting'' mechanisms to transparently distribute tokens via blockchain before platform launch, enabling token purchases, to credibly signal users' intentions to participate in the platform.

Token issuers must develop a deep understanding of market supply and demand dynamics to implement effective, dynamic token value control. This enables them to gracefully manage network expansion while avoiding inflationary pressures caused by excessive token issuance. Such adaptive adjustment accommodates the continuously evolving supply-demand landscape within DePIN ecosystems.

\begin{figure}[!t]
    \centering
    \includegraphics[scale=0.3]{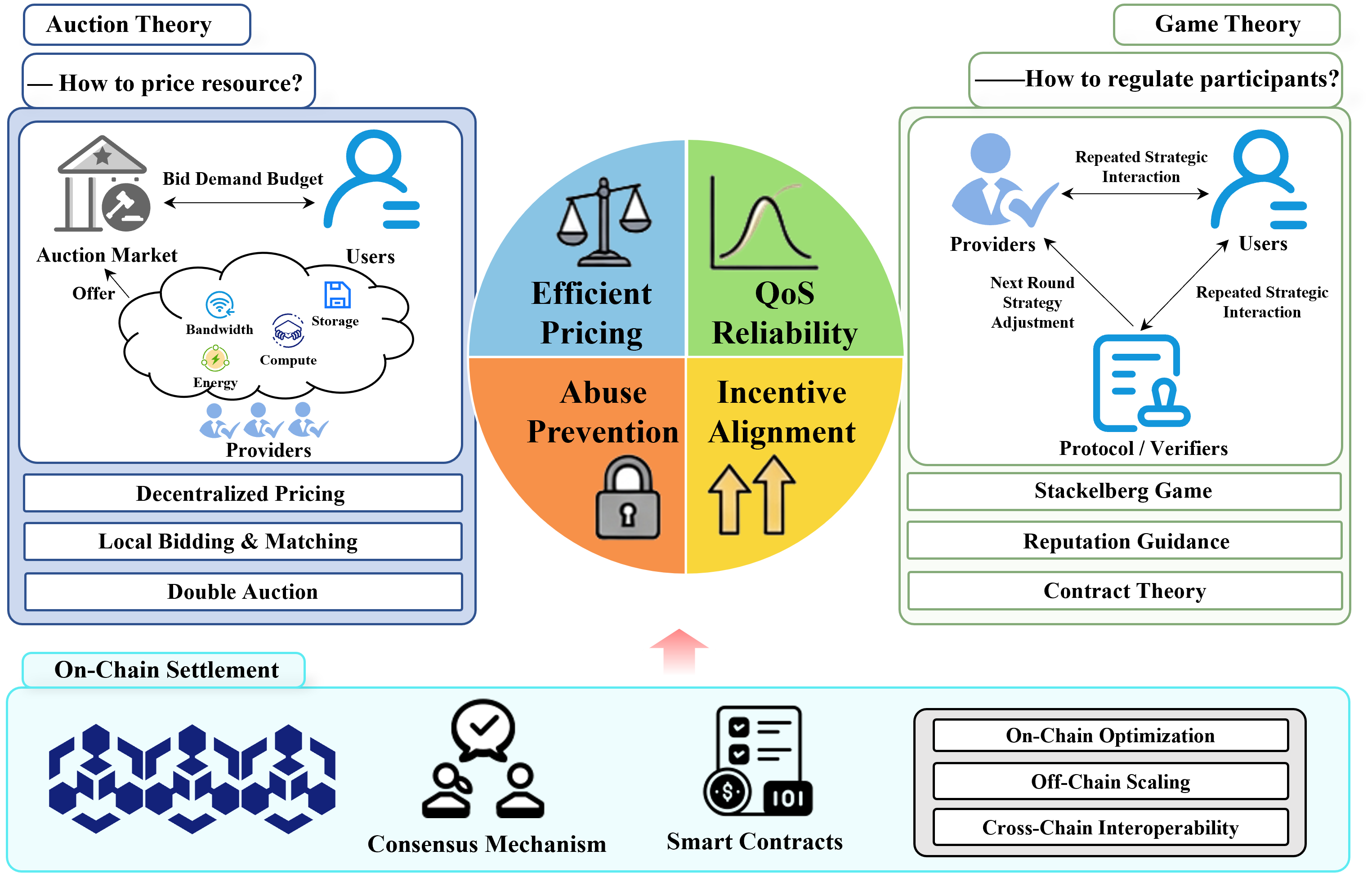}
    \caption{Illustration of the application of auction and game theory in DePIN systems.}
    \label{fig:Auction_Game}
\end{figure}

        \subsection{Auction and Game Theory}

        In the incentive layer, beyond the macro-level tokenomics architecture, micro-level resource pricing and multi-party interaction mechanisms are equally essential for sustaining autonomous system operation and long-term security.
        This section provides an in-depth discussion of the application of auction theory and game theory in the incentive layer of DePIN, as illustrated in Fig.~\ref{fig:Auction_Game}.

        \subsubsection{Auction Theory based Pricing Mechanisms}
        Auction theory offers a basis for DePIN pricing
        since it supports heterogeneous resource allocation under strategic behavior while considering truthfulness, budget balance, individual rationality, and economic efficiency~\cite{Qiu_auction_TCCN, Liu_auction_TCSS}.
        Guo \textit{et al.}~\cite{Guo_auction_TCS} proposed a decentralized auction framework in which buyers and sellers make local and independent decisions under budget, capacity, and matching constraints.
        This framework suggests that decentralized resource pricing can be achieved through local bidding and local decision-making, which is relevant to DePIN settings where heterogeneous resource providers and consumers interact without trusting a centralized market operator.
        This idea was further advanced in distributed double-auction mechanisms for large-scale D2D resource trading~\cite{Gao_auction_TN}, where neighboring buyers and sellers exchange local information, perform distributed matching, and determine prices for matched pairs.

        \subsubsection{Game-Theoretic Multi-Party Interaction and Regulation}
        DePIN involves large-scale physical infrastructures and autonomous heterogeneous participants, making QoS assurance and security regulation a multi-party strategic problem.
        Game theory combined with blockchain provides an economic framework for modeling strategic interactions and guiding participant behavior.
        Xu \textit{et al.}~\cite{Xu_game_TMC} adopted a Stackelberg game to balance benefits among providers and users, where providers dynamically set resource prices and users adjust demand according to satisfaction and cost. This leader–follower formulation is useful for DePIN scenarios involving dynamic resource pricing and demand response.
        In~\cite{Tang_game_SJ}, a credit-based verifier declaration game was introduced, and a worker attitude game was designed in the incentive layer. Reputation, completion time, effort cost, and task quality jointly yield a unique equilibrium that incentivizes high-effort participation and suppresses free-riding.
        Zhang \textit{et al.}~\cite{Zhang_game_FGCS} studied selfish node attacks through a blockchain-assisted RL framework, in which reputation is updated based on both immediate and historical behaviors, making attacks less attractive when long-term utility loss exceeds short-term gain.
        Zhou \textit{et al.}~\cite{Zhou_game_IoTJ} combined contract theory with blockchain-based reputation to support collaboration under heterogeneity and asymmetric information, using differentiated contracts and career-like reputation to screen participants and discipline interactions.

\subsection{Characterized Consensus Mechanism}

\begin{table*}[t]
    \centering
    \captionsetup{
    labelfont={color=black},
    textfont={color=black},
    font={scriptsize}
    }
    \caption{Comparison of Representative Consensus and Proof Mechanisms in DePIN.}
    \label{tab:Consensus}
    \fontsize{6pt}{7pt}\selectfont
    \begin{tblr}{
        width = \linewidth,
        colspec = {X[0.8]X[0.8]X[0.8]X[1.1]X[1.1]X[1.1]X[3.5]},
        column{1-6} = {c},
        column{7} = {l},
        row{1} = {font=\bfseries, c},
        cells = {m, fg=black},
        hlines = {black},
        vline{2-7} = {black},
    }
        Project 
        & Mechanism 
        & Refs. 
        & Comp. Cost 
        & Comm. Overhead 
        & Veri. Latency 
        & Main Advantages in DePIN \\ \hline

        Helium Network
        & PoC
        & \cite{Rammouz_Helium_WiMob}
        & Medium
        & Medium
        & High
        & \bluecircle \texttt{} Verify real-world wireless coverage \\

        Filecoin
        & PoRep
        & \cite{Giacomelli_Filecoin_CrypTorino}
        & High
        & Low
        & High
        & \bluecircle \texttt{} Prevent fake or duplicated storage \\

        Filecoin
        & PoSt
        & \cite{He_PoSt_arXiv, Heo_Filecoin_ICBC}
        & High
        & Medium
        & Task-dependent
        & \bluecircle \texttt{} Ensure continuous storage availability \\

        Internet Computer
        & CKC
        & \cite{Assmann_ICP}
        & Medium
        & Medium
        & Low
        & \bluecircle \texttt{} Enable fast cross-subnet verification \\

        Bittensor
        & PoI
        & \cite{Steeves_PoI_Bittensor, Mafrur_PoI_IET}
        & Task-dependent
        & Medium
        & Task-dependent
        & \bluecircle \texttt{} Reward useful AI contribution \\

        /
        & PoR
        & \cite{Xu_PoR_Network}
        & Medium
        & Medium
        & Medium
        & \bluecircle \texttt{} Link routing quality to incentives \\

        /
        & PoB
        & \cite{Abid_Lightx_ICNS}
        & Medium
        & Medium
        & Medium
        & \bluecircle \texttt{} Incentivize measurable bandwidth supply \\

        /
        & PoFL
        & \cite{Qu_PoFL_TPDS}
        & High
        & High
        & Medium
        & \bluecircle \texttt{} Convert consensus work into useful learning \\

        /
        & PoBT
        & \cite{Biswas_PoBT_IoTJ}
        & Low
        & Low
        & Low
        & \bluecircle \texttt{} Support lightweight IoT transaction validation \\
    \end{tblr}
\end{table*}

In traditional blockchains, consensus mechanisms ensure state consistency and fault tolerance, delivering decentralized trust \cite{Wang_consensus_Access}. In DePIN, consensus mechanisms will play a vital role in the incentive layer, directly influencing node participation, resource provisioning, and service execution \cite{Yu_consensus_ICPADS}.
Given the heterogeneity of resource types and trust assumptions, DePIN demands application-specific consensus designs that couple on-chain logic with real-world verification. Several representative examples are given in Table \ref{tab:Consensus}.

\subsubsection{Proof-of-Coverage (PoC)}
PoC, adopted by the Helium Network \cite{Rammouz_Helium_WiMob}, exemplifies how DePIN consensus must connect digital trust with physical verification. 
It confirms whether deployed hotspots truly provide wireless coverage, aligning token rewards with geographic network expansion.
Since coverage verification depends on witness reports, environment observations, and oracle validation, PoC incurs medium computational cost and communication overhead, but its verification latency can be high when physical evidence must be collected and cross-checked over time.
The core challenge lies in the difficulty and scalability of verifiably anchoring real-world physical truths on-chain. To improve efficiency, Helium has transitioned to the Oracle PoC \cite{HIP_70} mechanism. However, this shift introduces new issues related to balancing decentralization and efficiency, as well as ensuring security under reliance on trusted oracles.

\subsubsection{Proof-of-Replication \& Proof-of-Spacetime (PoRep \& PoSt)}
In Filecoin \cite{Giacomelli_Filecoin_CrypTorino}, PoRep and PoSt guarantee that physical storage resources are uniquely committed and continuously maintained. 
They enable the generation of fast and verifiable cryptographic proofs to mitigate risks specific to storage-based models, such as Sybil attacks and generation attacks \cite{He_PoSt_arXiv}.
Compared with lightweight consensus mechanisms, these storage proofs incur high computational cost due to encoding, proof generation, and repeated verification. 
Their communication overhead is low to medium because submitted proofs are compact; for example, a 48.82 KB proof submitted at a 25 minute challenge interval corresponds to about 0.26 kbps for a single proof stream~\cite{Zhang_epost_TIFS}.
However, verification latency can be high or task-dependent, especially when storage challenges and proof windows are involved. 
This reflects a typical DePIN trade-off: stronger physical resource verifiability often comes at the expense of efficiency and delayed reward settlement.

\begin{figure}[!t]
    \centering
    \includegraphics[scale=0.35]{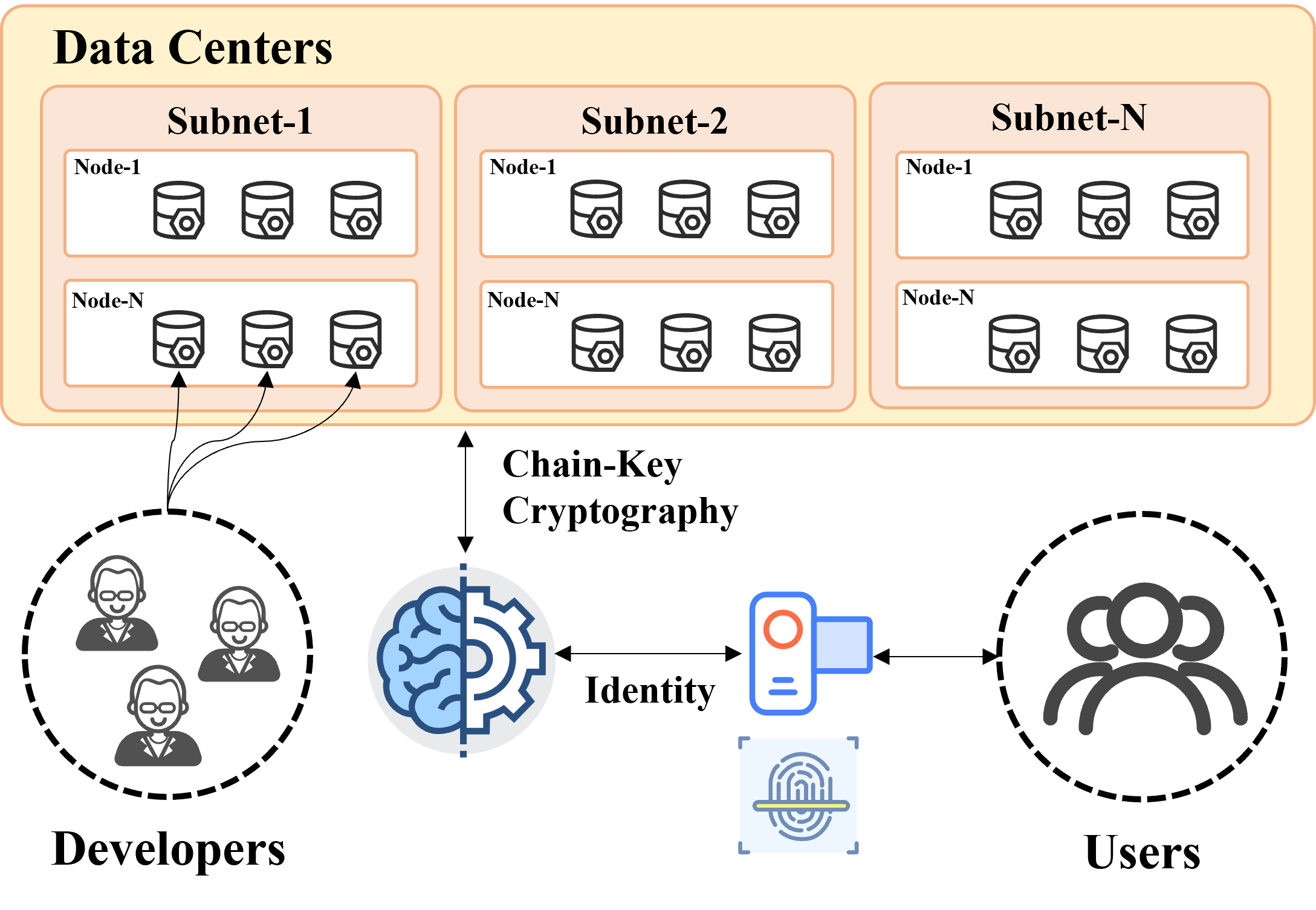}
    \caption{Illustration of CKC protocol. It provides consensus with secure signing and cross-subnet verification capabilities.
    }
    \label{fig:CKC}
\end{figure}

\subsubsection{Chain-Key Cryptography (CKC) Protocol}
CKC is employed in the Internet Computer project \cite{ICP}, which shares a cryptographic key method across subnets to ensure fast, secure state transitions, as depicted in Fig. \ref{fig:CKC}.
CKC represents a new pathway in consensus design, serving as a cryptographic primitive and consensus optimization aimed at enabling scalability and interoperability at the base layer of decentralized computation.
In terms of operational overhead, CKC involves medium computational cost and communication overhead because threshold signing, subnet coordination, and key management require cryptographic processing and inter-node interaction~\cite{Assmann_ICP}. 
Its verification latency is low, since compact chain-key signatures enable fast cross-subnet state verification without requiring applications to validate the full consensus process.
This approach exemplifies how consensus mechanisms in DePIN can be rooted in cryptographic progress, expanding the notion of ``customized consensus'' beyond the direct verification of physical resources.

\subsubsection{Proof-of-Intelligence (PoI)}
Bittensor's PoI mechanism \cite{Steeves_PoI_Bittensor} extends the DePIN paradigm into AI services by rewarding nodes for their computational and cognitive contributions. It employs the Shapley value framework \cite{Mafrur_PoI_IET} to objectively quantify each participant's marginal contribution to collective intelligence.
Unlike storage or bandwidth proofs, the computational cost and verification latency of PoI are task-dependent, because model evaluation may vary with model size, inference workload, benchmark complexity, and validator design. 
Its communication overhead is typically medium, as validators and miners need to exchange model outputs or task evidence rather than continuously transmit large raw datasets.
The main challenge is the objective and fair valuation of non-physical and complex tasks, ensuring that rewards are tied to the quality and accuracy of actual outputs while preventing potential system abuse.

\subsubsection{Others}
Various consensus algorithms have been developed and applied in the construction of decentralized ICT infrastructure within DePIN. For instance, Xu \textit{et al.} \cite{Xu_PoR_Network} proposed the Proof-of-Routes (PoR), a lightweight consensus mechanism centered on routing performance. It combines fractal network modeling, dynamic heartbeat verification, and hierarchical consensus group election to support performance-oriented decentralized network coordination.
El Abid \textit{et al.} \cite{Abid_Lightx_ICNS} introduced the Proof-of-Bandwidth (PoB), in which bandwidth fluctuation serves as a trust metric, combined with normalized harmonic averaging and damped iterative methods to strike a balance between efficiency and Byzantine robustness.
Qu \textit{et al.} \cite{Qu_PoFL_TPDS} proposed Proof-of-Federated Learning (PoFL), which integrates FL with blockchain consensus by redirecting Proof-of-Work (PoW) computation to useful FL tasks. In this way, PoFL reduces the energy inefficiency of conventional PoW systems.
Moreover, Proof-of-Block-and-Trade (PoBT) is a lightweight consensus algorithm designed for blockchain systems in IoT \cite{Biswas_PoBT_IoTJ}. PoBT addresses the bottlenecks of traditional blockchain consensus mechanisms in IoT by employing a dual-verification scheme that combines transaction-level and block-level validation.
These mechanisms have different DePIN suitability. 
Generally,
PoR and PoB impose moderate overhead through periodic routing or bandwidth measurements. PoFL incurs higher computation and communication for local training, parameter exchange, and contribution evaluation. 
PoBT is lightweight with low computation, communication, and verification costs, making it suitable for resource-constrained IoT DePIN systems.

\subsection{Summary and Lessons Learned}
The economic driving force of DePIN lies in turning users into resource contributors and network stakeholders. By aligning participant incentives with network growth, DePIN fosters a service ecosystem responsive to community needs, and conducive to innovation.
This structure improves the feasibility of scaling DePIN by lowering capital entry barriers and enabling organic, incentive-driven growth.

Despite the foundation for the incentive layer in DePIN, challenges and unresolved issues remain.
First, consensus mechanisms must reward verifiable physical effort fairly while avoiding centralized oracles.
Second, large-scale verification and reward distribution across heterogeneous devices create a scalability decentralization trade-off. Mechanisms, e.g., Oracle PoC, improve efficiency through partial re-centralization and may weaken DePIN’s decentralization principle.

\section{Application Layer: Decentralized Ecosystem}\label{sect:application}
This section examines the applications of decentralized ICT infrastructure from the perspective of DePIN across key vertical domains, along with the introduction of several typical projects, while exploring the synergistic integration between AI and decentralized ICT infrastructure.

\begin{table*}
    \centering
    \captionsetup{
    font={scriptsize}}
    \caption{Summary of different application scenarios of decentralized ICT infrastructure in DePIN } \label{tab:Application}
    \fontsize{6pt}{7pt}\selectfont
    \begin{tblr}{
        width = \linewidth,
        colspec = {X[1.5]X[2]X[0.8]X[8.7]},
        row{1} = {c},
        cell{1}{1} = {r=2}{c},
        cell{1}{2} = {r=2}{c},
        cell{1}{3} = {r=2}{c},
        cell{1}{4} = {r=2}{c},
        cell{3}{1} = {r=2}{c},
        cell{3}{2} = {r=2}{c},
        cell{3}{3} = {r=2}{c},
        cell{3}{4} = {r=2}{l},
        cell{5}{1} = {r=2}{c},
        cell{5}{2} = {r=2}{c},
        cell{5}{3} = {r=2}{c},
        cell{5}{4} = {r=2}{l},
        cell{7}{1} = {r=2}{c},
        cell{7}{2} = {r=2}{c},
        cell{7}{3} = {r=2}{c},
        cell{7}{4} = {r=2}{l},
        cell{9}{1} = {r=3}{c},
        cell{9}{2} = {r=3}{c},
        cell{9}{3} = {r=3}{c},
        cell{9}{4} = {r=3}{l},
        cell{12}{1} = {r=2}{c},
        cell{12}{2} = {r=2}{c},
        cell{12}{3} = {r=2}{c},
        cell{12}{4} = {r=2}{l},
        cell{14}{1} = {r=2}{c},
        cell{14}{2} = {r=2}{c},
        cell{14}{3} = {r=2}{c},
        cell{14}{4} = {r=2}{l},
        cell{16}{1} = {r=2}{c},
        cell{16}{2} = {r=2}{c},
        cell{16}{3} = {r=2}{c},
        cell{16}{4} = {r=2}{l},
        hlines,
        vline{2-4} = {},
        }
        \textbf{Aspects}       & \textbf{Typical Projects}                       & \textbf{Refs.}                                                       & \textbf{Core Concept}                \\ \\ \hline

        Smart City             & MapMetrics, Hivemapper                          & \cite{Xie_SmartCity_COMST, Chiu_CircularX_ISNCC}                             &
        Empower citizens with participatory rights and co-governance authority over urban infrastructure systems                                                                                       \\ \\

        Healthcare             & Pulse, HealthBlocks                             & \cite{Zaabar_HealthBlock_CN, Ettaloui_GDPR_ICAISE, Said_GDPR_ICMDE}          &
        Ensure robust patient data privacy and security while incentivizing collaborative participation in the construction, maintenance, and efficient utilization of medical databases               \\ \\

        Smart Home             & Tempest, EnviroBLOQ                             & \cite{Xu_Home_IoTJ, Ardagna_Home_TCPS}                                       &
        Achieve secure interoperability in decentralized environments with user sovereignty, while fostering open collaboration and resource sharing among interconnected systems                      \\ \\

        Wireless Communication & Helium, Dawn, Grass, Wicrypt, ROAM, Karrier One & \cite{Caprolu_DeWi_ICNS, Khor_DeWi_IoTJ}                                     &
        Incentivize individuals to deploy and operate wireless networking infrastructure and establish a more cost-effective, ubiquitous, and resilient decentralized wireless communication ecosystem \\ \\ \\

        Energy Sharing         & PowerLedger, Daylight, StarPower                & \cite{Wang_Energy_TSMCM, Kim_Energy_ICOIN}                                   &
        Leverage decentralized technologies for facilitating P2P trading and management of renewable energy, thereby enhancing energy utilization efficiency and sustainability                        \\ \\

        Data Storage           & Filecoin, Arweave, Ocean Protocol               & \cite{Giacomelli_Filecoin_CrypTorino, Williams_Data_Arweave, Ocean_Protocol} &
        Transform the centralized paradigm of traditional cloud storage by achieving enhanced data security, absolute privacy preservation, cost efficiency, and robust censorship resistance          \\ \\

        Compute                & IO.NET, Aethir, Render Network                  & \cite{Yuan_Compute, Bowe_Compute_SP}                                         &
        Establish an open, flexible, and efficient decentralized computing marketplace by aggregating underutilized computational resources \textemdash particularly GPU across global networks        \\ \\
    \end{tblr}
\end{table*}

\subsection{Vertical Applications}
DePIN is demonstrating application potential across vertical domains, as shown in Table \ref{tab:Application}. Representative projects validate the technical feasibility of DePIN, and highlight its unique value in building user-driven, open, and shareable infrastructure ecosystems. This subsection analyzes key vertical sectors, accompanied by an exploration of notable project implementations in each domain.

\subsubsection{Smart City}
For urban systems, DePIN introduces a decentralized service model that connects community contributed resources with energy management, waste management, and transportation applications \cite{Xie_SmartCity_COMST}. 
Decentralized ridesharing, food delivery, and mapping networks help reduce platform dependence and enable users to participate directly in service provision. 
Representative examples include \textit{MapMetrics}, which rewards users for real-time geospatial contributions, and \textit{Hivemapper}, which crowdsources dashcam footage to build dynamic maps with HONEY token incentives \cite{Chiu_CircularX_ISNCC}.

\subsubsection{Healthcare}
Within healthcare systems, DePIN shifts health-data management toward a user controlled and incentive compatible data sharing paradigm. Patients can retain ownership of their health data while satisfying regulatory requirements such as HIPAA \cite{Ettaloui_GDPR_ICAISE} and GDPR \cite{Said_GDPR_ICMDE}. Token incentives and P2P funding can reduce access barriers and motivate data contribution.
Current projects, e.g., \textit{Pulse} and \textit{HealthBlocks}~\cite{Zaabar_HealthBlock_CN}, aggregate wearable device data and allow users to share it with external third-party entities for real-world rewards. Secure permissioned sharing supports global medical collaboration while preserving patient confidentiality \cite{Sun_Federated_GBC}.

\subsubsection{Smart Home}
DePIN mitigates key limitations of traditional IoT systems, including privacy risks, single points of failure, and limited cross-vendor interoperability, through decentralized device control \cite{Rathore_Home_IoT}. Projects, such as \textit{Tempest} and \textit{EnviroBLOQ}, are targeting the vast amounts of data generated by IoT systems by issuing tokens to acquire usage rights, thereby incentivizing data sharing in a privacy-preserving and user-consented manner \cite{Xu_Home_IoTJ, Ardagna_Home_TCPS}.

\subsubsection{Wireless Communication}
Decentralized Wireless (DeWi) is one of the most dynamic areas within the DePIN application layer. DeWi offers a novel approach to wireless connectivity by leveraging blockchain-based incentives and community-operated hardware to overcome the limitations of traditional telecommunications models.
\textit{Helium Mobile} \cite{Helium} is one of the most prominent projects, followed by emerging projects including \textit{Dawn}, \textit{Grass}, \textit{Wicrypt}, \textit{ROAM}, and \textit{Karrier One} \cite{Caprolu_DeWi_ICNS, Khor_DeWi_IoTJ}. As deployments expand and supporting technologies mature, DeWi is expected to gain further practical and economic significance.

\subsubsection{Energy Sharing}
Decentralized energy networks form a key DePIN scenario for coordinating distributed energy resources and renewable energy markets. 
Moving beyond centralized supply models, projects such as \textit{PowerLedger}~\cite{PowerLedger}, \textit{Daylight}~\cite{daylight}, and \textit{StarPower}~\cite{starpower} explore decentralized energy management for distributed grids \cite{Hu_Grid_TGCN}. 
For instance, \textit{Daylight} enables users with solar panels or grid-connected devices to sell energy and related data to utility providers, supporting more responsive grid operation.

\subsubsection{Data Storage}
DePIN addresses critical challenges in traditional data ecosystems, e.g., limited accessibility, lack of transparency, and high privacy risks, through decentralized data storage and data marketplaces. Decentralized storage solutions like \textit{Filecoin}~\cite{Giacomelli_Filecoin_CrypTorino} and \textit{Arweave}~\cite{Williams_Data_Arweave} leverage distributed architectures to minimize single points of failure while enhancing cost efficiency. Meanwhile, projects such as \textit{Ocean Protocol} \cite{Ocean_Protocol} employ tokenization and decentralized exchange mechanisms to establish open channels for data exchange, thereby facilitating data valuation and liquidity.

\subsubsection{Compute}
In the domain of decentralized computing, idle GPU resources are aggregated to meet the demands of compute-intensive workloads \cite{Bowe_Compute_SP}. Projects, such as \textit{IO.NET}~\cite{IONET}, \textit{Aethir}~\cite{Aethir}, and \textit{Render Network}~\cite{render}, are advancing the scalability of decentralized GPU computation. This model harnesses underutilized chips from sources like gaming personal computers and data centers to create a permissionless compute marketplace.

\subsection{Decentralized Artificial Intelligence}
The vast amount of data generated by DePIN nodes provides resources for the fine-tuning and continual optimization of AI models. 
With the decentralization of computing resources and the introduction of blockchain-based incentives, the authority over model training, deployment, and monetization is shifting to users and developer communities \cite{Tang_TMC}. We explore decentralized AI model marketplaces and challenges associated with model training in DePIN.

\subsubsection{Decentralized AI Model Marketplaces}
The establishment of decentralized AI model marketplaces aims to address longstanding issues in traditional AI ecosystems, e.g., lack of transparency, high privacy risks, and misaligned incentive mechanisms.
By leveraging the decentralization, immutability, and programmability of blockchain, these platforms offer a trustworthy environment for data and model transactions \cite{Harris_DAI_ICBC}.

In decentralized marketplaces, data providers gain greater control over the lifecycle of their data while data consumers verify the provenance and authenticity of the data.
Several emerging projects have explored this. For example, \textit{Bittensor}~\cite{Steeves_PoI_Bittensor} is a decentralized AI network in which participants train ML models and earn TAO tokens in proportion to the value of their models' contribution to the collective intelligence of the network.
\textit{Effect AI}~\cite{effectAI} enables individuals to contribute to AI training by completing microtasks on a decentralized platform.
\textit{SingularityNET}~\cite{Anastasiu_SingularityNET} is a decentralized marketplace for AI services, where developers can publish, share, and monetize AI algorithms.

Moreover, Ouyang \textit{et al.} \cite{Ouyang_DAI_IoTJ} proposed a decentralized AI collaboration framework called the ``Learning Market''. This framework leverages smart contracts to implement core functionalities such as model verification and incentive quantification to provide a fair, transparent, and traceable marketplace for AI collaboration and model exchange.
Yu \textit{et al.} \cite{Yu_DAI_TNNLS} proposed a decentralized framework named IRONFORGE, which eliminates the need for a central coordinator by adopting a DAG-based architecture, enabling fully decentralized and asynchronous operations.

\subsubsection{Decentralized Model Training}
DePIN systems generate massive amounts of data that fuel model training. Challenges include data heterogeneity, regulatory compliance, and privacy.

To address data heterogeneity, the study in \cite{Chen_DAI_TDSC} proposed a robust personalized FL framework that combines $\alpha$-based hierarchical normalized similarity with local collaborative training. This approach constructs personalized aggregated models while extracting key features, enabling the unlearning of corrupted data.
Considering diverse data structures, heterogeneous tasks, and imbalanced sample distributions,
Baccour \textit{et al.} \cite{Baccour_DAI_IoTJ} introduced federated meta-learning and a blockchain-based cooperative coalition formation game. This game leverages reputation metrics, user similarity, and incentive mechanisms to cluster meta-learners to address heterogeneity. Similar issues are discussed in \cite{Wan_WFL_COMST, Qu_FL_TNSE, Feng_WFL_TIFS, Xiao_FL_TIFS}.

Regulatory compliant data usage remains a significant challenge, particularly in privacy-sensitive domains like healthcare.
Lian \textit{et al.} \cite{Lian_DAI_TSC} introduced blockchain to enable decentralized control and full traceability of data access, ensuring that original data remains local during model training and is not exposed.
Liu \textit{et al.} \cite{Ren_DAI_TNSE} leveraged IoT edge computing and local modeling to ensure patient data remains on local devices, participating in collaborative training only through low-rank features or model parameters.
With blockchain support~\cite{Sun_Federated_GBC}, decentralized training can be coordinated at a global scale, while smart contracts automate task allocation and model aggregation to reduce the risk of data leakage.

Noisy data can leave persistent adverse effects in trained models, which may spread through P2P networks and decentralized model marketplaces. The work \cite{Yu_DAI_TNNLS} leverages DAG-based structures in a blockchain network to enable model training traceability. It employs a noise-aware Proof-of-Learning mechanism to detect and penalize poisoning behavior, and model unlearning to mitigate the impact of noisy data.

\subsection{AI4ALL}
\begin{figure}[!t]
    \centering
    \includegraphics[scale=0.4]{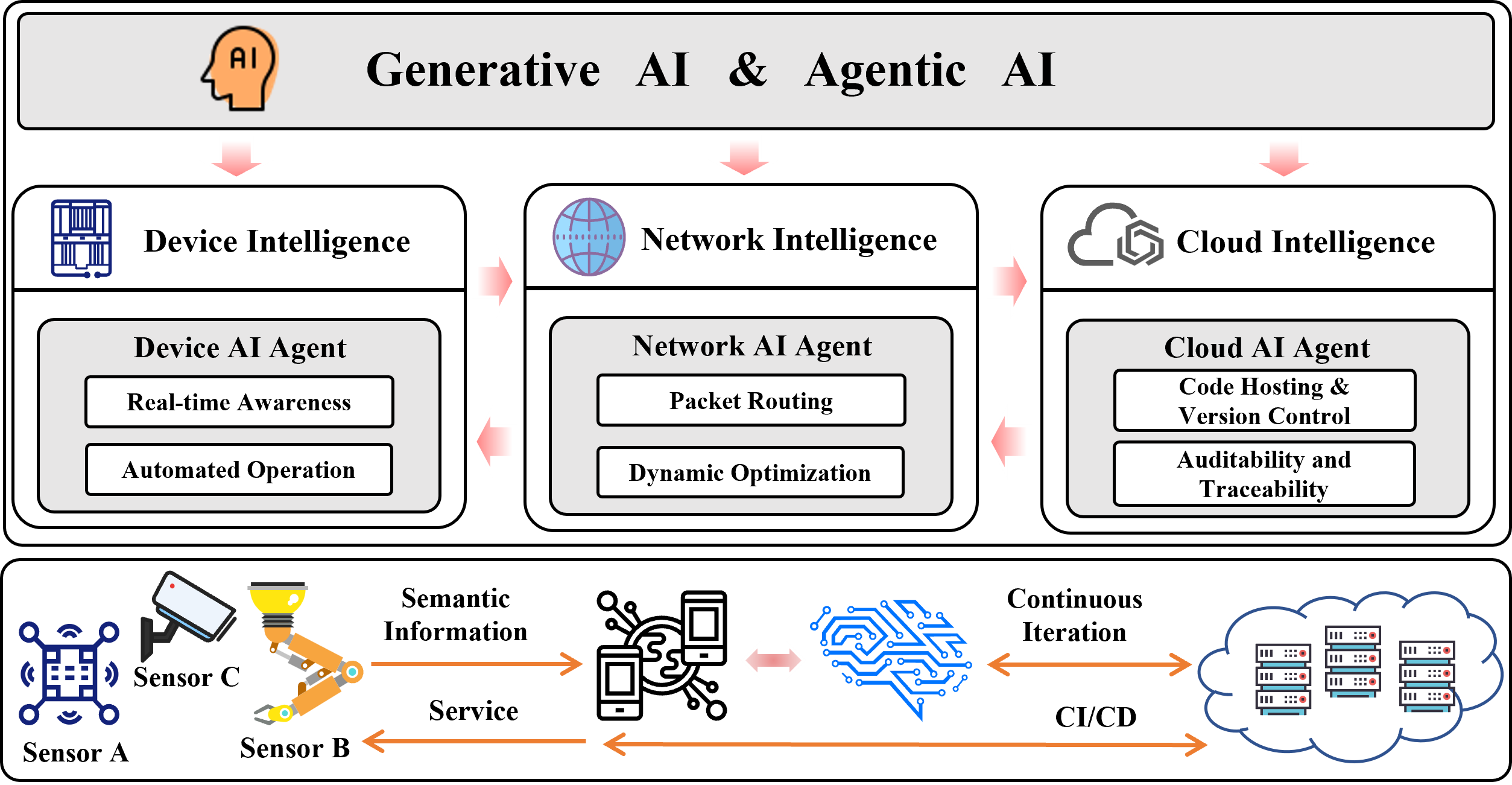}
    \caption{AI4ALL in DePIN: Generative and Agentic AI for Device, Network, and Cloud Agents.}
    \label{fig:AI4ALL}
\end{figure}

In the DePIN ecosystem, AI is becoming a pervasive capability spanning devices, networks, and clouds, enabling decentralized infrastructure to evolve from passive resource coordination to autonomous perception, decision-making, and service delivery~\cite{Manalastas_DAI_CCNC}; see Fig. \ref{fig:AI4ALL}.
Generative AI and agentic AI are key enablers, but their integration also raises risks related to hallucinations, erroneous decisions, edge deployment costs, and security attacks.

\subsubsection{Generative AI and Agentic AI}
Generative AI and agentic AI are no longer auxiliary application modules in DePIN; they form the enabling intelligence stack that underpins device-level perception, network-level orchestration, and cloud-level lifecycle automation.
Generative AI can provide semantic understanding and content or policy generation~\cite{Lv_GAI_CR}, while agentic AI supports planning driven by user objectives, tool invocation, memory, and closed-loop execution across heterogeneous DePIN resources~\cite{schneider_AAI_arXiv}.
Future DePIN systems may leverage multimodal models to interpret heterogeneous sensor streams, network telemetry, and operational logs; lightweight small language models deployed through distillation or fine-tuning methods such as Low-Rank Adaptation (LoRA) for low-latency edge inference~\cite{Lin_LLM_TMC}; and Retrieval Augmented Generation (RAG) combined with decentralized knowledge bases to ground outputs in real-time conditions~\cite{Zhu_RAG_COMS}.
Built on this semantic layer, agentic orchestration~\cite{Ruan_Aorchestra_arXiv} can coordinate cross-domain infrastructure operations across network, device, and cloud domains, supporting adaptive resource control, autonomous decentralized Continuous Integration/Continuous Deployment (CI/CD) and service lifecycle management.
In addition, verifier models and self-reflection mechanisms may be introduced to validate plans before execution, improving the reliability and safety of cross-layer autonomous operations in DePIN.

\subsubsection{Network Agent for Intelligent Routing and Allocation}
Network agents, enabled by generative AI for intent understanding and by agentic AI for goal-driven planning and execution, can optimize packet routing in decentralized networks, reducing latency and improving overall network performance \cite{Jiang_HAMARL_TCCN}. 
They can interpret high-level service intents, adapt to changing conditions, route traffic through alternative paths when failures occur~\cite{Valcher_DAI_ICSTCC}, and dynamically distribute tasks to different device agents based on real-time network states.

The integration of network agents with smart contracts introduces advanced capabilities, e.g., dynamic decision-making, real-time adaptability, and predictive analytics. Smart contracts can analyze both historical and real-time data to enhance the accuracy of future decisions, while enabling self-correction and adaptation to new conditions \cite{Duan_MSL_TIFS}. This reduces the need for manual intervention and increases operational efficiency.

\subsubsection{Device Agent for Intelligent Monitoring and Scheduling}
In manufacturing, IoT sensors gather task-relevant data in response to requests issued by network agents. This data is encoded and transmitted using joint source-channel coding to produce semantic representations \cite{Zhang_DAI_TWC}. Device agents reconstruct the semantic information, e.g., voice commands, and execute the corresponding tasks.
Device agents dynamically adapt the modality and granularity of data collection based on environmental conditions. Semantic forwarding allows for minimal bandwidth consumption without sacrificing task accuracy and communication efficiency, especially in extreme communication environments \cite{Lin_SLACC_TVT}.

\subsubsection{Cloud Agent for Decentralized CI/CD}
CI and CD are two practices in modern software engineering that enhance development efficiency and code reliability \cite{Zampetti_DAI_ICSME}. In Web 3.0 and dApps, these principles face new challenges.
Decentralized CI/CD relies on cloud agents to translate generated artifacts and deployment intents into verifiable actions while maintaining consistency between on-chain states and off-chain logic.
The absence of centralized coordination and the reliance on trustless environments introduce gaps in automation, verification, and governance \cite{Ugwueze_DAI_IJCATR}.

\begin{itemize}
    \item \textit{Code Hosting and Version Control}: A decentralized code hosting system should support on-chain recording and verifiable auditing of code submission and merge operations. Current systems, e.g., Radicle, face limited scalability, weak version traceability, and insufficient conflict resolution. 
    AI agents can assist code review, conflict detection, and audit record generation, improving governance efficiency while preserving verifiable accountability.

    \item \textit{Auditability and Traceability Support}: Despite the immutability of blockchain, the current system lacks end-to-end traceability that cryptographically links source code, build artifacts, and deployed binaries. Establishing a unified and tamper-evident pipeline that enables full accountability and rapid recovery across all development stages remains an unresolved challenge \cite{Saleh_DAI_CIoT}.
\end{itemize}

\subsubsection{Risks}
While generative AI and agentic AI enhance operational efficiency in DePIN, they introduce risks.

\begin{itemize}
    \item \textit{Hallucinations and Erroneous Decision} represent a primary risk of generative AI and agentic AI in DePIN~\cite{Chakraborty_hallucination_ACM}. 
    Fabricated intents and flawed reasoning may propagate across device, network, and cloud layers, leading to incorrect behavior and degraded QoS. 
    Such risks are harder to verify, attribute, roll back, and recover from in decentralized environments due to model opacity and non-determinism.

    \item \textit{Edge Deployment Costs}: The deployment of AI capabilities at the edge remains costly. Many DePIN nodes are resource-constrained and cannot support the computation, memory, energy, and communication overhead required by generative models. Frequent model updates, real-time inference, and cross-layer agent coordination may further increase latency and operational expenditure (OPEX), while creating participation barriers for smaller nodes and weakening the inclusiveness of DePIN.

    \item \textit{Security Attacks}: The tight coupling among generative AI, autonomous agents, smart contracts, and physical devices in DePIN ecosystems gives rise to severe cross-layer vulnerabilities~\cite{schneider_AAI_arXiv, Liu_DAI_TKDE}. Frameworks such as OpenClaw amplify this concern by exposing low-level physical interfaces to AI agents, making prompt injection or logic manipulation capable of triggering unauthorized physical operations~\cite{Chakraborty_hallucination_ACM}. 
    In decentralized CI/CD pipelines, AI-generated misconfigurations may be propagated from software logic to physical nodes at scale, creating systemic risks that are difficult to isolate, roll back, and recover from once deployed.
\end{itemize}

\subsection{Summary and Lessons Learned}
As the DePIN ecosystem expands across diverse domains, ICT infrastructure may evolve along four directions: 
1) semantic and task-oriented communication, which emphasizes meaning preservation rather than bit-level accuracy; 
2) decentralized data governance, which supports data sovereignty, value circulation, and resistance to leakage or model inversion; 
3) AI-blockchain co-design, which enables trustworthy on-chain reasoning and transparent model lifecycle management; and 
4) heterogeneous multi-task collaboration, which requires adaptive task allocation across diverse devices and user demands to support resilient autonomous systems.

\section{Cross Layer System Consistency in DePIN}\label{sect:crosslayer}
The preceding six-layer architecture provides a structured view of DePIN.
Yet, practical DePIN systems are not a simple stack of independent layers. 
Physical infrastructure, blockchain settlement, interaction protocols, trust mechanisms, token incentives, governance rules, and application requirements are interdependent.
The rest of this section analyze this coupling from five dimensions.

\subsection{Reality--Ledger Consistency}
In DePIN, on-chain state advances faster than the physical evidence needed to justify it.
Physical contributions depend on delayed measurements or corroboration before their validity is established~\cite{choi_DePIN_ICBC}.
This mismatch can be handled in three stages:
(1) Signed service receipts or proof commitments are recorded without waiting for complete validation~\cite{Chen_sharding_IoTJ}.
(2) The underlying evidence is verified asynchronously through witnesses, oracles, or dispute procedures, with the challenge period adapted to evidence latency and verification cost~\cite{Sober_rollups_DAPPS}.
(3) Rewards remain provisional until verification is completed, while fraud challenges and compensation mechanisms discourage false claims~\cite{Rammouz_Helium_WiMob}.

\subsection{Trust--Value Alignment}
DePIN ties reward eligibility to a unique physical device.
Manufacturer-issued credentials can establish device provenance, while PUF fingerprints and TEE/TPM attestation bind the digital identity to a hardware root and approved firmware~\cite{Avellaneda_DID_COMSM,Duan_Light_TIFS}.
Then, periodic attestation and key rotation can maintain this binding and exclude compromised devices from the reward eligible set~\cite{Cirne_TEE_COMST}.
Reward settlement will combine current attestation results with service evidence. 
Game-based Reputation updates are expected to reduce the expected gains from credential reuse, low effort participation, and verifier collusion~\cite{Tang_TMC,Zhou_game_IoTJ}.
Correspondingly, the incentive mechanism can compensate contributors for ensuring trusted hardware and continuous attestation, linking 
verification costs with participation rewards.

\subsection{Service--Proof Coherence}
In DePIN, service verification begins with path construction rather than after service completion. 
Route selection jointly considers QoS requirements and the availability of independent evidence sources~\cite{Cao_BRAN_TWC,Guo_route_IoTJ}. 
As the service is delivered, participating nodes generate signed, timestamped receipts that record forwarding actions and service performance.
These records are then checked against the requested QoS, route announcements, and settlement conditions.
The verification outcome updates node reputation and subsequent routing, reducing the reuse of paths that provide favorable short-term performance but insufficient auditability~\cite{He_P2P_TPDS}.

\subsection{Privacy--Accountability Balance}
DePIN preserves privacy during routine verification while enabling controlled accountability in cases of fraud or non-compliance.
Techniques such as ZKPs reveal only eligibility or compliance results~\cite{Rosenberg_zkcreds_SP}.
Once fraud is reported, the default privacy state shifts to selective disclosure, limited to the attributes required for resolution.
To prevent misuse of disclosure authority, each disclosure request requires approval from multiple governance entities and is recorded together with the resulting enforcement action~\cite{Dai_PASSP_TNSM, Battah_MPA_Access}. 
Only evidence disclosed and verified through this procedure can serve as the basis for penalties, while unrelated identity and service data remain protected.

\subsection{Policy--Market Compatibility}
Token design in DePIN is constrained by the regulatory status of the service~\cite{Freni_Token_BRA}. 
The function of a token determines its issuance and transfer requirements, which are further incorporated into participant eligibility and settlement rules under domain-specific regulations~\cite{Cappiello_blockchain_Book}.
Within these constraints, token issuance and reward allocation can be calibrated to underlying service economics rather than token circulation alone~\cite{ Akcin_Token_arXiv}.
Regulatory violations or deteriorating provider viability can then trigger revisions to eligibility, reward, and settlement rules, maintaining alignment between token utility, compliant service delivery, and sustained demand.

\section{DePIN Feasibility Assessment Framework}\label{sect:feasibility}
A structured feasibility assessment is needed to move DePIN beyond conceptual promise toward practical deployability~\cite{zheng2025navigating}. 
In the current early-stage of DePIN systems, this assessment does not define universal pass/fail thresholds. 
Instead, it identifies key feasibility dimensions, examination criteria, and representative indicators that can be adapted to different service requirements, deployment scenarios, and regulatory contexts. 
Table~\ref{tab:feasibility} summarizes the framework from technical, regulatory and governance, and economic dimensions using a \textit{Question-Criteria-Indicators} logic.
This framework can serve as a practical reference for evaluating DePIN projects. 
It helps researchers and practitioners identify which dimensions need evidence and how qualitative assessment can be connected with measurable indicators in a specific DePIN deployment context.

\begin{table*}[t]
    \captionsetup{
    labelfont={color=black},
    textfont={color=black},
    font={scriptsize}
    }
\caption{A feasibility assessment framework for DePIN deployment. The marker `\bluecircle' represents assessment criteria, `\redcircle' denotes representative indicators.}
\label{tab:feasibility}
\fontsize{6pt}{7pt}\selectfont
\begin{tblr}{
    width=\linewidth,
    colspec={X[1,l,m] X[1.8,j,m] X[1.2,l,m] X[2.4,l,m] X[2.8,l,m]},
    row{1} = {c, font=\bfseries},
    cells = {fg=black},
    cell{2}{1} = {r=5}{},
    cell{2}{2} = {r=5}{},
    cell{7}{1} = {r=2}{},
    cell{7}{2} = {r=2}{},
    cell{9}{1} = {r=2}{},
    cell{9}{2} = {r=2}{},
    hlines = {black},
    vline{2-5} = {black},
}
Feasibility & Guiding Question & Dimension & Assessment Criteria &  Representative Indicators \\ \hline

Technical
& Can the DePIN system verifiably deliver the promised service at scale?
& Verifiability
& \bluecircle \textit{} Third-party cross-validation; \newline 
  \bluecircle \textit{} Auditable evidence; \newline
  \bluecircle \textit{} Standardized proof formats
& \redcircle \textit{} Service-proof success rate; \newline 
  \redcircle \textit{} QoS proof fulfillment rate; \newline
  \redcircle \textit{} Contrib. cross-validation consistency; \newline 
  \redcircle \textit{} Disputed or unverifiable ratio \\

& & Service Performance
& \bluecircle \textit{} Application-level QoS; \newline
  \bluecircle \textit{} Cross-provider consistency
& \redcircle \textit{} Coverage probability; \newline 
  \redcircle \textit{} Task completion rate; \newline 
  \redcircle \textit{} Service fulfillment ratio; \newline
  \redcircle \textit{} Verifiable QoS-delivery ratio \\

& & Reliability
& \bluecircle \textit{} Resilience to churn/failures; \newline
  \bluecircle \textit{} Critical-region redundancy; \newline 
  \bluecircle \textit{} Closed-loop recovery; \newline 
  \bluecircle \textit{} Stable version governance
& \redcircle \textit{} Interruption frequency; \newline 
  \redcircle \textit{} MTBF, MTTR; \newline 
  \redcircle \textit{} Node-churn resilience ratio \\

& & Scalability
& \bluecircle \textit{} Hierarchical scaling; \newline 
  \bluecircle \textit{} Overhead control; \newline 
  \bluecircle \textit{} Settlement expansion/efficiency
& \redcircle \textit{} Convergence time; \newline 
  \redcircle \textit{} Proof generation latency; \newline
  \redcircle \textit{} Cross-domain sync. overhead \\

& & Security and Anti-Cheating
& \bluecircle \textit{} Key/device protection; \newline 
  \bluecircle \textit{} Cheating detection \& penalty; \newline 
  \bluecircle \textit{} Vulnerability response
& \redcircle \textit{} Attack detection success rate; \newline 
  \redcircle \textit{} Punishment execution rate; \newline 
  \redcircle \textit{} Contribution fraud detection ratio \\

Regulatory \& Governance
& Is it feasible to deploy and govern the DePIN system compliantly?
& Accountability and Data Governance
& \bluecircle \textit{} Role identifiability; \newline 
  \bluecircle \textit{} Explicit data boundaries; \newline 
  \bluecircle \textit{} Governance for data operations 
& \redcircle \textit{} Compliance incident rate; \newline 
  \redcircle \textit{} Non-attributable incident ratio; \newline
  \redcircle \textit{} Authorization record completeness \\

& & Policy Enforcement
& \bluecircle \textit{} Upgrade/pause/rollback; \newline 
  \bluecircle \textit{} Cross layer coordination; \newline 
  \bluecircle \textit{} Rule-to-action closure
& \redcircle \textit{} Policy proposal execution rate; \newline 
  \redcircle \textit{} Violation closure rate; \newline 
  \redcircle \textit{} Rollback success rate; \newline
  \redcircle \textit{} Decision implementation latency \\

Economic
& Will the network sustain supply and demand over time?
& Unit Economics
& \bluecircle \textit{} Hardware/OPEX burden; \newline 
  \bluecircle \textit{} Resilience to weak incentives
& \redcircle \textit{} Payback period; \newline 
  \redcircle \textit{} Service-revenue-to-OPEX ratio \\

& & Demand-side Sustainability
& \bluecircle \textit{} Genuine usage share; \newline 
  \bluecircle \textit{} Post-incentive demand survival; \newline 
  \bluecircle \textit{} User retention
& \redcircle \textit{} Service-driven revenue share; \newline 
  \redcircle \textit{} Demand survival rate; \newline 
  \redcircle \textit{} Post-bootstrapping user retention \\

\end{tblr}
\end{table*}

\subsection{Technical Feasibility}
\textit{Question}: Can the DePIN system verifiably deliver the promised service at scale with acceptable performance, reliability, security, and operational continuity~\cite{Assen_performance_ICBC}?

\subsubsection{Verifiability}
In DePIN systems, verifiability gives physical contributions a form that can be trusted beyond the contributor's claim.
A reported event or action should be accompanied by evidence that other parties can inspect and when necessary, dispute.
This dimension is less about producing proofs, but more about making proofs auditable and corroborated across validators.
Metrics such as service-proof success rate~\cite{Anand_Service_ACM}, QoS proof fulfillment~\cite{Rammouz_Helium_WiMob}, cross-validation consistency, disputed proof rate, and unverifiable contribution ratio~\cite{Wang_QoS_TSC} can be used to reflect how often submitted claims pass the verification.

\subsubsection{Service Performance}
Service performance needs to be evaluated according to the specific application. 
For communication DePIN projects, this may include coverage, latency, throughput, availability, and handover performance. 
For storage or computing DePIN projects, it may include retrieval delay, task completion rate, execution time, and service availability. The key point is whether decentralized providers can offer service quality that is stable enough for real users~\cite{Wang_QoS_TSC}.
Representative indicators include coverage probability, task completion rate, service fulfillment rate, and verifiable QoS delivery ratio~\cite{Cao_BRAN_TWC}.

\subsubsection{Reliability}
Reliability is critical as DePIN infrastructure is operated by distributed and often independent participants.
Nodes may leave, fail, update software, suffer attacks, or behave inconsistently, so the service must depend on redundancy, recovery mechanisms, and adaptive coordination.
Evaluation should look at how service quality changes during churn or failure events~\cite{Dotan_reliability_ACM}, with interruption frequency, mean time between failures (MTBF), mean time to repair (MTTR)~\cite{Mauro_reliability_COMST}, regional redundancy, and post-churn service quality~\cite{Carlos_Resiliency_COMST} serving as supporting indicators.

\subsubsection{Scalability}
Scalability in DePIN is not a matter of adding more nodes; the growth of the network increases the burden of proof generation, synchronization, coordination, and reward settlement~\cite{Duan_Cross_JAS, Han_atomicity_DLT}.
A large or highly dynamic network, e.g., a mobile network, may be less efficient if physical proofs accumulate faster than they can be verified, or if cross-domain interactions introduce excessive latency and overhead.
The evaluation should capture how the system behaves as participation expands across heterogeneous devices. Decentralized control convergence time~\cite{Du_scalability_TCNS}, proof generation latency~\cite{Bai_ZK_IoTJ}, synchronization overhead, settlement latency, and overhead growth rate~\cite{Fan_scalability_ACM} can serve as indicators.

\subsubsection{Security and Anti-Cheating}
Open participation makes DePIN attractive, but it creates opportunities for various attacks and cheating.
Security evaluation should connect cyber protection with physical accountability, e.g., by validating coverage claims against independent physical measurements.
This is important for attacks, e.g., fabricated wireless coverage or replayed service records, where the system may appear functional unless contribution claims are challenged~\cite{Rammouz_Helium_WiMob}.
Relevant evidence can be false-positive and false-negative rates~\cite{Qiu_Security_IoTJ}, punishment execution rate, attack detection success rate, fraudulent contribution detection ratio~\cite{Shastri_Security_arXiv}, and vulnerability response time.

\subsection{Regulatory and Governance Feasibility}
\textit{Question}: Is it feasible to deploy and govern DePIN systems with identifiable responsibilities, governable data, and legally executable settlement across deployment scenarios?

\subsubsection{Accountability and Data Governance}
Accountability and data governance link decentralized participation with clearly assigned responsibility. In DePIN systems, data is produced by physical devices, interpreted by service providers, verified by validators, and used for reward allocation or governance decisions. Unclear data boundaries may lead to ambiguous accountability~\cite{Mazzocca_DID_COMST}.
Once data moves across physical devices, off-chain services, and on-chain records, responsibility will not follow a single administrative chain; hence, data governance is needed to transform distributed actions into attributable system behavior.
Supporting indicators include compliance incident rate, non-attributable incident ratio~\cite{Deng_regulation_TC}, auditable data access ratio~\cite{Wang_regulation_IoTJ}, authorization record completeness, and revocation execution rate.

\subsubsection{Policy Enforcement}
Policy enforcement concerns the gap between governance as a decision-making process and governance as an operational capability.
When a DePIN community approves a protocol, the decision is meaningful after it is deployed into practice. This makes enforcement complex, since on-chain rules and off-chain infrastructure do not automatically evolve at the same pace.
Evaluation should pay attention to implementation delays, inconsistent execution across domains, unresolved violations, failed rollbacks, and the ability to restore system stability when governance decisions produce unintended consequences.
Proposal execution rate, violation closure rate~\cite{Ma_policy_TSE}, rollback success rate, and decision implementation latency~\cite{Kiayias_policy_ACM} can serve as supporting indicators.

\subsection{Economic Feasibility}
\textit{Question}: Will the network sustain participation, service supply, and infrastructure maintenance beyond the bootstrapping phase under realistic demand and market conditions?

\subsubsection{Unit Economics}
For resource providers, DePIN participation is constrained by the economics of operating physical infrastructure.
Unlike digital validators, physical contributors must bear both capital expenditure and ongoing operational costs before earning service revenue~\cite{han_incentive_ACM}.
This dimension looks at the provider-side cost revenue balance behind each unit of contributed resource.
Indicators can include payback period, maintenance cost, service-revenue-to-OPEX ratio~\cite{ Cong_economy_arXiv}, infrastructure cost recovery ratio, and the share of token-independent revenue~\cite{Freni_Token_BRA}.

\subsubsection{Demand-side Sustainability}
The demand side raises a different question: whether users and applications assign sufficient value to the service.
In early-stage DePIN systems, token incentives may create apparent demand through reward-driven transactions, device participation, or service usage.
Sustainable demand should be evaluated by repeated use, external payment, post-bootstrapping retention, and continued service consumption after internal incentive circulation declines.
Service-driven revenue share, repeat use rate, active user ratio, demand survival rate, and post-bootstrapping user retention can be used as supporting indicators~\cite{Chami_economy_arXiv}.

\section{Challenges and Future Research Directions}\label{sect:future}
Although DePIN demonstrates the potential to reshape the future ICT infrastructure, it remains an emerging paradigm that faces a series of critical challenges. This section examines these challenges and outlines key research directions that merit further exploration.

\subsection{Secure and Trustworthy Physical Hardware}
\textit{Challenges}:
The value of DePIN lies in its ability to securely and credibly map physical resources and services into the digital world. The inherent mutability of physical devices and the complexity of their environments challenge the assumption that ``data on-chain is trust'' \cite{Cirne_challenges_COMST}. Ensuring the authenticity of physical actions, e.g., sensor readings, network coverage proofs, and computational contributions, represents the ``last mile'' trust problem that DePIN must address.

Future research directions lie in the development of ``security-by-design” DePIN devices, where the core of potential solutions is hardware-level security techniques.
\begin{itemize}
    \item \textit{Integration of TEE}: Future research should focus on extending TEE to diverse edge devices, e.g., routers and sensors, to establish isolated secure enclaves at the chip level. Critical task code and cryptographic keys can be executed and stored within the TEE. Even if the device operating system is compromised, the core logic and sensitive data remain intact and confidential.
    \item \textit{PUF as Hardware Fingerprint}: PUF could be explored to generate a unique and unclonable hardware fingerprint for each device, serving as the root for the device DID.
    \item \textit{Cryptographic Agility and Secure Firmware Upgrade}: DePIN devices should be designed with built-in cryptographic agility and secure firmware upgrade capabilities, which enable the devices to evolve from legacy primitives to stronger schemes without hardware replacement.
    This is critical for authenticated updates, key rotation, rollback protection, and long-term trustworthiness across heterogeneous edge DePIN devices.
    \item \textit{Secure Design and Standardization}: Adopting the ``security-by-design'' principle, efforts should be directed to developing security certification standards and unified protocols for DePIN devices to establish a consistent security baseline across heterogeneous manufacturers.
\end{itemize}

Several issues remain to be investigated, including the performance overhead and feasibility of deploying TEE on  resource-constrained edge devices, the stability and uniqueness of PUF under large-scale deployment, and the effectiveness of countermeasures against side-channel attacks.

\subsection{Cross-Chain and Cross-Domain Collaboration Mechanism}
\textit{Challenges}:
DePIN spans multiple chains and domains, where blockchain data silos hinder cross-domain flows of data, assets, and users.
At the same time, the topology, state, and performance of physical infrastructure must be efficiently and reliably synchronized with the on-chain logical world \cite{Hazra_challenges_ACM}.

The potential research directions are outlined as follows.
\begin{itemize}
    \item \textit{Physical-Logical Collaboration Topology Management}: Develop frameworks capable of dynamically and reliably mapping physical infrastructures, such as hotspot locations and sensor networks, into their on-chain digital twins. This requires integration of decentralized oracle mechanisms and IoT to enable real-time verification and on-chain synchronization of physical states.
    \item \textit{Heterogeneous Network and Device Abstraction Layer}: Given the diversity in network protocols and device capabilities, future research should design a unified abstraction layer and self-organizing network architecture. This layer can shield upper-layer applications from underlying heterogeneity, provide standardized interfaces, and enable automatic resource discovery and coordination.
\end{itemize}

To ensure trustworthy integration between the physical and digital domains and achieve interoperability, coordinated standardization and alignment efforts are crucial. 
In particular, attention should be directed towards 1) Defining standards for cross-chain interoperability protocols; 2) Establishing data formats and verification frameworks for the on-chain integration of physical device information; and 3) Developing unified models for resource scheduling and accounting across heterogeneous networks.

\subsection{Post-Quantum Security and Cryptography }
\textit{Challenges}:
    Post-quantum security is an important consideration for DePIN, as adversaries with access to quantum computing may eventually weaken the classical cryptographic foundations of decentralized authentication and trust establishment.
    For DePIN, however, the key challenge lies in making post-quantum cryptographic (PQC) mechanisms practical for large-scale decentralized infrastructures with heterogeneous and unreliable nodes.
    Future research should focus on how PQC mechanisms can be made verifiable, lightweight, formally secure, and smoothly deployable.
    \begin{itemize}
        \item \textit{Quantum-Resistant Authentication and Signing}: Given that future large-scale quantum computing may break the public-key cryptographic foundations of identity authentication and transaction authorization~\cite{Wang_quantum_TDSC}, quantum-resistant authorization and signing will become critical for DePIN. This requires suitable post-quantum signature and key establishment schemes for repeated node authentication, examining their compatibility with existing DID and off-chain device credentials, and mitigating the practical overhead caused by larger-sized keys and more expensive cryptographic operations.
        \item \textit{Lightweight Distributed Post-Quantum Verification}: Particularly relevant directions include batch verification, compact proof structures, lower key exchange overhead, caching of session keys for repeated authentication, and optimization tailored to edge environments with limited bandwidth or strict latency requirements~\cite{Kumar_quantum_IoTJ}.
        \item \textit{Formal Security Analysis}: Rather than assuming that PQC primitives alone guarantee end-to-end security, more efforts should be devoted to the formal analysis of protocol correctness, composability, replay resistance, downgrade resilience, and state consistency across both on-chain and off-chain interactions.
        \item \textit{Hybrid PQC Framework}: A fully post-quantum migration is unlikely to occur abruptly in DePIN, since its devices and protocols are heterogeneous and often remain deployed for long periods. Hybrid classical post-quantum frameworks are a practical necessity for preserving service continuity and trust compatibility during transition~\cite{Turnip_quantum_COMST}.
    \end{itemize}

    Post-quantum security and cryptography in DePIN still face several unresolved challenges:
    The long-term feasibility of quantum-resistant trust in heterogeneous DePIN ecosystems,
    the absence of standardized PQC benchmarks and
    quantum-secure consensus primitives.

\subsection{Future Communication and Network Intelligence}
\textit{Challenges}:
The traditional communication paradigm based on Shannon's theorem faces limitations in energy and spectrum efficiency when applied to massive, heterogeneous, and dynamic DePIN nodes. Moreover, the absence of centralized control entities makes global resource optimization and autonomous network management exceedingly challenging.

As shown below, to enable large-scale, self-organizing, and intelligent DePIN ecosystems, several techniques may be considered for integration into future communication networks.
\begin{itemize}
    \item \textit{AI-Native Networks}: Moving beyond the auxiliary role of ``AI-empowered networking,'' future research is likely to treat AI as a core and intrinsic capability in network architecture design \cite{Rossi_challenges_TNSM}.
    Future DePIN networks will rely on distributed learning and reasoning to achieve intent-driven operations, resource scheduling, and topology optimization without centralized orchestration.
    The emergence of LLM and agentic AI can serve as the network's ``cognitive engine'', interpreting and translating high-level intents into configuration commands for distributed agents~\cite{Boateng_LLM_COMST}.
    Future directions include on-chain coordination for AI-driven network slicing \cite{Li_slice_TCOM}, real-time anomaly detection and self-healing mechanisms, with the goal of achieving Level-5 fully autonomous network operations, a prerequisite for viable and scalable DePIN.

    \item \textit{Semantic and Task-Oriented Communications}: This direction explores new paradigms that transcend bit-level transmission \cite{Ma_TTE_COMST, Zou_SC_TNET}. Semantic communication enables efficient coordination among DePIN nodes by transmitting only task-relevant information, thereby reducing communication overhead and improving robustness in dynamic environments.
    LLM and agentic AI are ideal candidates for realizing semantic communication \cite{Li_IoTJ, Jiang_LLM_arxiv}. LLMs can extract and encode the semantic essence of data, while agentic AI can interpret semantic cues for context-aware decisions.
    This can empower DePIN to integrate communication, computation, and sensing  efficiently across the physical infrastructure layer.

    \item \textit{Quantum-Enhanced Communication}: For DePIN architectures involving cross-domain collaboration among untrusted nodes, quantum technologies offer a potential leap in communication security and reliability \cite{Zhou_Quantum_COMST}.
    As a forward-looking research direction, efforts should be devoted to how quantum phenomena, e.g., entanglement or key distribution, can be leveraged to provide DePIN with unconditional secure transmission guarantees.

\end{itemize}

The integration of LLMs and agentic AI raises challenges in edge computational overhead, lightweight fine-tuning and distillation, and the stability and interpretability of AI-native decision making in large-scale distributed DePIN environments. 
Other open issues include the generalizability of semantic communication across heterogeneous tasks and modalities, and the engineering feasibility of quantum communication in practical systems.

\subsection{Token Value Management and Economic Sustainability}
\textit{Challenges}:
The inherent high volatility of cryptocurrencies poses a threat to the stability of DePIN networks. Large fluctuations in token prices can distort incentive models, affecting user participation and the pricing stability of network services.

To address this challenge, potential mitigation strategies may merit further exploration.
\begin{itemize}
    \item \textit{Functional Decoupling and Stablecoin Integration}: Investigate the separation of functions, e.g., payment medium, governance rights, and value storage, across different tokens \cite{Goldstein_Token_JF}. Explore hybrid economic models that incorporate fiat-backed or over-collateralized stablecoins as payment mechanisms \cite{Mita_challenges_IIAI}, providing transactional stability while maintaining incentive effectiveness.
    \item \textit{Adaptive Community Governance}: Design advanced DAO governance frameworks that enable communities to dynamically adjust key economic parameters based on on-chain data and voting outcomes, facilitating the continuous evolution of economic models.
\end{itemize}

Given the regulatory gaps in decentralized networks and the volatility of cryptocurrency values, it is essential to educate participants on the design principles and risks of token economics, and provide project teams with best-practice guidelines for compliant token issuance and management.

\subsection{DePIN-as-a-Service (DaaS) and Developer Experience}
\textit{Challenges}:
The current interaction with DePIN protocols involves high technical barriers and complex procedures, hindering large-scale adoption by traditional developers and enterprises.
To enable an open, accessible, and developer-friendly DePIN ecosystem, several aspects remain to be improved.
\begin{itemize}
    \item \textit{Unified APIs}: Develop a mature DaaS platform that abstracts the complexity of underlying multi-chain protocols, token payments, and resource scheduling through unified APIs or SDKs. Developers can access decentralized storage, computing, or other network services via simple interfaces without the need to understand the underlying implementation details.
    \item \textit{Composable Service Stack}: Investigate methods for encapsulating diverse DePIN services, e.g., Filecoin storage, Helium networking, Render computing, into composable digital assets, enabling one-click deployment and elastic scalability across heterogeneous infrastructures.
    \item \textit{User Awareness and Education}: Develop intuitive development tools, comprehensive documentation, and simulation test environments to reduce the learning curve for developers. For end users, design seamless Web2-like interaction interfaces that abstract the underlying blockchain complexity and enhance usability.
\end{itemize}

Beyond these directions, several key areas require standardization: Interface standards for DaaS are needed to enable interoperability and service compatibility across different DePIN platforms; cross-DePIN protocol metrics for QoS and SLA are needed to provide a consistent framework for performance evaluation and service assurance.

\subsection{Real-World Deployment Feasibility}
\textit{Challenges}:
The real-world deployment of DePIN is constrained by the broader challenge of converting decentralized participation into reliable infrastructure operation.
In practice, the central challenge is moving DePIN beyond feasible architectures and pilot demonstrations toward sustained deployment, where technical scalability, accountable operation, regulatory viability, and economic sustainability must hold under real-world conditions~\cite{ZHENG202551}.

\begin{itemize}
    \item \textit{Hardware Deployment and Long-Term Maintenance}: A key requirement for real-world deployment is to prevent device heterogeneity from translating into fragmented maintenance responsibility, rising operational expenditure, and unstable service quality. Future work should focus on modular maintainability, remote lifecycle management, automated upgrade mechanisms, and operational models that reduce the cost of maintaining large-scale heterogeneous devices.
    \item \textit{Performance Guarantees and Service Continuity}: Loose coordination is insufficient for practical DePIN once services are expected to operate with infrastructure-level reliability. This calls for research on multi-source QoS measurement and proof, fault tolerance through redundancy, layered SLA mechanisms, and coordinated designs in which control remains off-chain while settlement is anchored on-chain.
    \item \textit{Regulation and Policy}: DePIN often spans heterogeneous legal and institutional environments with varying data governance, communications regulation, and accountability requirements. Compliance in small pilots may not transfer smoothly to cross jurisdictional deployment or large-scale commercialization. Future work should therefore develop modular compliance, domain-specific governance, accountable identity and reputation systems, and transparent dispute resolution to preserve legal enforceability and accountability clarity.
    \item \textit{Long-Term Economic Sustainability}: 
    Early token incentives may help DePIN bootstrap, but they do not by themselves guarantee sustainable deployment. If service supply and user retention cannot be sustained by real demand, node participation may decline as rewards weaken, undermining both coverage and continuity.
    This points to the need for a tighter coupling between verifiable service value and revenue capture, as well as sustainable feedback loops linking service demand, infrastructure maintenance, and economic return.
\end{itemize}

The real-world deployment feasibility is the stage where DePIN must prove its maturity as an infrastructure paradigm. Its long-term value depends on the ability to organize distributed physical contributions into a stable socio-technical system, in which service quality, governance execution, regulatory adaptation, and economic incentives can reinforce one another over time.
Its broader significance lies in showing how decentralized networks may evolve from incentive-driven expansion toward dependable infrastructure that can be deployed, trusted, and maintained in the real world.

\section{Conclusion}\label{sect:conclusion}
This survey provides a holistic view of DePIN, spanning its multi-layer design (physical, blockchain, interaction, trust, incentive, and application) and highlighting the interplay between technological innovation and socio-economic mechanisms.
From the hardware layer to the application ecosystem, DePIN can potentially transform resource utilization, enhance user sovereignty, and stimulate community-driven innovation.
The evolution of AI-native communication networks, quantum-secure channels, and compliance-aware governance frameworks can further shape DePIN.
The remaining challenges include the security and verifiability of physical devices, the interoperability of heterogeneous networks, the stability of tokenized economies, and the usability of developer tools.
Future research on DePIN can focus on cross-layer co-design, standardization, and real-world validation to bridge the gap between conceptual design and operational reality. By integrating cryptographic primitives, intelligent networking, and sustainable economic models, DePIN can serve as a cornerstone for Web 3.0, empowering individuals, communities, and enterprises to co-create an open, resilient, and intelligent digital-physical ecosystem.

\section*{Acknowledgment}

The authors would like to express their sincere gratitude to Prof. Dusit Niyato from Nanyang Technological University for the constructive comments and valuable suggestions that helped improve the quality of this paper.

\bibliographystyle{IEEEtran}
\bibliography{ref_modified}

\end{document}